\documentclass{aa}

\usepackage{graphicx, txfonts}

\newcommand{\msolyr}{{M$_{\odot}$}\,yr$^{-1}$ }

\begin{document}

%%% referee version
%\bdouble

\title{Near-infrared spectroscopy of AGB star candidates in
Fornax, Sculptor and NGC\,6822\thanks{Based on observations collected at the 
European Southern Observatory, Paranal, Chile (ESO programme 75.D-0152).
Figure~6 is available in electronic form via http://www.edpsciences.org
}
}

\author{
M.A.T.~Groenewegen
\inst{1} 
\and
A.~Lan\c{c}on
\inst{2} 
\and
M. Marescaux
\inst{3}
}

\offprints{Martin Groenewegen (marting@oma.be)}

\institute{
Koninklijke Sterrenwacht van Belgi\"e, Ringlaan 3, B--1180 Brussels, Belgium \\ \email{marting@oma.be}
\and
Observatoire Astronomique, Universit\'e de Strasbourg \& CNRS (UMR 7550), 11 rue de l'Universit\'e, F--67000 Strasbourg, France
\and
Instituut voor Sterrenkunde, Katholieke Universiteit Leuven, Celestijnenlaan 200 D, B--3001 Leuven, Belgium 
}

\date{received: 2009,  accepted:  2009}

%
%\authorrunning{Groenewegen}
\titlerunning{Near-infrared spectroscopy of AGB star candidates}

\abstract {The Asymptotic Giant Branch (AGB) phase is characterised by
  substantial mass loss that is accompanied by the formation of
  dust. In extreme cases this will make the star no longer visible in
  the optical. For a better understanding of AGB evolution it is important 
  to identify and characterise these very red AGB stars.}
{The first aim of this article is to improve the census of red AGB stars 
  in three Local Group galaxies, based on near-IR spectroscopic
  observations of new candidates with red 2MASS $(J-K)$ colours. 
  The opportunity is taken to compare the near-IR spectra with
  those of Milky Way stars.}
{We used ISAAC on the ESO VLT to take $J$ and $H$-band spectra of 36 targets
  in Fornax, Sculptor and NGC 6822.}
{Twelve new C-stars are found in Fornax, and one is confirmed in Sculptor. 
All C-stars have $(J-K) > 1.6$, and are brighter than $-3.55$ in
bolometric magnitude.
Ten new oxygen-rich late-type giant stars are identified in Fornax,
but none is extremely red or very luminous. Five luminous O-rich AGB
stars are identified in NGC 6822, of which 3 show water absorption,
indicative of spectral type M. Again, none is as red as Milky Way
OH/IR stars, but in this galaxy the list of candidate AGB stars is
biased against very red objects.
In some C-stars with $(J-K)>2$ an extremely strong 1.53 $\mu$m absorption 
band is found. These stars are probably all Mira variables and the feature 
is related to the low temperature, high density chemistry that 
is a first step towards dust formation and mass loss.
}
{}


\maketitle

\section{Introduction}

Asymptotic Giant Branch (AGB) stars play an important role in stellar
population studies (e.g. Lan\c{c}on, 1999) and galactic structure
studies (e.g. Dejonghe \& Caelenberg, 1999), and through their mass
loss they contribute significantly to the enrichment of the
interstellar medium (Habing, 1996; see Kerschbaum et al. 2007
for recent reviews).  The AGB phase is the final
evolutionary stage for more than 95\% of all stars that leave the main
sequence (MS) in a Hubble time.  Their intrinsically high luminosities
and well defined evolutionary stage make them important constituents
and probes of extragalactic systems.  Due to their old age they define
highly relaxed subsystems.  At the same time, the extragalactic
studies are important for our understanding of AGB-evolution itself.

Even though the AGB evolution is relatively well understood, there
remain a number of important unsolved questions, such as the MS mass
interval for which carbon star formation is possible (as a function of
metallicity), or the mass loss dependence on time, mass, metallicity.
By observing AGB populations at known distances and of different
metallicity light can be shed on these and other related questions.

One of the most efficient methods for the identification of AGB stars
in extragalactic systems uses two narrow-band filters at 7800 and 8100
\AA, centred on a CN-band in carbon stars (and near-continuum in
oxygen-rich stars), and a TiO band in oxygen-rich stars (and continuum
in C-stars), respectively, and two broad-band filters from the set
$V,R,I$. In an [78-81] versus $[V-I]$ (or $[R-I]$) colour-colour plot,
carbon stars and late-type oxygen-rich stars clearly separate redwards
of $(V-I) \approx$ 1.6.  For an illustration of this, see Cook \&
Aaronson (1989) or Nowotny \& Kerschbaum (2002). This narrow-band
filter system was introduced by Wing (1971) and Palmer \& Wing (1982)
and then first applied by Richer et al. (1984) and Aaronson et al. (1984). 
In recent years several groups have used this technique to survey many
galaxies of the Local Group (LG), see e.g., Battinelli \& Demers (2005a,b) 
for a summary of results of their surveys of more than 10 LG
galaxies over the last 6 years, Rowe et al. (2005), Nowotny et al. (2001, 2003) 
or Harbeck et al. (2004).

The narrow-band surveys identify AGB stars in the optical. However, as
evolution on the AGB proceeds, the mass loss rate increases
(culminating in the so called ``superwind'' phase) and hence the
obscuration of the central star by dust formation becomes increasingly
important. In fact, at some point, AGB stars become ``invisible'' in
the optical. This phenomenon has been relatively well studied in the
Galaxy, LMC, and SMC ever since it was first highlighted by the IRAS
satellite, and with subsequent observation by the ISO satellite
(e.g. Groenewegen 1995; Trams et al. 1999; Whitelock et al. 2003), and
Spitzer (e.g. Groenewegen et al. 2007). However, in none of the other
LG galaxies has this effect been properly studied, although Menzies et
al. (2002) have discovered 5 very red objects in Leo {\sc i} which
they believe to be dust-enshrouded AGB stars, and recently other
ground-based NIR studies have appeared (Cioni \& Habing (2005a,b),
Davidge (2005), Kang et al. (2005, 2006) and Sohn et al. (2006)) that
identify a candidate AGB star population in a few LG galaxies based on
colours, and Matsuura et al. (2007) presented Spitzer IRS spectra for
a few AGB stars in Fornax.  
{\bf The present paper tries to lift some of these optical biases by
  considering a sample based on 2MASS data, that will allow the detection of mass-losing stars.}

In this paper we present NIR medium resolution spectroscopy 
of new AGB star candidates in the LG galaxies Fornax, Sculptor
and NGC\,6822, selected from 2MASS. 
Very few such spectra are available in the literature. 
In Sect.~2 the sample selection is outlined, and
in Sect.~3 the colour-magnitude diagram is discussed. Section~4
discusses the observations, and Sect.~5 describes the spectra and
classification. Section~6 presents a brief discussion of the 
stellar properties in the framework of the galaxy histories. 
We also highlight the behaviour of the interesting 1.53\,$\mu$m
feature observed in the $H$-band spectra of some of the carbon stars.

\section{Sample selection}

The all-sky release of the 2MASS $JHK$ survey (Cutri et al. 2003)
offers a unique opportunity to detect the reddest AGB stars 
in Local Group galaxies. 
%as the limiting magnitudes of that survey are such that AGB stars
%can be found to very large distances. 
%One of the authors (MM) 
%as part of his undergraduate thesis,  investigated 
%
We searched the 2MASS point source catalog in the direction of all LG
galaxies within 1\,Mpc of the Milky Way (excluding the SMC, the LMC,
M31, M32 and M33), and selected sources with de-reddened colour index
$(J-K)_0>1.22$. This ensures that most carbon stars are selected. Also
selected are unobscured O-rich AGB stars with spectral type $\sim$M5 and 
later (C. Loup 2003, private communication based on the 
M-star models of Fluks et al. 1994), or reddened objects.
Because spectral type changes rapidly over a small interval in (J-K),
uncertainties in the 2MASS colours are likely to lead to the inclusion
of stars of earlier spectral types than M5 in the sample, especially near the 
completeness limit of 2MASS.
% COMMENT by AL : I prefer to avoid the discussion on spectral types.
% From (longish) searching, I believe that indeed low metallicity
% giants of a given J-K will have a much earlier spectral type than M5,
% but that involves playing around with unpublished low metallicity models
% and making assumptions on what spectral types mean at low metallicity... Too
% difficult to describe briefly.
%The 2MASS colour-magnitude diagrams of the Large and
%Small Magellanic Clouds (Nikolaev \& Weinberg 2000,
%Marigo et al. 2008) show that the colour-cut also retains
%some of the stars of the optically visible O-rich AGB sequences.
%which are broadened by pulsation, differential reddening and observational errors.

We further restricted the sample to objects fainter than
$M_{\rm bol}=-7$ in order to exclude foreground stars. For this
purpose, first estimates of the bolometric luminosities were obtained
using the bolometric correction at $K$ derived by Bessell \& Wood (1984)
from star observations in 47\,Tuc and the SMC.
We rejected sources with errors larger than
$0.12$ magnitudes on either $J$ or $K$, as well as objects listed
as non-stellar in the SIMBAD database\footnote{Operated at
CDS, Centre de Donn\'ees Astronomiques de Strasbourg}. 
One of the main purposes of a first spectroscopic follow-up being
the obtention of a more complete census of all the AGB stars 
in the target galaxies, we flagged spectroscopically confirmed
AGB stars from the literature and kept only the remaining objects
as new AGB star candidates. A total of 109 candidate AGB stars
were found in 16 galaxies, as summarised in Groenewegen (2006).

%We further restricted the sample to objects fainter than $M_{\rm bol}= -7$ 
%in order to exclude foreground objects
%% [AL- not also brighter than the tip of the RGB?]. 
%%  MG, nope. The early-AGB can be fainter than TRGB
%% AL : but in the following section you restrict yourself to brighter than TRGB
%% !!!!!!!!!!!!!!!!???????????
%This first estimate of the bolometric 
%luminosity used the bolometric correction for O-rich stars at $K$ from Bessell
%\& Wood (1984). Finally, we rejected 2MASS sources with errors larger 
%than $0.12$ magnitudes on either $J$ or $K$, and we double-checked
%with the SIMBAD database\footnote{Operated at CDS, Centre de Donn\'ees 
%Astronomiques de Strasbourg, France} for known non-stellar objects. 
%A total of 109 candidate AGB stars were found in 16 galaxies. 
%The results are summarised in Groenewegen (2006). 

Originally, spectroscopic observations were planned on 
the 34 AGB star candidates in Fornax only, 
but due to the high airmass of Fornax at the start of the
scheduled nights we were allowed to observe the AGB star candidates in
NGC 6822 and Sculptor. The AGB star candidates in these three galaxies 
and their near-IR magnitudes are listed in Table~1 with
their 2MASS identifications. In the second column we list the ``internal'' 
names used hereafter in the figures and text.

In Table~\ref{TAB-targets}, bolometric magnitudes are also given.
These values take the near-IR spectral classification of 
Sect.\,\ref{observations.sec} into account. The relationship
between bolometric correction and $(J-K)$ is taken from
Bessell \& Wood (1984) for O-rich stars, and from Whitelock et al. (2006)
for C-stars (the latter correction is systematically larger than the former by
about 0.2\,magnitudes). 
% Whitelock et al. 2006:
% BCK= +0.972 + 2.9292 x -1.1144 x^2 +0.1595 x^3 -9.5689(-3) x^4
% Bessell \& Wood (1984):
% 0.72 +2.65 x -0.67 x^2
%
To Fornax a distance modulus (DM) of 20.72 $\pm$ 0.04 (Rizzi 2007) is adopted, based on the TRGB,
Red Clump (RC) and the Horizontal Branch (HB).
To Sculptor a DM of 19.66 is adopted, the mean of 19.67 $\pm$ 0.12
based on RR Lyrae stars (Pietrzy\'nski, et al. 2008), 19.64 $\pm$ 0.08
based on the TRGB (Rizzi 2002) and 19.66 $\pm$ 0.15 from the HB (Rizzi 2002).
To NGC 6822 a DM of 23.34 is adopted, the mean of 23.31 $\pm$ 0.06 based
on Cepheids (Gieren et al. 2006), 23.36 $\pm$ 0.17 based on RR Lyrae
stars (Clementini et al. 2003), and 23.34 $\pm$ 0.12 based on the TRGB (Cioni \& Habing 2005b).
The photometry was de-reddened using $A_{\rm V}$ = 0.067
(Fornax), 0.059 (Sculptor), 0.784 (NGC 6822) (Schlegel et al. 1998),
and ratios $A_{\rm J} / A_{\rm V}$ = 0.27, and $A_{\rm K} / A_{\rm  V}$ = 0.11.

\section{The colour-magnitude diagram}

Near-IR colour-magnitude diagrams are shown here for the target
galaxies and the nearby field.
The observations described in the next section lack the spectral
resolution to determine radial velocities. The question whether a source 
is part of the galaxy or is a contaminant is therefore of concern. 

\begin{figure}
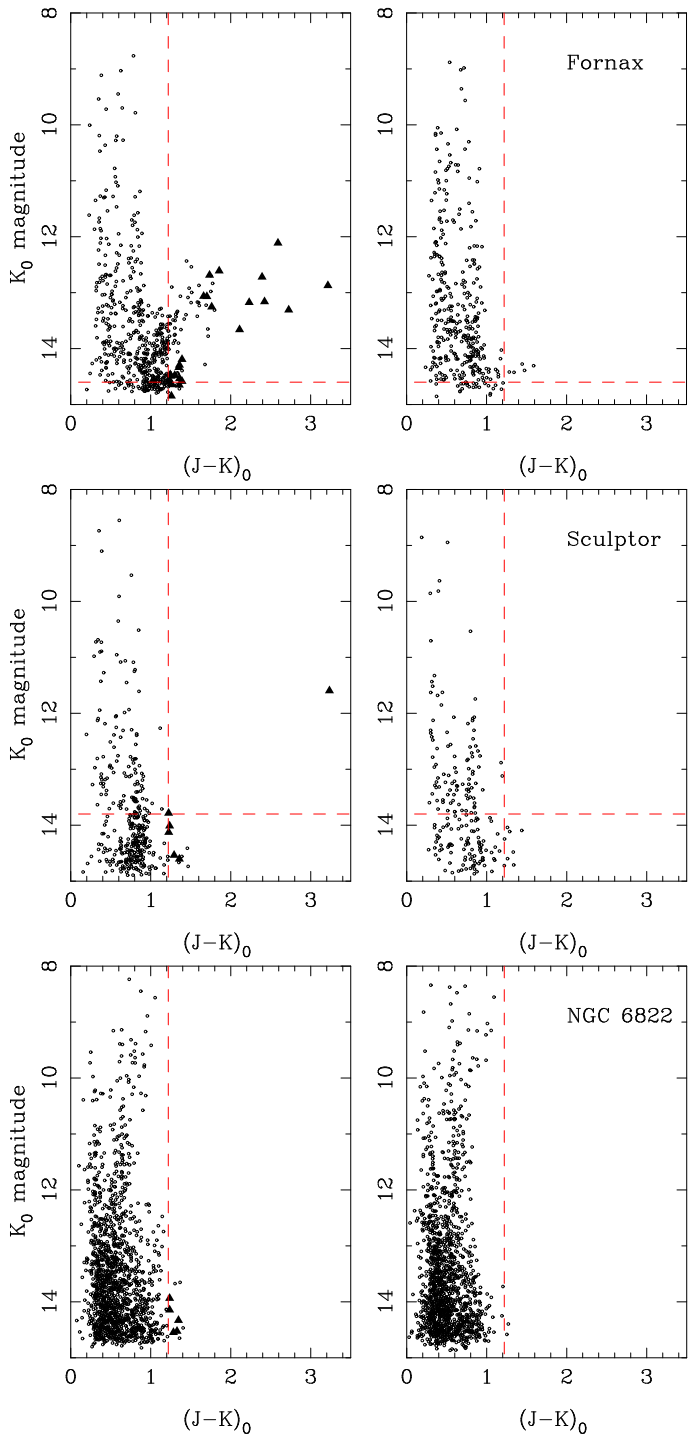


%\begin{minipage}{0.49\textwidth}
%\resizebox{\hsize}{!}{\includegraphics{cmd_for.ps}}
%\end{minipage}
\includegraphics[clip=,width=0.49\textwidth]{cmd_for.ps}
% NB : the "clip=" tells includegraphics that yes, it should use the 
%      BoundingBox provided in the postscript file.
% You can add an angle=90 (or the like) before the width command if you
%      want to rotate your figure. You can also set the height instead of
%      or in addition of the width, but that's usually less convenient.

\begin{minipage}{0.49\textwidth}
\resizebox{\hsize}{!}{\includegraphics{cmd_scl.ps}}
\end{minipage}

\begin{minipage}{0.49\textwidth}
\resizebox{\hsize}{!}{\includegraphics{cmd_n6822.ps}}
\end{minipage}

\caption[]{
Colour-Magnitude diagram for Fornax, Sculptor and NGC 6822 (left) and their surroundings (right).
Stars for which a spectrum was obtained are marked by the filled triangles.
The dashed vertical line indicates the lower limit of $(J-K)_0= 1.22$ used in selecting late-M and C-stars.
The dashed horizontal line indicates the tip of the RGB. For NGC 6822 it is located at $K_0$= 17.0, beyond 
the limit of the plot. (see text).
}
\label{Fig-cmd}
\end{figure}

Sources were extracted from the 2MASS catalog at the central position
of the galaxies, and at two locations 1.5 degrees above and below in
declination. Radii of 25$'$, 20$'$ and 17.5$'$ were used respectively
for Fornax, Sculptor and NGC 6822, both for the central and for the
neighbouring positions (about 2-3 core radii following Mateo, 1998),
as used for the selection described in the previous section.
Figure~\ref{Fig-cmd} shows the colour-magnitude diagrams for the
central and the off-position (only every second star is shown
in the combined diagram of the two off-positions for each galaxy). 
Stars with photometric errors above 0.12 mag in either
$J$ or $K$ are rejected. 

Stars brighter than the tip of the RGB (TRGB) are those of 
interest in considering the AGB nature of the targets observed.
The TRGB is located at $K=$\,14.61 $\pm$ 0.04 (Gullieuszik et al. 2007) 
in Fornax, or $K_0\,=$\,14.60.
In Sculptor it is located at $K_0=$ 13.8 (Babusiaux et al. 2005).  In
NGC 6822 it is found at $K=$ 17.10 $\pm$ 0.01 (Cioni \& Habing 2005b),
or about $K_0=$ 17.0.  
{\bf This implies a selection bias in the case of NGC 6822 as only the
  very brightest AGB stars can be observed given the limiting magnitude of 2MASS. }

Counting the number of sources brighter than the TRGB and redder than
1.22 in $(J-K)$, the statistical contamination is about 7\%
for Fornax (5/71). It is higher in NGC 6822 (2/9) and basically undetermined
in Sculptor (0/2).

%% Fornax 80 - 5 
%% Scl    10 - 8
%% N6822   9 - 4

%%checked all stars J-K>1.22 in the off-position with SIMBAD, none is known.

\section{Observations}
\label{observations.sec}

The observations took place in visitor mode on the nights of 23 - 27
September 2005 using ISAAC (Moorwood et al. 1998) on ESO's VLT UT1 telescope.

Apart from the AGB star candidates we targeted a few late-type stars
of known spectral type in Fornax and Sculptor (see Table~1 for the
references), and 2 known M-dwarfs.
The observations were taken in low-resolution mode (i.e. a resolution
of about 500 for a 1\arcsec\ slit) centred at 1.25 and 1.65 $\mu$m,
respectively, to cover the entire $J$ and $H$-band atmospheric windows.
The on-chip integration time was 120 seconds for the science targets
and 5 seconds or less for the telluric standard star observations. 
A standard nodding technique was used with a throw of 20\arcsec. To
build up signal-to-noise 1-6 nod cycles were taken depending on the
brightness of the target. A slit of 1\arcsec\ was used.
A telluric standard was observed immediately after each science target. 
It was taken from the Hipparcos database in the direction of
the science target and was preferentially a hot star (B8 or earlier
spectral type), or otherwise a G2 solar analog.

The basic data reduction was done using the ESO offered ISAAC data
reduction pipeline (version 5.4.2) for flat-fielding, wavelength
calibration (using the sky lines), correction for the slit curvature
and the star trace, combination of the nod positions and cycles and
the detection and extraction of the final spectrum, that was written
to an ascii file.

The task of removing the telluric features was done in a separate
Fortran code. The basic idea is to divide the observed spectrum of the
science target by that of the telluric standard and multiply the result by the
intrinsic spectrum of the standard.

A high-resolution near-infrared spectrum of the Sun was taken from
Castelli et al. (1997), and those for early-type stars were kindly
calculated and made available by Peter Hauschildt.  An approximate
absolute flux level was achieved by scaling the theoretical spectrum
for the telluric standards to a flux level based on the 2MASS $J$ and
$H$ magnitudes, but no corrections for slit losses were made.

The high-resolution spectra of the standard stars were convolved to
lower resolution matching the ISAAC observations.  To further judge
the quality of the removal of the telluric feature we convolved the
high-resolution telluric spectrum available at the ESO
website\footnote{http://www.eso.org/sci/facilities/paranal/instruments/isaac/tools/spectroscopic\_standards.html}
to the resolution of the observations and compared it to the final
science spectrum in particular in parts of the spectrum with strong
telluric features.

The final spectra are shown in Figure~\ref{Fig-ISAAC}, which is
available in its entirety in electronic form via
http://www.edpsciences.org. The ordinate is in W/m$^2/\mu$m and is on
an approximate absolute level. 
%The carbon stars in Fornax are plotted first, 
%ordered by increasing $(J-K)$ colour, then the C-stars in
%Sculptor (including the 2 previously known reference objects).
%Subsequently the O-stars are plotted by galaxy, and by increasing
%$(J-K)$ colour, followed by the two galactic dwarfs. Finally outliers are
%appended for completeness: one object
%with an odd $H$-band spectrum (Fornax 16), and confirmed or 
%likely low-redshift galaxies.

% awk '{print "2MASS"$12, " & Fornax"$4, "&", $13, "&", $17, "&", $21}' fp_2mass.fp_psc11085.tbl        

% awk '{print $2-$3, $0}'  2mass.dat

\begin{table*}
\caption{AGB star candidates and targets observed. } \label{TAB-targets}

\setlength{\tabcolsep}{1.6mm}

\begin{tabular}{rlrrrrrl} \hline
2MASS identifier & other  & $J$ & $H$ & $K$ & $M_{\rm bol}$ & spectral & remarks; (references) \\
 & identifier &  & & &  & type &  \\
\hline
2MASS02380397-3420196  & Fornax1 & 15.764 & 14.836 & 14.499 & -3.24 & K3-K4: & \\ 
2MASS02401824-3428412  & Fornax2 & 15.914 & 15.136 & 14.542 & -3.10 & G8-K4: & \\
2MASS02401650-3415374  & Fornax3 & 16.122 & 15.137 & 14.853 & -2.88 & G8-K4: & \\
2MASS02414248-3421284  & Fornax4 & 15.797 & 14.912 & 14.485 & -3.21 & K2-K3 & \\  % K2 better when data is dereddened with Av=1
2MASS02400772-3425156  & Fornax5 & 16.195 & 15.460 & 14.845 & & - & not observed  \\
2MASS02394347-3413379  & Fornax6 & 15.981 & 15.024 & 14.292 & & - & not a point source, no spectrum taken \\
2MASS02401171-3431187  & Fornax7 & 16.027 & 15.131 & 14.658 & -2.99 & K2-K3 & \\
2MASS02403857-3428344  & Fornax8 & 15.789 & 14.813 & 14.551 & -3.21 & K3-K4: & \\
2MASS02405606-3431273  & Fornax9 & 15.917 & 15.064 & 14.541 & & - & not observed  \\
2MASS02395732-3431211  & Fornax10 & 15.907 & 15.095 & 14.666 & & - & not observed \\
2MASS02405333-3412130  & Fornax11 & 15.034 & 13.981 & 13.261 & -4.01 & C & 1.53 absent; (7) \\ %SR 230
2MASS02402188-3406309  & Fornax12 & 15.729 & 14.843 & 14.474 & -3.27 & K3-K4 & \\
2MASS02400666-3423222  & Fornax13 & 14.485 & 13.377 & 12.618 & -4.63 & C & 1.53 weak; (1,2,7) \\ %variable w/o period
2MASS02405288-3428303  & Fornax14 & 15.706 & 14.784 & 14.343 & -3.31 & K2-K3 &\\
2MASS02400946-3406256  & Fornax15 & 15.790 & 14.556 & 13.668 & -3.56 & C & 1.53 strong\\
2MASS02401349-3426192  & Fornax16 & 15.930 & 14.794 & 14.652 & -3.07 & not C & H band not usable  \\
2MASS02403123-3428441  & Fornax17 & 14.745 & 13.689 & 13.072 & -4.23 & C & 1.53 weak; (3,5,7) \\ %variable w/o per
2MASS02390184-3427528  & Fornax18 & 15.672 & 14.987 & 14.302 & -     & non stellar & 2MASX source \\
2MASS02405840-3408022  & Fornax19 & 15.994 & 15.030 & 14.591 & -3.03 & K2-K4 & \\
2MASS02401207-3426255  & Fornax20 & 15.131 & 13.732 & 12.728 & -4.53 & C & 1.53 weak; (7) \\ % SR 375
2MASS02401778-3427357  & Fornax21 & 15.424 & 14.122 & 13.182 & -4.05 & C & 1.53 weak; (7) \\ % SR 340
2MASS02400252-3403272  & Fornax22 & 15.855 & 15.157 & 14.608 & -3.15 &  K1-K2 & \\
2MASS02404718-3438510  & Fornax23 & 15.602 & 14.911 & 14.199 & -     & non stellar & 2MASX source \\
2MASS02395421-3438368  & Fornax24 & 15.601 & 14.162 & 13.167 & -4.09 & C & 1.53 weak; (7) \\ % variable w/o period
2MASS02391232-3432450  & Fornax25 & 14.722 & 13.262 & 12.120 & -5.18 & C & very red, features weak, dust emission?; (5,6,7) \\%mira 470
2MASS02394528-3431581  & Fornax26 & 15.941 & 14.989 & 14.667 & & - & not observed \\
2MASS02410355-3448053  & Fornax27 & 14.441 & 13.365 & 12.694 & -4.59 & C & 1.53 absent; (3,4,7) \\ %mira 280
2MASS02402039-3437345  & Fornax28 & 15.930 & 14.875 & 14.649 & & - & not observed \\
2MASS02402554-3441585  & Fornax29 & 15.784 & 14.926 & 14.439 & & - & not observed \\
2MASS02410145-3437154  & Fornax30 & 15.803 & 14.737 & 14.401 & & - & not observed \\
2MASS02380618-3431194  & Fornax31 & 16.052 & 14.483 & 13.315 & -4.03 & C & 1.53 medium\\
2MASS02385700-3446340  & Fornax32 & 14.789 & 13.664 & 13.076 & -4.21 & C & 1.53 absent; (7) \\ %variable w/o period
2MASS02392952-3449431  & Fornax33 & 15.905 & 15.144 & 14.644 & & - & not observed \\
2MASS02385056-3440319  & Fornax34 & 16.106 & 14.525 & 12.879 & -4.70 & C & 1.53 extreme; (5,6,7) \\ % mira 350
                       & Fornax-S66  & 14.768 & 13.901 & 13.642 & & M2S & \# 66  from Stetson et al. (1998) \\
                       & Fornax-S71  & -      & -      & -      & & K5  & \# 71  from Stetson et al. (1998) \\
                       & Fornax-S99  & -      & -      & -      & & S   & \# 99  from Stetson et al. (1998) \\
                       & Fornax-S116 & 15.004 & 14.041 & 14.028 & & SC  & \# 116 from Stetson et al. (1998) \\

\hline
\end{tabular}
\end{table*}

\setcounter{table}{0}
\begin{table*}
\caption{continued }

\setlength{\tabcolsep}{1.7mm}

\begin{tabular}{rlrrrrrl} \hline
2MASS identifier & other  & $J$ & $H$ & $K$ & $M_{\rm bol}$ & spectral & remarks; references \\
                 & identifier &  & & &  & type &  \\
\hline

2MASS01002467-3352012  & Scl1          & 15.024 & 14.280 & 13.791 & -2.92 & non stellar & 2MASX source \\
2MASS01005598-3352344  & Scl2          & 15.977 & 15.290 & 14.610 & -1.97 & K2-K4 & H band curvature odd (background?) \\
2MASS00590481-3343297  & Scl3          & 16.118 & 15.283 & 14.688 & & - & not a point source, no spectrum taken \\
2MASS01004469-3327134  & Scl4          & 15.367 & 14.643 & 14.133 & -2.57 & non stellar & 2MASX, 2dFGRS S503Z077 (z=0.11) \\
2MASS00595925-3337091  & Scl5          & 15.963 & 15.284 & 14.565 & & - & not observed \\
2MASS00595367-3338308  & Scl6          & 14.846 & 13.144 & 11.603 & -4.93 & C & 1.53 strong; (4) \\
2MASS00593953-3332545  & Scl7          & 15.264 & 14.501 & 14.018 & -2.67 & non stellar & 2MASX source \\
2MASS01005213-3343136  & Scl8          & 15.845 & 15.171 & 14.543 & -2.10 & non stellar & 2MASX source \\
                       & Scl-Az1-C     & 14.713 & 14.040 & 13.871 & & C  & ALW1 from Azzopardi et al. (1985) \\ %J00583615-3340258  
                       & Scl-Az3-C     & -      & -      & -      & & C  & ALW3 from Azzopardi et al. (1985) \\ %00 59 58.9 -33 28 34 
                       & Scl-scms-209  & 13.628 & 13.080 & 12.823 & & M4 & from Schweitzer et al. (1995) \\ %J00595802-3345174  
                       & Scl-scms-1448 & 15.167 & 14.497 & 14.340 & & M3 & from Schweitzer et al. (1995) \\ %J00594404-3341158
                       &      \\

2MASS19445931-1446010  & N6822-1  & 16.222 & 15.148 & 14.737 & & - & not observed  \\
2MASS19444196-1440072  & N6822-2  & 15.889 & 14.766 & 14.417 & -5.93 & K4-M & \\
2MASS19450019-1446040  & N6822-3  & 15.189 & 14.234 & 13.760 &  & - & not observed \\
2MASS19443399-1446237  & N6822-4  & 15.593 & 14.519 & 14.231 & -6.22  & M &  H$_2$O absorption\\
2MASS19445849-1448037  & N6822-5  & 16.142 & 15.314 & 14.785 & & - & double star, no spectrum taken \\
2MASS19444529-1451252  & N6822-6  & 16.083 & 15.059 & 14.553 & & - & not observed \\
2MASS19445689-1442381  & N6822-7  & 16.064 & 15.104 & 14.610 & -5.76 & M & H$_2$O absorption \\
2MASS19444914-1449051  & N6822-8  & 16.042 & 14.913 & 14.633 & -5.78 & K4-M & \\
2MASS19444900-1444334  & N6822-9  & 15.230 & 14.204 & 13.738 & & - & not observed \\
2MASS19452775-1453299  & N6822-10 & 15.383 & 14.415 & 14.022 & -6.43 & M & H$_2$O absorption\\
                       &      \\

%2MASS22421460-6426092  & Tuc1     & 15.671 & 15.022 & 14.325 & & spectrum taken, but not point like \\
%                       &      \\

                       & LHS 517   &  6.510 &  5.899 &  5.594 & & M3.5 dwarf &  \\ %redflag=111
                       & LHS 3788B &  9.459 &  8.840 &  8.531 & & M5 dwarf &  \\

\hline
\end{tabular}

1= Demers \& Kunkel (1979);
2= Westerlund et al. (1987);
3= Bersier \& Wood (2002);
4= Mauron et al. (2004); 
5= Matsuura et al. (2007);  
6= Lagadec et al. (2008); 
7= Whitelock et al. (2009)

\end{table*}

\section{Description of the spectra and classification}

\subsection{Classification criteria}

\begin{figure}
\includegraphics[clip=,width=0.49\textwidth]{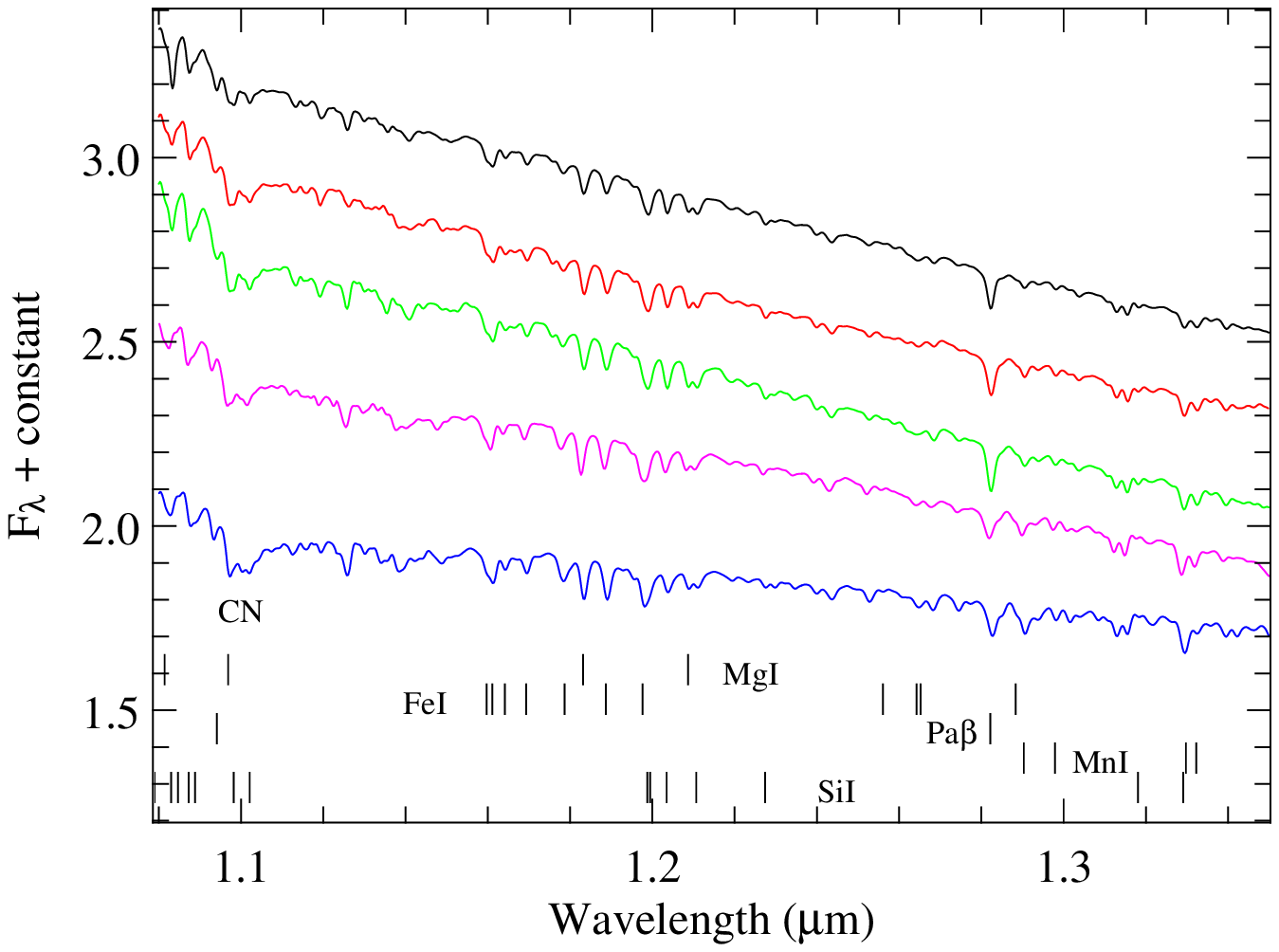}
\includegraphics[clip=,width=0.49\textwidth]{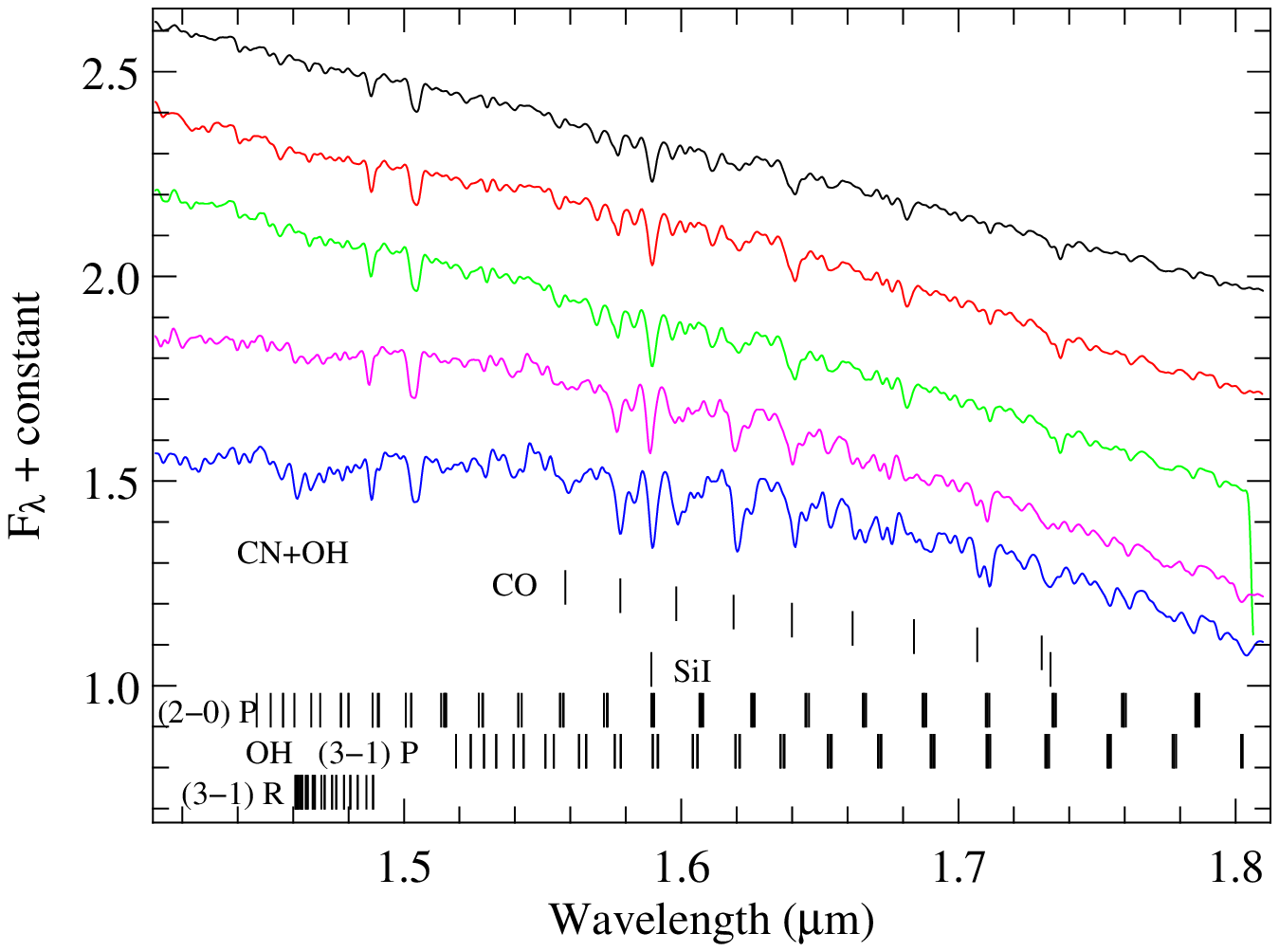}
\caption[]{Spectra of metal-poor giants (Vacca, priv. comm.). Estimated
metallicities from the literature are scattered between [Fe/H]=-0.4 to -1.
Spectral types from top to bottom are G3 (HD\,88639), G8 (HD\,135722), 
G8.5 (HD\,16139), K2 (HD\,2901) and K3.5 (HD\,99998).
OH line identifications are taken from Maillard et al. (1976).}
\label{Vacca.fig}
% Done with plots.i in subdirectory Vacca/
\end{figure}

At the resolution of our observations, the main features seen in the
spectra of luminous red stars are molecular bands, although line and
line blends due to neutral atoms are also present. Spectra useful for
rapid band identifications and classification can be found among
others in Lan\c{c}on et al. (1999) or Lan\c{c}on \& Wood (2000; 
LW2000 hereafter),
in Gautschy-Loidl et al. (2004) for carbon stars, in Joyce et al. (1998) 
for S stars, in McLean et al. (2003) or Cushing et al. (2005) for 
M dwarfs and brown dwarfs.
% see also the ARA\&A review of Kirkpatrick (2005)
% see also Kirkpatrick et al. 1993 for the H-band of M dwarfs.
In the case of O-rich stars, the sensitivity of the spectra to
metallicity is larger than for carbon stars. Our comparison sample
therefore includes six near-IR spectra of Population II giants (type
G2 to K3.5), of which five were kindly provided by W.D. Vacca 
(Fig.\,\ref{Vacca.fig}). Finally,
we may exploit the theoretical and empirical spectra of red
supergiants analysed in Lan\c{c}on et al. (2007), and the 
SpeX/IRTF library of near-IR spectra of Milky Way stars of all types
(Vacca et al. and Rayner et al., in preparation)\footnote{http://irtfweb.ifa.hawaii.edu/$\sim$spex/}.

The features most useful for the classification of our spectra of
O-rich giants are the molecular bands and line blends in the $H$-band
(mostly CO and OH, but also metal lines; see e.g. Origlia et al. 1993,
Fr\'emaux et al. 2006), and the curvature of the spectrum across the
$H$-band (first used for classification purposes by Terndrup et al. 1991). 
As seen in Fig.\,\ref{Vacca.fig}, the (negative) curvature of the
$H$-band increases with decreasing temperature between types G2 and
K4, as a result of the combined effects of the varying importance of
H$^-$ opacity (which has a minimum around 1.5\,$\mu$m) and of CN
absorption (which has a broad band on the short wavelength side of the
H-window). The CN bandhead at 1.1\,$\mu$m, which had been a motivation
for our $J$-band observations, turns out to be difficult to use because
it lies at the edge of the detected spectral range and has poor
signal-to-noise ratios. H$_2$O, when present, is expected to curve the
H-window spectrum even more because of absorption bands on either
side. It also displays a sharp bandhead at 1.34\,$\mu$m. H$_2$O in red
giants is interpreted as a signature of Mira-type long period
variability (LW2000, Tej et al. 2003). It is also
present in low mass dwarfs, which in our case would be foreground
contaminants.  VO and TiO features will show in the $J$-band spectra of
stars very low effective temperatures (late M type).  Evolutionary
tracks do not predict such temperatures at low metallicity, but long
period variability can extend atmospheres enough to produce at least
the VO band at 1.04\,$\mu$m (see e.g. the population II Mira-type
variable S\,Carinae, LW2000).

Carbon stars are characterised by strong bands of the red system of
cyanogen ($\Delta v=0$ bandhead of CN at 1.09\,$\mu$m, $\Delta v=-1$
band responsible for structure throughout the $H$-window, 
e.g. the (1,2) band at 1.45 $\mu$m)
% Ref :Wing & Spinrad 1970
and by the Ballik-Ramsey system of C$_2$ (sharp $\Delta v=0$ bandhead
at 1.77\,$\mu$m, and $\Delta v=+2$ bands at 1.17, 1.20, 1.23 $\mu$m).
% Ref : Ballik & Ramsey 1963, but plotted at more useful resolution in
%       Wing & Spinrad 1970
Extreme carbon stars (in a sense that will be further discussed later) 
display a band around 1.53\,$\mu$m, for which the carrying molecule
has been a matter of debate (e.g. LW2000, Joyce 1998 and
references therein) and that is currently assigned mainly to
C$_2$H$_2$ (Gautschy-Loidl et al. 2004). 
Carbon stars also show the second overtone ro-vibrational CO absorption
band series longwards of 1.56\,$\mu$m. This
series has a particularly regular appearance in S and some
S/C stars (see Fornax-S99 and Fornax-S116 in the present paper and the
S/C star BH\,Crux from LW2000).

\subsection{Background galaxy contaminants}

A few of our initial targets appeared clearly extended 
on the acquisition image and no spectrum was taken as indicated in 
Table~\ref{TAB-targets}.

Six spectra (4 in Sculptor, 2 in Fornax) clearly show a positive
curvature in the $H$-band. One more (Scl2) displays some similarity with
the latter, but the curvature is less pronounced.  At the data
reduction stage the acquisition image of all targets were inspected
and the FWHM of the target and, when available, other stellar sources
in the field, were measured. The evolution of the FWHM of the carbon
stars and the spectral reference sources through the night gave an
indication of the evolution of the image quality, and this showed that
the 6 first odd-spectrum sources are extended.  One was found listed
in the 2dF Galaxy Redshift Survey with a redshift of 0.11: the dip
seen in the $H$-band spectra is due to the combined effects of the H$^-$
opacity minimum near rest wavelength 1.6\,$\mu$m, and molecular bands
(mainly CN) and the Brackett jump near rest wavelength 1.46\,$\mu$m.
All 6 objects are likely low-redshift galaxies.  Scl2 might be one as
well, although it is not resolved (see Morris et al. 2007).

\subsection{Fornax data}

The initial search of 2MASS objects in the direction of Fornax
fulfilling the criteria mentioned in Sect.~2 resulted in a list of 155
known late-type stars, all from the work by Stetson et al. (1998) and references therein. 
Out of the ones with a known spectral type, 19 have spectra type M (or MS), 7 S (or SC), and 39 C
Out of the 34 new AGB stars candidates in Fornax, 25 have been observed. 
We provide classification for 22 stars in Table~\ref{TAB-targets}. 
Of the 3 remaining targets, two are likely low-redshift galaxies, and
the third has such a peculiar, noisy $H$-band spectrum that we do not classify it.
% AL: One has a corrupted H band spectrum, the two others have an unexplained
%  positive curvature in the H band... No explanation... Maybe a star
%  in the offset position of the nodding cycle??
We do not include them in our discussion.

Out of the 22 useful program stars, 11 are O-rich giants and 12 are
carbon stars.  Little was previously known about these objects. Some
of the stars we now classify as C-stars had been previously identified
as very red objects. Since the submission of the observing proposal, one 
Fornax star has been identified as a C-star by Mauron et al. (2004) 
based on optical spectra in a survey for C-stars in the Galactic Halo. 
Matsuura et al. (2007) obtained Spitzer IRS
spectra for 3 objects, and Lagadec et al. (2008) obtained mid-IR
images for 2 objects.
Recently, Whitelock et al. (2009) present the result of a
near-infrared survey of the inner 42$\times$42 arcmin. 
Several of the C-stars in our sample are in their lists of variable
stars (see Tab.\,\ref{TAB-targets}).  None of our targets is among
their selection of non-variable upper AGB candidates.

There is a clear distinction between the C- and the non-C-stars 
we observed in Fornax.  
All stars redder than $(J-K)_0$ = 1.63 are C-stars, all stars bluer
than $(J-K)_0$ = 1.40 are O-rich. A similar distinction exists in 
the bolometric magnitudes: all O-rich stars in our sample are
fainter than $M_{\mathrm{bol}}=-3.42$, while C-stars are brighter than $-3.55$, 
reaching $-5.2$ for the reddest one.

% ./Proposals/esoform-75A/P2PP/Stetson_table1_ALL.dat 

% awk '{if ($13 <= 2.88 && $13 > -1.0) print $12, $13, $14,$15, $1}' Stetson_table1_ALL.dat | sort -k2n                   
% awk '{if ($13 <= 3.25 && $13 > 2.88) print $12, $13, $14,$15, $1}' Stetson_table1_ALL.dat | sort -k2n 
% awk '{if ($13 <= 3.85 && $13 > 3.25) print $12, $13, $14,$15, $1}' Stetson_table1_ALL.dat | sort -k2n  
% awk '{if ($13 <= 9.99 && $13 > 3.85) print $12, $13, $14,$15, $1}' Stetson_table1_ALL.dat | sort -k2n 

%              C  S/SM M/MS  ? ==>  C S  M ==> C  S  M
%>3.85        17   0    0    5      5 0  0    22  0  0 
%>3.25 <=3.85 14   2    0    8      7 1  0    21  3  0
%>2.88 <=3.25  8   4   11   28     10 5 13    18  9 24
%<=2.88        3   2    6   52     14 9 29    17 11 35
%                                             =========
%                                             78 23 59  C/O= 0.95

{\bf What is interesting is the apparent difference between the ratio of C- to
  O-rich stars found by Stetson et al., 39/26= 1.5, and in this study,
  12/11= 1.1. The true ratio is probably even smaller as the $(J-K)$
  colours of the stars {\it not} observed indicate they should be
  O-rich, in which case the C/O ratio drops to 12/19= 0.63 ($\pm$0.22
  if one considers Poisson errors).  Many of the stars in Stetson et
  al. do not have a spectral classification and it appears that these
  are preferentially O-rich stars. To investigate this, we have
  divided the Stetson et al. sample in 4 bins of $(B-R)$ colour and
  counted the number of known C, S/SC and M/MS stars per colour bin. The
  non-classified stars were then classified on \it pro rata 
  \bf basis. 
  The estimated number of C, S/SC and M/MS in the Stetson et
  al. sample of red stars is, respectively, 78, 23 and 59,
  corresponding to a C/O ratio of 0.95 ($\pm$0.14 if one considers
  Poisson errors).  This ratio does not depend on any reasonable
  choice of the colour bins.  The conclusion is that the ratio of C to
  non-C stars in the optical and 2MASS sample are the same within the
  errors.}

%[AL - How does the recent paper of Battinelli, Demers, Mannucci, 2007,
%A\&A 474, 35 fit in? They find a large number of 
%C stars in NGC 6822 with J-K as blue as 1.1. I'm not sure we should keep
%this here: the section is about Fornax, and here information about
%a variety of galaxies is mixed in.]
%The cut in colour is consistent with previous work. For instance, 
%Cioni \& Habing (2005b) and Cioni et al. (2008) adopt a borderline 
%$(J-K)_0$ = 1.36, for stellar samples in NGC\,6822 and M33.
%Groenewegen (2007) argued that a slightly purer sample of M- and
%C-stars could be obtained (in NGC 6822) using a slightly redder cut 
%by comparing the infrared selected C-star sample in 
%Cioni \& Habing (2005b) with that of C-stars found by a narrow-band 
%filter survey by Letarte et al. (2002), 
%in agreement with a limit of $(J-K)_0$ = 1.4 (and $(H-K)_0 >$ 0.45) 
%adopted by Davidge (2005).

In order to provide a  tentative spectral type for the O-rich giants,
we have compared them individually with the reference spectra available. 
{\bf The spectral features in the data, such as the CO
bands, are weak. Poor matches are therefore obtained  with high 
luminosity stars and with most of the spectra of LW2000. On the 
contrary, good matches are obtained with the spectra of Fig.\,\ref{Vacca.fig}. 
We caution that spectral 
types were originally defined based on optical
spectra of solar neighbourhood stars: little is known about the systematics
that affect the relation between spectral type and near-IR properties for cool,
metal poor stars (R.~Gray, 2009).} 
%Based on comparisons with stars of LW2000 and of the IRTF spectral library, 
As an example, the K3.5 giant of our reference sample,
with [Fe/H]$\sim -1$, is about 0.2 magnitudes redder in $(J-K)$
than solar neighbourhood stars of the same spectral type (LW2000 and 
IRTF spectral library). It has very similar near-IR colours and features to
a solar neighbourhood star of type M1III. {\bf Clearly, the 
spectral types assigned here are preliminary and 
metallicity-related systematics would require dedicated studies with large 
samples of optical and near-IR spectra.}  

\begin{figure*}
\includegraphics[clip=,width=0.49\textwidth]{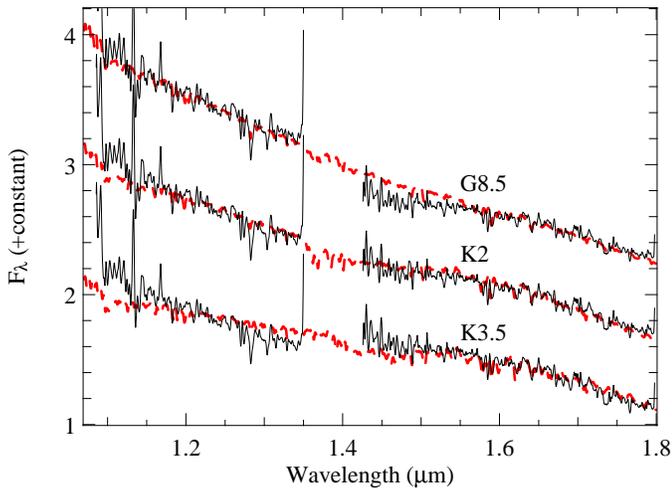}
\caption[]{Illustration of the comparison between a Fornax star spectrum
and the spectra of metal poor giants from Fig.\,\ref{Vacca.fig}. 
The case of Fornax 22 is shown (solid black) with three comparison spectra
(dashed red).  The flux zero point holds for the bottom spectra; a constant has 
been added to the other sets of spectra for display.}
\label{Fx22.fig}
\end{figure*}

In the process of comparison, the $J-$ and $H-$band ISAAC spectra were 
rescaled separately to match the flux
levels of the reference spectra. The scaling factors never differed by
more than 20\,\%, which indicates that slit losses did not vary too
severely between our observations. {\bf An illustration is given
in Fig\,\ref{Fx22.fig}. The curvature of the $H$-band spectrum together
with the slope within the $J$-band are the most discriminant features 
in this limited set of comparisons. At the signal-to-noise ratio of the
data, individual narrow spectral features provide no significant additional 
discrimination. We also explored to what extent reddening
corrections could affect the classification (using the extinction 
law of Cardelli et al. 1989). In about half the cases, equally good
fits were obtained with the data dereddened by up to $A_V \sim 1$, but with 
slightly earlier spectral types. The ranges of spectral types given in
Table~\ref{TAB-targets} are all inclusive.} 
%We note that there could be systematic offsets in this classification,
%due to that fact that the reference sample is not quite as metal poor as 
%Fornax.  .... AL: from what I just read, Fornax isn't that metal poor.

Assigned spectral types extend from G8 to K3.5, 
K3.5 being the limit of our comparison sample of metal
poor giants.  We identify no obvious
candidate for a much later spectral type than K4.  The absence of
H$_2$O features is in agreement with this statement. It argues
against contamination by foreground M dwarf stars. In addition, it
suggests the absence of large amplitude variables in the O-rich sample.

% [AL - I'm wondering about colours. Fluks finds that in
%the Milky Way (J-K)$>$1.2 corresponds to type M5. How can we have stars
%as early as K in our sample?? Is there something funny about the 2MASS
%J and K filters, or those of Fluks? Fluks uses ESO J and K.
%Carpenter 2001 gives colour transformations from 2MASS to 
%all sorts of systems. (J-K)(2MASS) is within 5% of (J-K)(ES0). 
%He cites papers by Bouchet et al. or van der Bliek et al. on ESO 
%photometry, in which it is said that the ESO system is pretty
%similar to others except CIT. So I find nothing to explain more
%than a few percent difference in J-K. Maybe something is wrong with
%the spectral types of Fluks, or with those of B. Vacca...
%Or we see some funny metallicity effect: for a given energy distribution
%(and J-K), metal poor stars could have fainter bands, and therefore 
%earlier spectral types... Yes, that's probably it!
%Bill Vacca's K4III giant at [Fe/H]=-1 is 0.24 magnitudes redder in J-K
% than a solar neighbourhood K3III giant. It compares very well to an M1III
% giant from either LW2000 or the IRTF library. cf tests
% done with both in subdirectory IRTFlib/.]

\subsection{Sculptor data}

The initial search of 2MASS objects in the direction of Sculptor
fulfilling the criteria mentioned in Sect.~2 resulted in a list of 14
knows AGB stars: 8 C-stars from Azzopardi et al. (1985), and 3 C- and
3 M-stars from Schweitzer et al. (1995).
 
Of the 8 candidate AGB stars, 6 targets have been observed in the
Sculptor field, and four are selection errors (low-redshift galaxies
with a characteristic rise in the $H$-band spectrum; one has a known
redshift of 0.11).  One is either an O-rich giant or an unresolved
background galaxy.  One is a carbon star. This is a remarkable object
as it is redder than all 12 C-stars detected in Fornax and the second
brightest object of our whole sample in bolometric magnitude. Its
spectrum is also exceptional, as it displays the strongest
1.53\,$\mu$m feature of all our C-star observations (see
Sect.\,\ref{1.53muFeature.sec}).

%
%The other objects observed in Sculptor have colours ($1.22 < (J-K)_0 < 1.42$) 
%consistent with them being O-rich stars. Their classification
%is subject to caution, as all these spectra display an unexpected rise
%beyond 1.6\,$\mu$m when compared to those of W.Vacca or of giants in Fornax
%(in Fornax, a similar rise has only been found for Fornax\,23). 
%
%[AL - I have no explanation for this. Never seen it before. Any 
%observational error that could have occurred systematically for Sculptor,
%and in two cases for Fornax?]

\subsection{NGC\,6822 data}
\label{subsection.ngc6822}

The initial search of 2MASS objects in the direction of NGC 6822
fulfilling the criteria mentioned in Sect.~2 resulted in a list of 130
known AGB stars. With the exception of one S-star (Aaronson et al 1985)
and one M-star (Elias \& Frogel 1985), they all come from the C-star
survey presented by Letarte et al. (2002) which contains a total of 904 C-stars.

Of the 10 new candidate AGB stars, five were observed in NGC 6822 and all are O-rich, 
although hundreds of C-stars have been identified earlier in this dwarf galaxy
(Letarte et al. 2002, Demers et al. 2006). They are also very bright: 
with bolometric magnitudes smaller than $-$5.76, they lie near or above
the bright end of the C-star luminosity function for NGC\,6822 presented by 
Groenewegen (2007), based on the $R,I$ photometry of Letarte et al. (2002).

Three of the five stars are likely Mira-like long period variables as they
display a sharp bandhead at 1.34\,$\mu$m due to H$_2$O (numbers 4, 7 and
to a slightly lesser extent 10). We cannot exclude that the other two are 
also of this type, since Mira-like stars do not always show H$_2$O (e.g.
LW2000). When compared with the O-rich stars observed
in Sculptor and in Fornax, or with the reference low metallicity
giants of Fig.\,\ref{Vacca.fig}, the O-rich stars
in NGC\,6822 display stronger features of CO and OH across the $H$-band. 
The strength of these features is known to 
increase with luminosity. Considering the bolometric magnitudes
of our targets, this result is not surprising.

\begin{figure}
\includegraphics[clip=,width=0.49\textwidth]{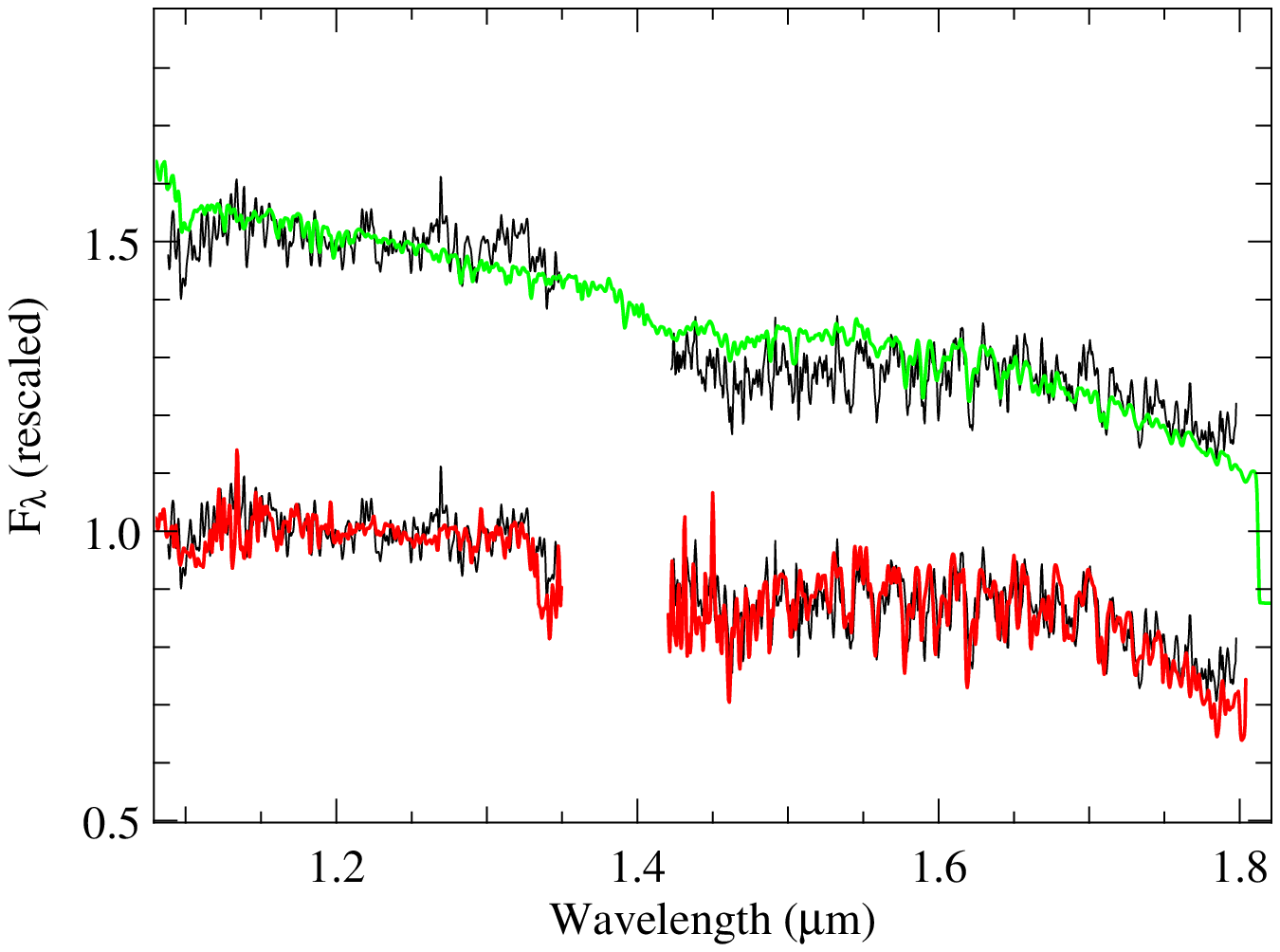}
\includegraphics[clip=,width=0.49\textwidth]{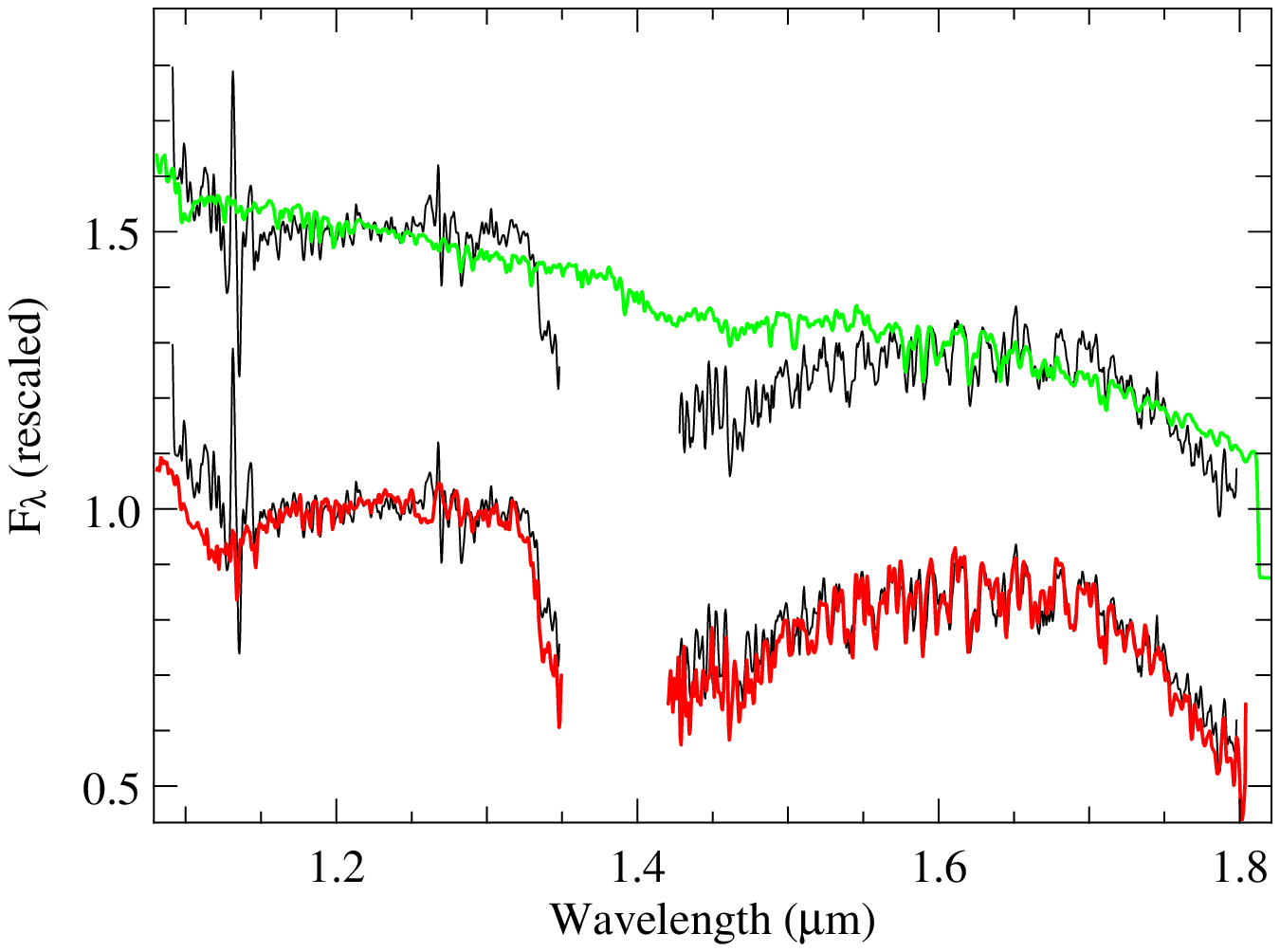}
\caption[]{Upper panel: Comparison of NGC\,6822-8 (black lines) 
with the coolest spectrum of Fig.\,\ref{Vacca.fig} (top) 
and with the May 1996 spectrum of the Milky Way long period variable S\,Phe 
(bottom). Lower panel: same as above for NGC\,6822-4, 
but with the April 1995 spectrum of
U\,Crt as a Milky Way template. In both panels, 0.5 flux units have
been added to the upper spectra for display.
% Made with tryLW in isaac/groen.i. See good cases just below
% next paragraph.
}
\label{N6822fits.fig}
\end{figure}

The dereddened ($J-K$) colours are in the range 1.23 to 1.40, when a
mean foreground extinction is applied.  We note that foreground
extinction is variable across the field of NGC\,6822 with known
differences of about 0.4 magnitudes in A$_V$ (Wyder 2003).  For the
five observed stars, we find a better match to the coolest metal-poor
spectrum of Fig.\,\ref{Vacca.fig} (type K3.5) with an $A_{\rm V}$ of 0.5
to 2. This suggests circumstellar extinction, or a spectral type later
than K4. Considering how much the near-IR energy distribution of a
Mira variable can change, it is difficult to find the perfect
template, and thus to argue one way or the other. We performed
comparisons with the Milky Way Mira type spectra that had similar CO
band strengths and $H$-band curvatures (LW2000), and
found no need for extra extinction with those templates.  Of course,
these Milky Way stars could be reddened themselves, and they are
likely to be more metal rich than those in NGC\,6822. Two examples are
given in Fig.\,\ref{N6822fits.fig}.

\section{Discussion}

\subsection{Evolutionary considerations}

Because our samples are not complete, they can not be used to put strict
constraints on the star formation histories of galaxies. Therefore,
we restrict this section to a brief verification of consistency with theoretical
predictions, using history determinations in the literature.
For the effects of metallicity on the location of the AGB in colour-magnitude
diagrams and the formation of carbon stars, we have used the work
of Mouhcine (2002), Mouhcine \& Lan\c{c}on (2003), Marigo \& Girardi (2007) and
Marigo et al. (2008). 
% Found most useful: Fig. 1 of Mouhcine & Lancon + Fig 1 of Marigo et al. 2008.

$\bullet$ {\bf Fornax}

As summarised by Coleman et al. (2005), based in particular on the
work of Stetson et al. (1998), Buonanno et al. (1999), Saviane et al. (2000), 
Tolstoy et al. (2003) and Pont et al. (2004), star formation has
occurred in Fornax over long timescales. An old low metallicity giant
branch coexists with an important young population, born over the last
few Gyr with metallicities centered on [Fe/H]$=-1$\,dex. Many carbon
stars testify of this intermediate age population (Aaronson \& Mould 1980, 
Azzopardi et al. 1999). The youngest stars may be as young as a
few 100\,Myr, and as metal rich as [Fe/H]$\,=-$0.4\,dex.

The C-stars in our sample have bolometric magnitudes between $-$3.56
and $-$5.12. Such luminosities are consistent with model
predictions. The luminosity of the brightest C-star translates into an
age younger than 1\,Gyr. The faintest C-star is compatible with the
quiescent H-burning luminosity along the TP-AGB only for [Fe/H]$<-$1
(Marigo \& Girardi, 2007).  Alternatively, this star could be evolving
through one of the luminosity dips that follow helium flashes, or be
the result of binary evolution.

We find that the O-rich stars in our sample are systematically fainter
than the C-rich stars, with bolometric magnitudes between $-$2.9 and
$-$3.4. These luminosities lie just below the transition between the
early-AGB and the thermally pulsing AGB. Early spectral types and 
the lack of evidence for large amplitude pulsation are therefore not
surprising.

%Assuming that the effective temperature scale of Schmidt-Kaler (1982)
%applies to stars at the metallicity of Fornax ([Fe/H]= -1.2), spectral
%types G8 to K4 correspond to effective temperatures of 4800 to 4000\,K.
%The evolutionary tracks of Fagotto et al. (1994) extend down to
%3500\,K for [M/H]=-0.7 and down to 4300\,K for [M/H]=-1.7.
%But I should use Mustapha's AGB tracks instead, because the Fagotto
%tracks don't have an AGB... I should also check that his AGB tracks
%predict only a very small proportion of O-rich AGB stars with respect to
%the C-stars.

$\bullet$ {\bf Sculptor}

The fraction of intermediate age stars is smaller in Sculptor than
in Fornax. Sculptor is believed to have formed most of its stars more 
than 10\,Gyr ago, although star formation then continued at a lower level 
until it stopped several Gyr ago (Tolstoy et al. 2003). The metallicities
of the oldest stars are clustered around [Fe/H]$=-$2\,dex 
(Tolstoy et al. 2004, Clementini et al. 2005); a more centrally
concentrated and kinematically distinct stellar population 
has a typical [Fe/H] of $-$1.4\,dex (Tolstoy et al. 2004, 
Babusiaux et al. 2005).

The single O-rich star in the Sculptor sample is relatively
faint (bolometric magnitude of $-$2). This star is likely 
to belong to the early-AGB. The one C-rich star we observed,
on the contrary, is bright ($M_{\mathrm{bol}}\simeq$\,-4.9). 
Stellar evolution predicts that such a luminosity is reached in
the quiescent H-burning phases of the TP-AGB only for progenitor 
masses near 2\,M$_{\odot}$ or higher, which translates into
ages of $\sim$2\,Gyr or less. This star might be experiencing a
helium flash, or it may have followed a non-standard evolutionary path
(a few isolated examples of old C-rich TP-AGB isochrones are 
shown in the low metallicity grids of Marigo et al. 2008). Because
it is a known variable, it is unlikely to 
be a dwarf carbon star in the foreground (Mauron et al. 2004).

$\bullet$ {\bf NGC\,6822}

NGC\,6822 is a Magellanic irregular galaxy with a bar, 
an ellipsoidal component known to host
carbon stars, and an extended HI polar ring apparently devoid of old
and intermediate age stars (Demers et al. 2006, de Blok et al. 2006).
The galaxy is thought to have formed stars over a large fraction of the
age of the universe, with possible fluctuations such as 
the differences  between the star formation rates seen in the bar
and around it over the past few hundred million years 
(Gallart et al. 1996\,a,b, Wyder 2003). The metallicities of the 
younger stars are 2 or 3 times below solar (e.g. Venn et al. 2001).
%Models suggest that metallicity reached about 1/10 solar in the first few
%Gyr, and increased steadily from there (Carigi et al. 2006).

The stars observed with ISAAC are bright. Evolutionary tracks for
the thermally pulsating AGB reach those
luminosities for initial masses above 3.5\,M$_{\odot}$, or ages younger
than about 500\,Myr.
Stars in this range spend no or very little time as C-stars before loosing
their envelope, because of combined effects of hot-bottom burning (which
burns C into N at the base of the convective envelope, when sufficiently
hot) and mass loss (which radically reduces TP-AGB lifetimes when luminosity
is high). The observed O-rich atmospheres are fully consistent with the
estimated luminosities.

\subsection{The 1.53\,$\mu$m feature of carbon stars and mass loss}
\label{1.53muFeature.sec}

The observations of stars with 2MASS colours $(J-K) > 1.22$ has
resulted in the confirmation of a dozen carbon stars in Fornax and Sculptor.
The presence of the sharp C$_2$ bandhead at 1.77\,$\mu$m was
adopted as a sufficient condition for a classification among 
carbon stars. Most of the carbon stars also display the characteristic
C$_2$ band that spreads between 1.15\,$\mu$m and 1.3\,$\mu$m, as well as
the CN absorption bandhead at 1.1\,$\mu$m.

The most surprising result of the present work is the frequent detection
of the so-called 1.53\,$\mu$m-feature (Goebel et al. 1981, Joyce 1998) 
in the sample of carbon stars. 
This band was previously noticed in a few low temperature carbon stars
of the Milky Way (Goebel et al. 1981, Joyce 1998).
The strength of the band was shown to depend on pulsation phase in 
C-rich Mira variables such as V\,Cyg or R\,Lep 
(Joyce 1998, LW2000). After some debate, it is 
now considered likely that the main carrier
of this feature is the C$_2$H$_2$ molecule, although the predicted
shape of the band in dynamical models of cool carbon stars does
not quite match the observed shapes as yet and HCN may contribute
(Gautschy-Loidl et al. 2004).
The 1.53\,$\mu$m band would then be an overtone of one of the bands 
contributing to the much stronger, and more common, 3.1\,$\mu$m feature
of carbon stars (see e.g. van Loon et al. 2006, 2008, for a sample in the LMC).

\begin{figure*}
\includegraphics[clip=,scale=0.55]{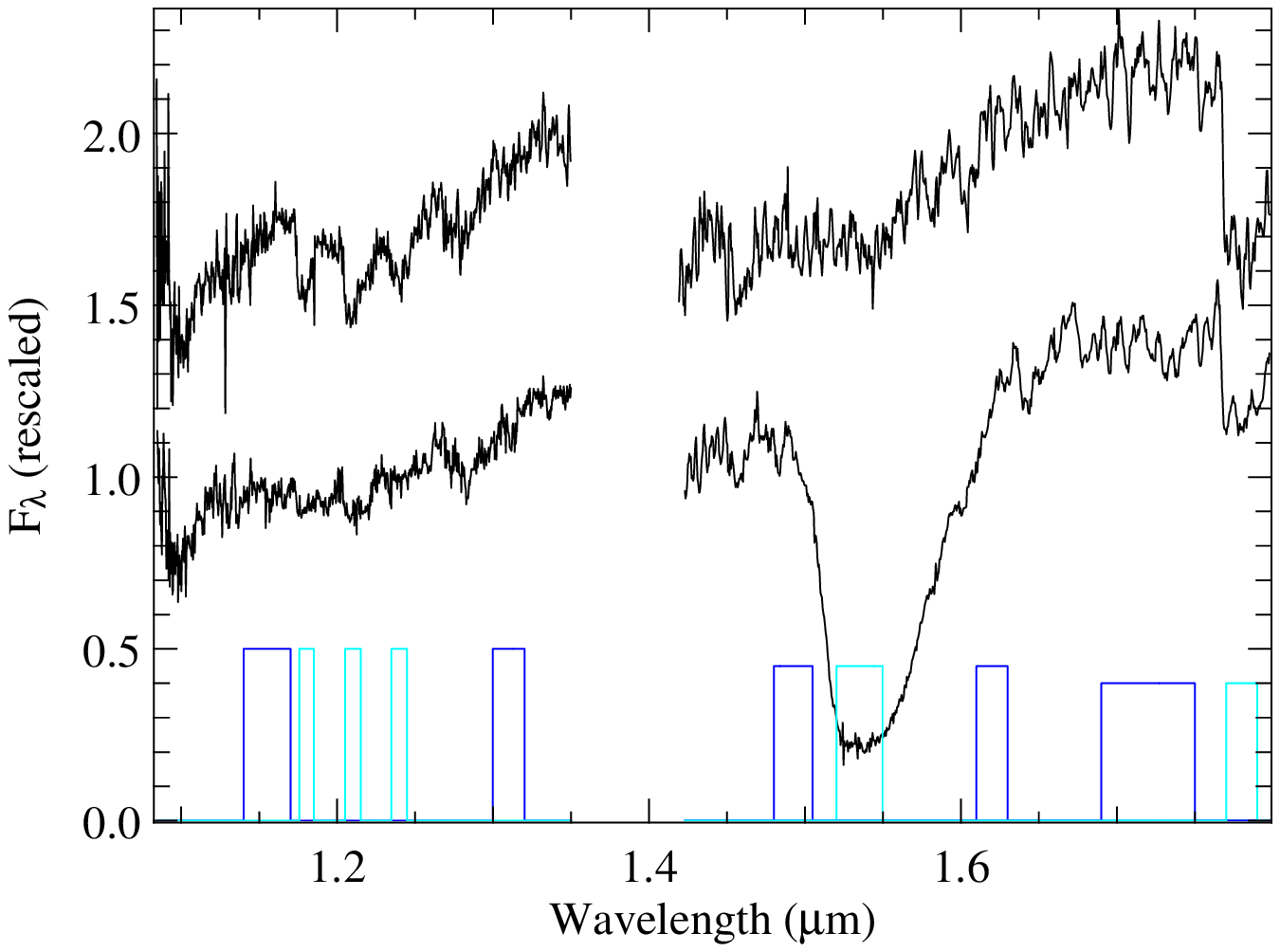}
\includegraphics[clip=,scale=0.55]{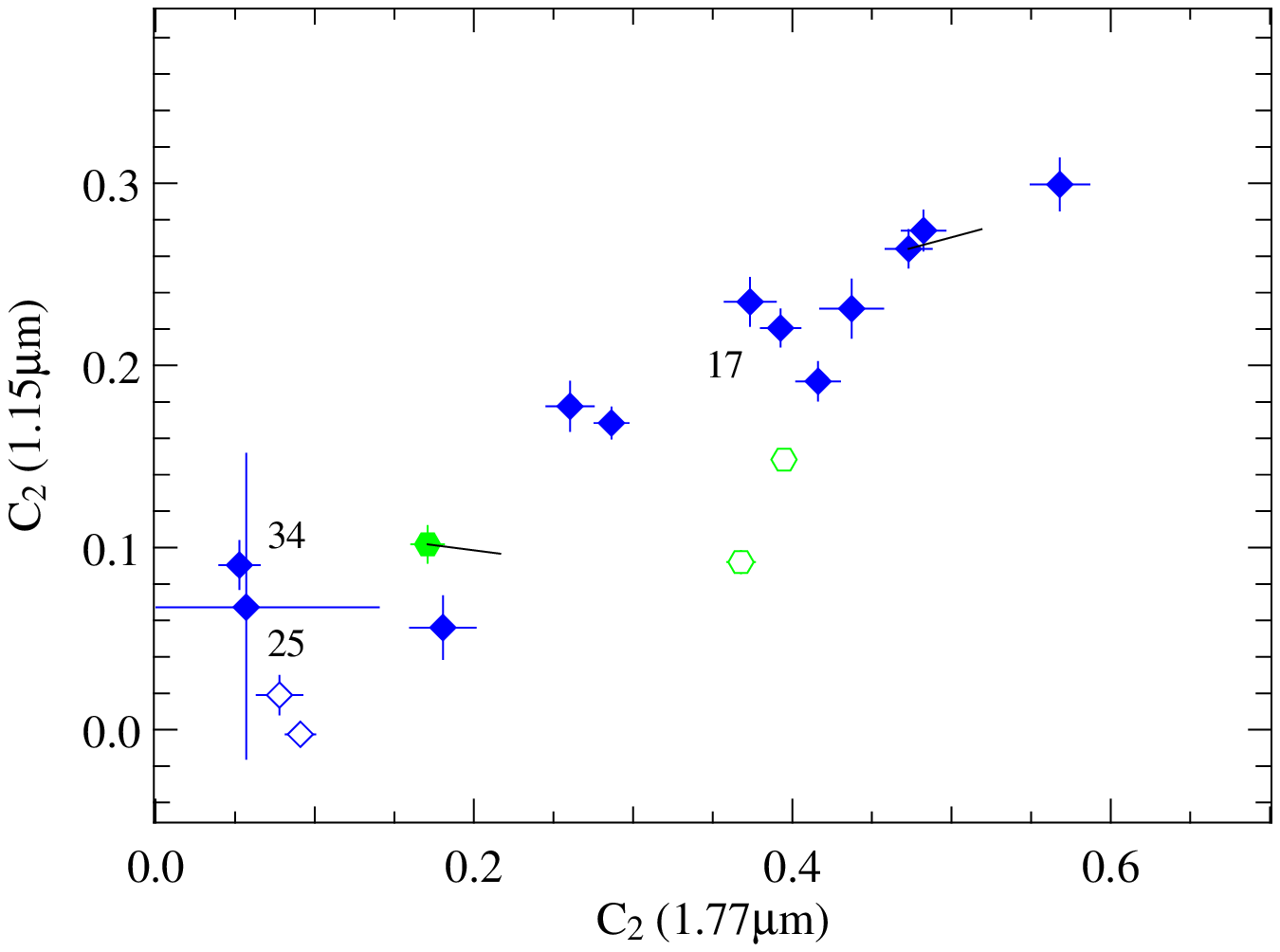}\\
\includegraphics[clip=,scale=0.55]{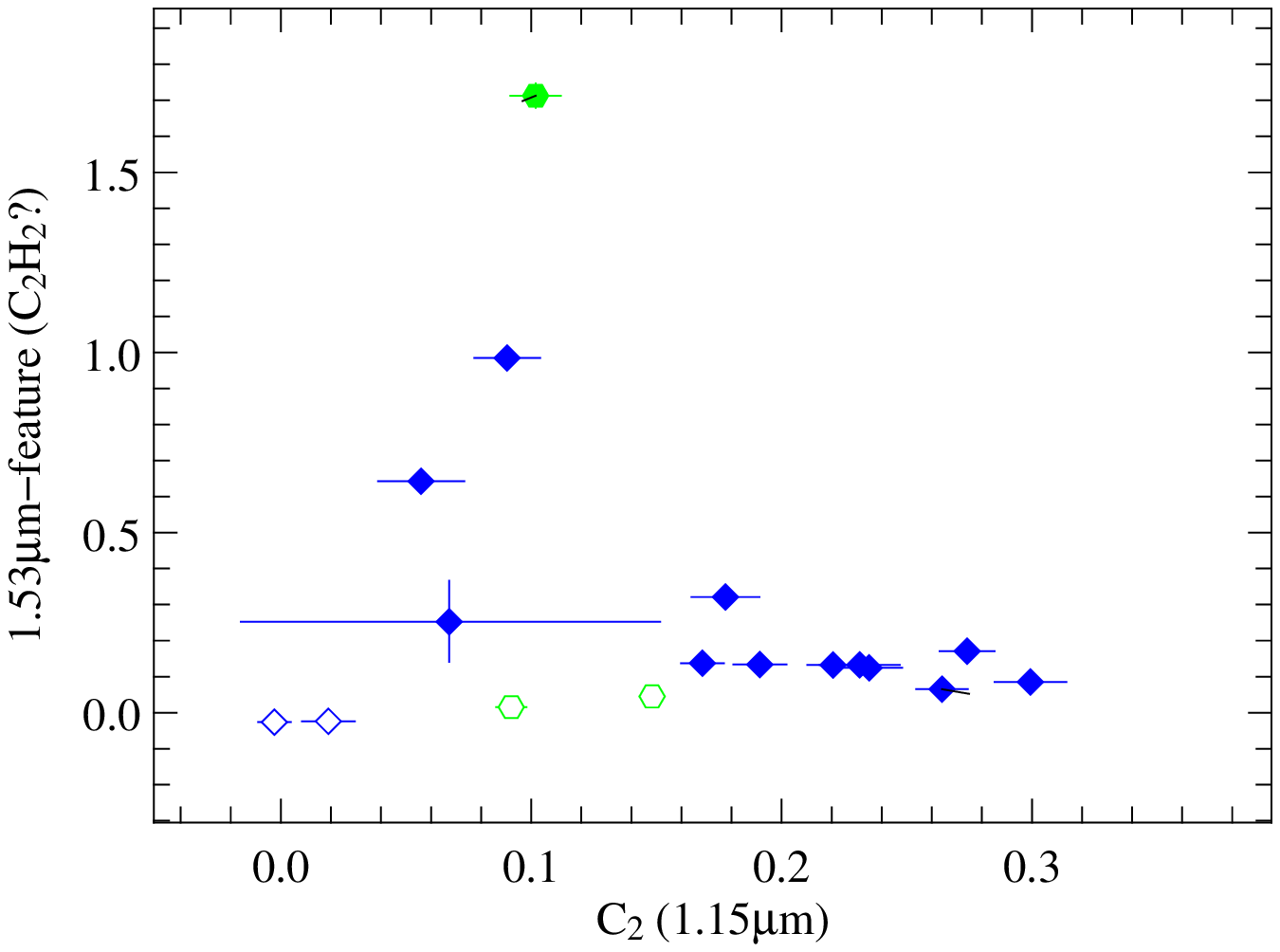}
\includegraphics[clip=,scale=0.55]{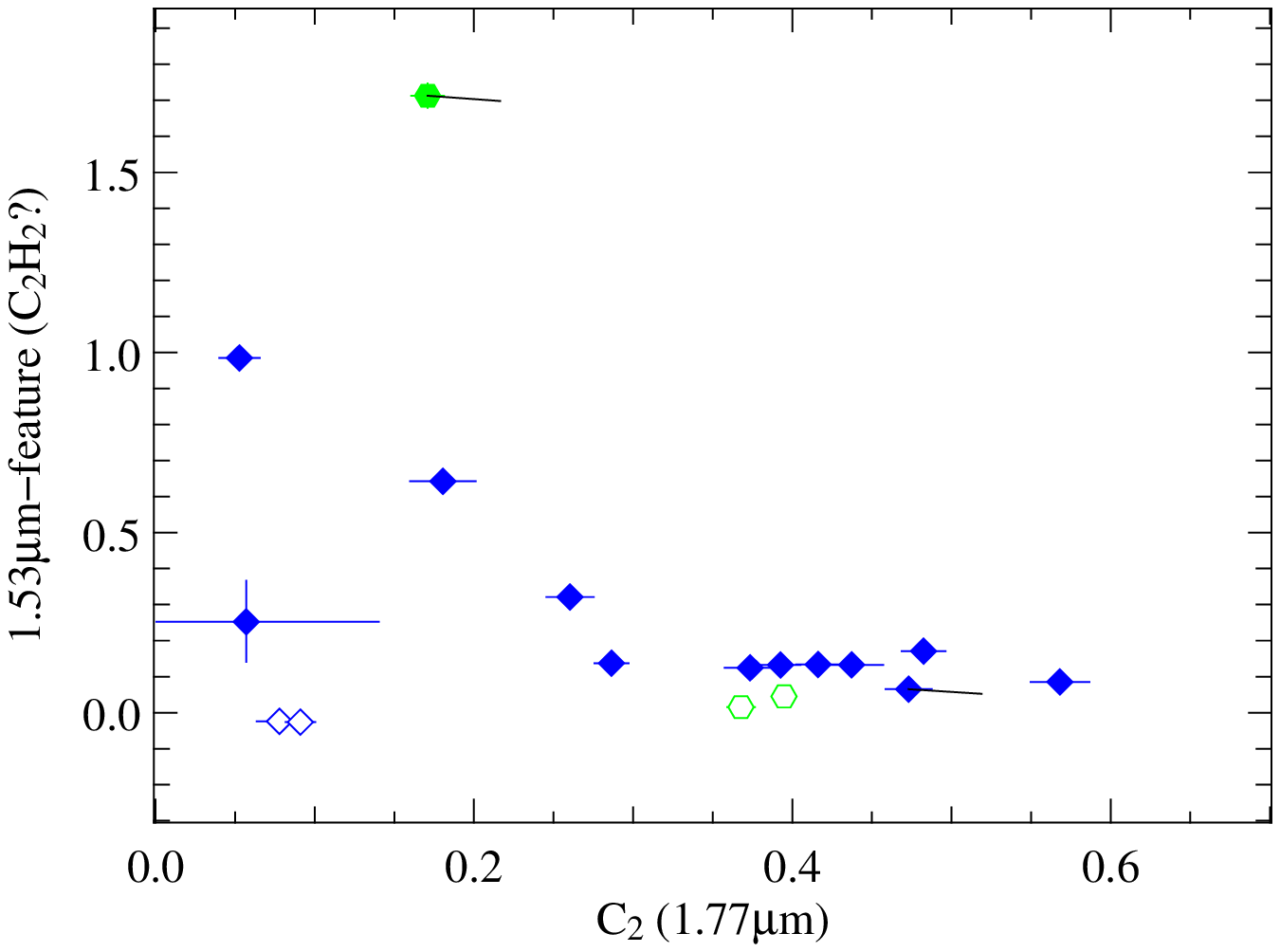}\\
\includegraphics[clip=,scale=0.55]{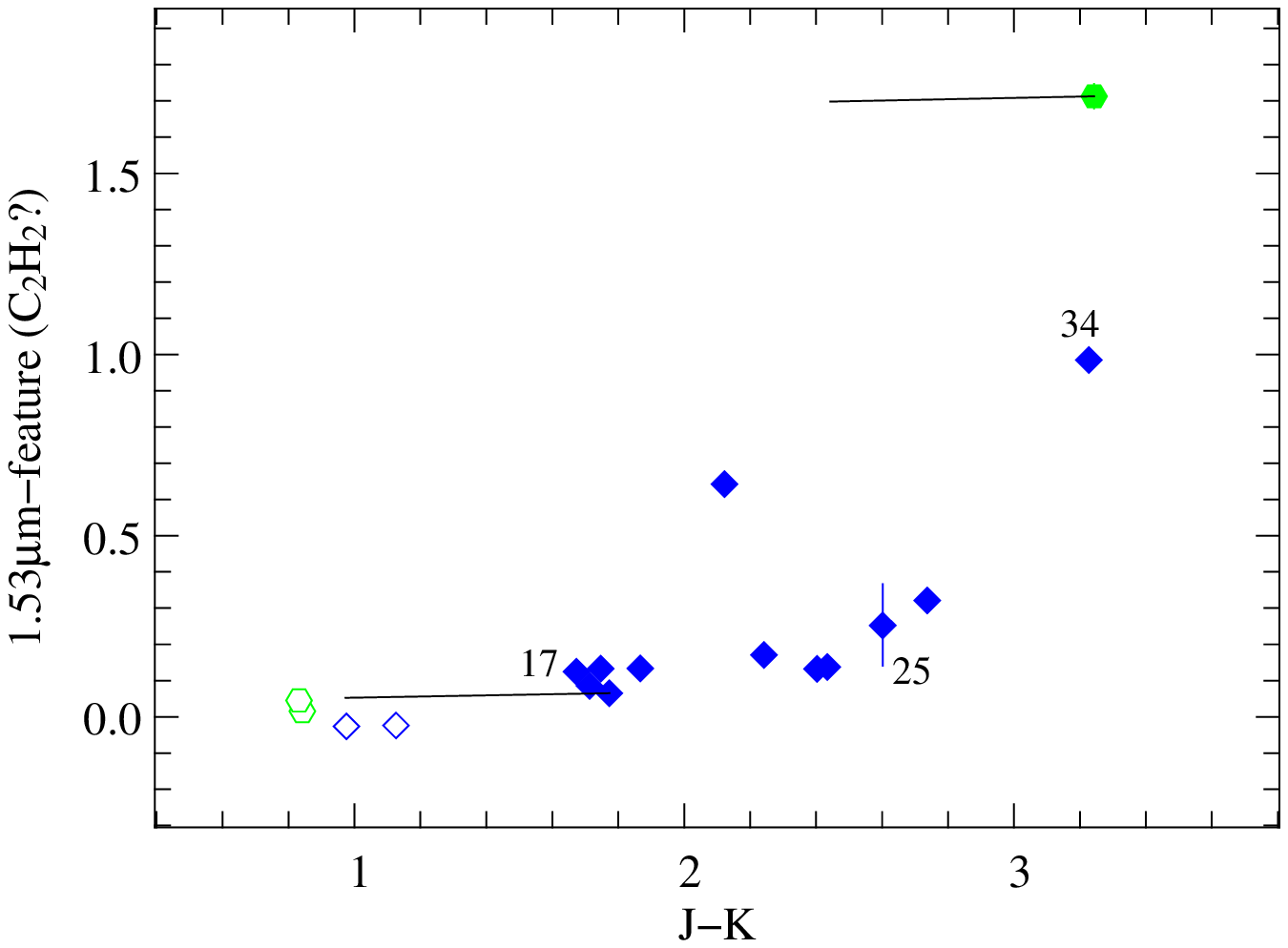}
\includegraphics[clip=,scale=0.55]{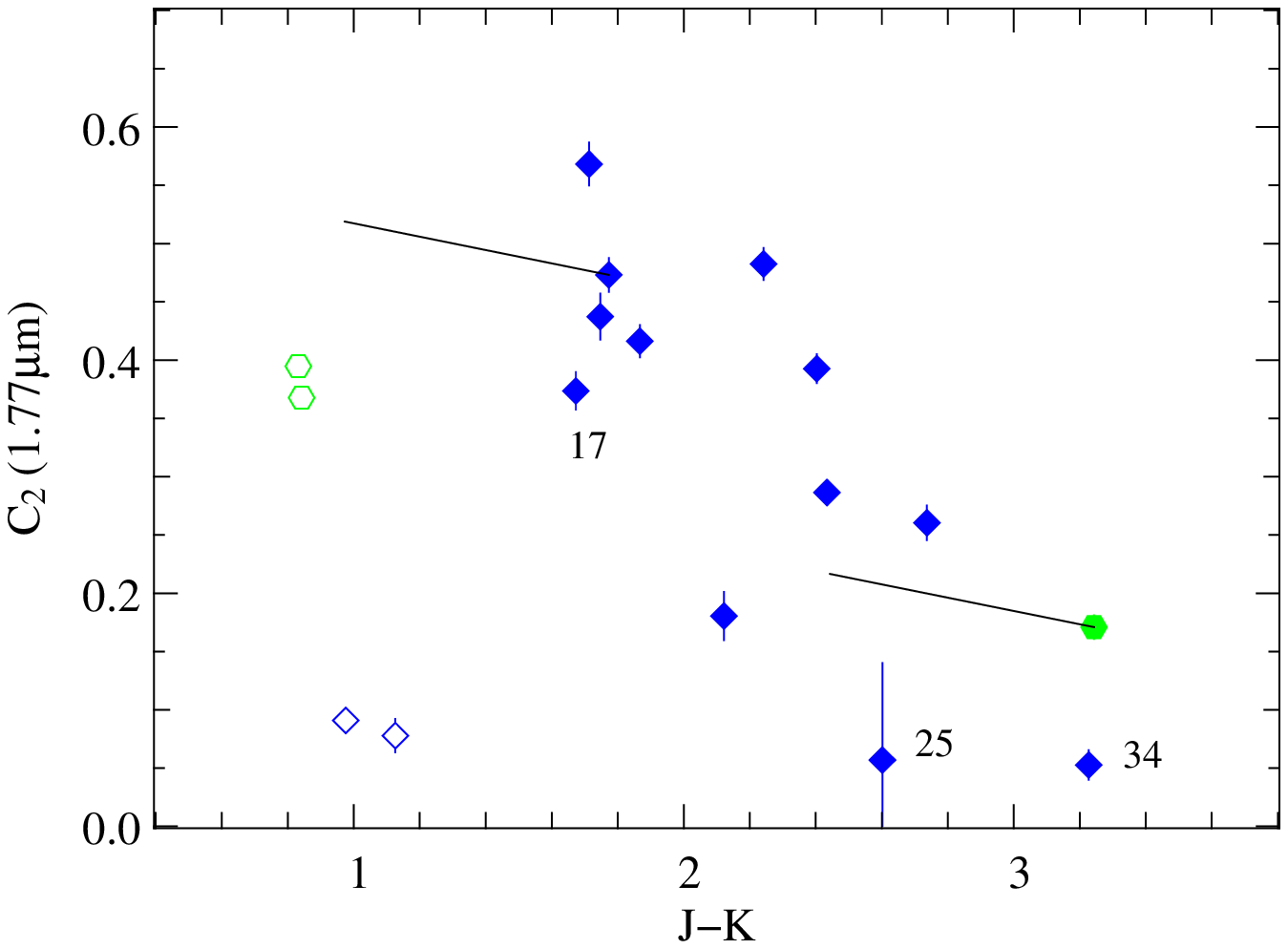}
\caption[]{ {\bf Upper left:} Spectral index definition passbands. The
  (arbitrary) passband heights indicate passbands used for measuring
  C$_2$ around 1.15$\mu$m (highest), the 1.53\,$\mu$m feature
  (medium), and C$_2$ at 1.77$\mu$m (lowest).  ``Continuum" passbands
  are marked in blue, ``feature" passbands in cyan. The spectra shown
  are those of Fornax 21 (top) and Sculptor 6.
% Done with function "bandids" in "groen.i", in subdirectory isaac/.
%
{\bf Other panels:} index measurements on the carbon star spectra.
Stars in Fornax are shown as dark blue diamonds, stars in Sculptor as
green hexagons. Open symbols are reference stars, solid symbols are
programme stars. The three stars labelled are those in common with
Matsuura et al. (2007). The de-reddening vector (for A$_V$=5\,mag) is shown
for two stars: it depends on the spectral type and is small for the
molecular index measurements.
% Done with function "cstars" in "groen.i", in subdirectory isaac/.
% I took the law of Cardelli et al. 1989 with Rv=3.1; for J-K I assumed
% Av/E(J-K)=0.16, for consistency with the text in Sect. 3.
% For the *indicative* error bars, I measured the rms between 1.3 and 1.35 mu
% only. Then I divided by the sqrt of the number of pixels in each passband.
% Min and max are Indexmin=2.5 log(continuum-rms)/(feature+rms)
%                 Indexmax=2.5 log(continuum+rms)/(feature-rms)
}
\label{indexdata.fig}
\end{figure*}

Although incomplete, our sample may help characterise when the 
1.53\,$\mu$m feature appears. 
The upper left panel of Fig.\,\ref{indexdata.fig} 
shows the passbands used to measure the strengths of various bands. 
When possible, two ``continuum" regions\footnote{The real continuum
is nowhere accessible in carbon stars with observations at the resolution
of our data.}  were
selected on either side of the region of interest, in order to reduce
sensitivity to extinction or to broader overlapping absorption 
features. This was not possible for C$_2$ (1.77\,$\mu$m), and 
large uncertainties near the short wavelength edge of the $J$-band
spectra made us abandon the measurement of the CN bandhead (1.10\,$\mu$m).
One or several regions were used for the measurement of the 
flux within the molecular bands of interest. The strength of any absorption
band $X$ is measured as 
\begin{equation}
I(X)=2.5\,\log \left( F_{\rm c}(X)/F_{\rm b}(X) \right)
\end{equation}
where $F_{\rm c}(X)$ (resp. $F_{\rm d}(X)$) is the mean energy density received in the 
wavelength bins of the ``continuum" region(s) (resp. in the absorbed regions).

Figure\,\ref{indexdata.fig} shows the measurements obtained. The following
comments can be made.

\begin{itemize}
\item The programme stars are offset from the S and C stars used for reference,
  because these are intrinsically bluer.
\item The strengths of the C$_2$ features measured in the $J$-band and in the 
  $H$-band correlate well.
\item The 1.53\,$\mu$m feature is found only when the C$_2$ features are 
  relatively weak. This is consistent with a picture in which carbon
  C$_2$H$_2$ formation happens at the expense of C$_2$ in the outer
  atmosphere, when layers of sufficiently cool temperature and 
  sufficiently high density exist (Gautschy-Loidl et al. 2004; 
  U. Jorgensen priv. comm.).
\item The 1.53\,$\mu$m features shows a positive correlation  with $J-K$,
  while the strength of the C$_2$ features in our sample shows the opposite
  trend. The extinction vectors show that a few magnitudes of (optical) 
  circumstellar extinction will not modify the trends.
  Incompleteness is more of a concern. The 
  trends with $J-K$ are consistent with the picture above, if $J-K$ can
  be taken as a first order indication of the temperatures present in
  the emitting layers of the atmosphere. 
\end{itemize}

%Three of the C-stars in Fornax have been observed using Spitzer-IRS
%and have their mass-loss rates determined from a fit to the SED
%(Matsuura et al. 2007).
%%
%Fornax 34 (their Fornax 13-23) and Fornax 25 (their Fornax 12-4) have
%similar mass-loss rates of $\sim4 \times 10^{-6}$ \msolyr, but the
%1.53 $\mu$m band is very strong in Fornax 34 and weak in Fornax 25.
%Fornax 17 (their Fornax 6-13) has a mass-loss rate $< 1.3 \times
%10^{-6}$ \msolyr and also a weak band. It appears that there is no
%obvious relation between the strength of the 1.53 $\mu$m band and the
%mass-loss rate based on these few stars. Although the Spitzer-IRS
%observation and our NIR spectra are not taken simultaneously there
%does seem to be a relation between the strength of the band at 1.53
%$\mu$m and that at 7.5 $\mu$m (due to HCN + C$_2$H$_2$) seen in the
%IRS spectra. Fornax 34 has a strong 7.5 $\mu$m band, while Fornax 25
%and Fornax 17 show a similarly weak one.

%The one C-star observed in Sculptor, which displays strong 1.53 $\mu$m
%absorption, is a known variable (Mauron et al. 2004).  

The dynamical
model atmospheres of Gautschy-Loidl et al. (2004) show that the 1.53
$\mu$m band appears in their coolest models
($T_{\rm eff} < 2800$K), which start developing a wind, near minimum light.  
The band strength shown in Gautschy-Loidl et al. is much less than
observed in some of the extreme cases observed here, but this
particular model has a mass-loss rate of $6.6 \times 10^{-7}$ \mbox{\msolyr.}
The picture that is emerging based on the available observational data
and theoretical model atmospheres is that the 1.53 $\mu$m band is a
function of phase in the pulsation cycle, strongest at minimum light,
and that its strength increases with mass-loss.

{\bf We have attempted to test this further by cross-identifying our sample
with the photometric multi-epoch survey of Fornax by Whitelock
et al. (2009) and with Spitzer-IRS data, also on Fornax, presented
with derived mass loss estimates by Matsuura et al. (2007).
All the objects Whitelock et al. marked as showing low amplitude semi-regular
or irregular variations display no or a weak 1.53 $\mu$m feature.  
Unfortunately, very small number statistics prevent 
of from identifying any other trends. The diversity in behaviours
suggests a large dispersion in any such trend, as a natural
result of the complexity induced by pulsation.}

\section{Conclusions}

Based on the 2MASS point source catalog, we have selected candidate
AGB stars in Local Group dwarf galaxies by requiring a large intrinsic
$(J-K)$ colour index as well as a AGB-like bolometric luminosity
(Groenewegen 2006).  The candidates in Fornax, NGC\,6822 and Sculptor
{\em not} previously confirmed spectroscopically are listed in
Table\,\ref{TAB-targets}, and $J$ and $H$-band spectra were obtained
for about two thirds of the candidates.  These spectra allow us to
identify carbon stars and to estimate the spectral types of
oxygen-rich AGB stars, as well as to recognise and reject
contaminating background galaxies.

In Fornax, 12 of the 22 stars observed are carbon stars, and this can
be compared to the list of 104 carbon stars identified previously
(Azzopardi et al. 1999).
In NGC\,6822 on the other hand, all 5 stars observed are O-rich and
not very red ($J-K \sim 1.4$).  This is likely a selection effect.
Because of distance the $J$ and $K$ magnitudes are close to the
detection limit.  Any obscured C- or O-rich star would likely not be
detected in $J$ and hence not be in the sample of candidate AGB stars.
As mentioned in Sect.~5.5 the luminosity of the O-rich stars is close
to or brighter than the bright end of the C-star LF as determined from
the Letarte et al. sample. It would be interesting to observe these
stars at high resolution in the optical and look for signatures of Hot
Bottom Burning. Our work did not add any new C-stars, but to exclude
the presence of mass-losing O- and C-stars would require a survey to
fainter magnitudes.
In Sculptor, our initial sample
is strongly contaminated by background galaxies. The only carbon star
we find is the extremely red object also mentioned by Mauron et al. (2004). 
The chemical natures and luminosities of the detected AGB stars are
compatible with current stellar evolution models, combined with the
star formation histories of the host galaxies, with some fine tuning
required for the extreme C star of Sculptor and for the faintest C
star found in Fornax. As usual for AGB stars, a variability study
would be useful to characterise the sources more completely.

Among the observed stars with $(J-K)>1.5$, we have not found any
O-rich star. In other words, we have not detected new heavily
dust-enshrouded M-type AGB stars. Carbon stars can have a $(J-K)$
index of about 2 even without circumstellar dust. We find eight carbon
stars with $(J-K)>2$ (7 in Fornax, 1 in Sculptor), which are most
likely all dust-enshrouded. The absence of O-rich dust-enshrouded AGB
stars in our sample may be the random result of small numbers; but it
is not too surprising a result considering the large range of initial
masses for which the final AGB stages are carbon rich at sub-solar
metallicities. 
The low mass AGB stars ($<1$\,M$_{\odot}$) spend a very short fraction
of their final O-rich AGB life in phases of heavy mass loss and are
also less luminous than their higher mass counterparts; the most
massive AGB stars spend a larger fraction of their final O-rich AGB
life in dust but are intrinsically rare due to the stellar IMF.

$J$ and $H$-band spectra of carbon stars are still rather rare in the
literature.  Our data confirms the well-behaved correlation between
the two main bands of C$_2$, near 1.15\,$\mu$m and 1.77\,$\mu$m.  In
some of our spectra, the 1.53\,$\mu$m feature probably due to
C$_2$H$_2$ is extremely deep. This feature would deserve more
attention, as it is remains difficult to reproduce with models.  From
our small sample, it seems that $(J-K)>2$ is a requirement for the
presence of a strong 1.53\,$\mu$m feature, but is not a sufficient
condition. When the 1.53\,$\mu$m feature is strong, C$_2$ bands are
weak, as expected from chemistry in sufficiently cool and dense
layers.  All this suggests that the 1.53\,$\mu$m feature is related to
the processes that drive mass loss from carbon stars. Confirmation of
the trends from repeated observations and a larger sample will be welcome.

%\end{document} 

\acknowledgements{ 
The authors would like to thank Peter Hauschildt and Joachim Puls for
providing the high-resolution B-star model spectra in the infrared.
Drs. Whitelock and Menzies are thanked for checking some stars on their IRTF images.
Drs. Yves Yung and Wolfgang Hummel (both ESO) are thanked for introducing the
ISAAC pipeline.
The telluric spectrum was taken based on NSO/Kitt Peak FTS data produced by NSF/NOAO.
This research has made use of the SIMBAD database, operated at CDS, Strasbourg, France.

{}

\Online

%xxxxxxxxxxxxxxxxxxxxxxxxxxxxxx

% ogle_1.tex:  /media/disk/disk84/groen/ISAAC/PRODUCTS/REDUCED

% 23.07.09 replaced ``:'' by ``.'' in filenames

\onlfig{6}{
 \begin{figure*}[H]
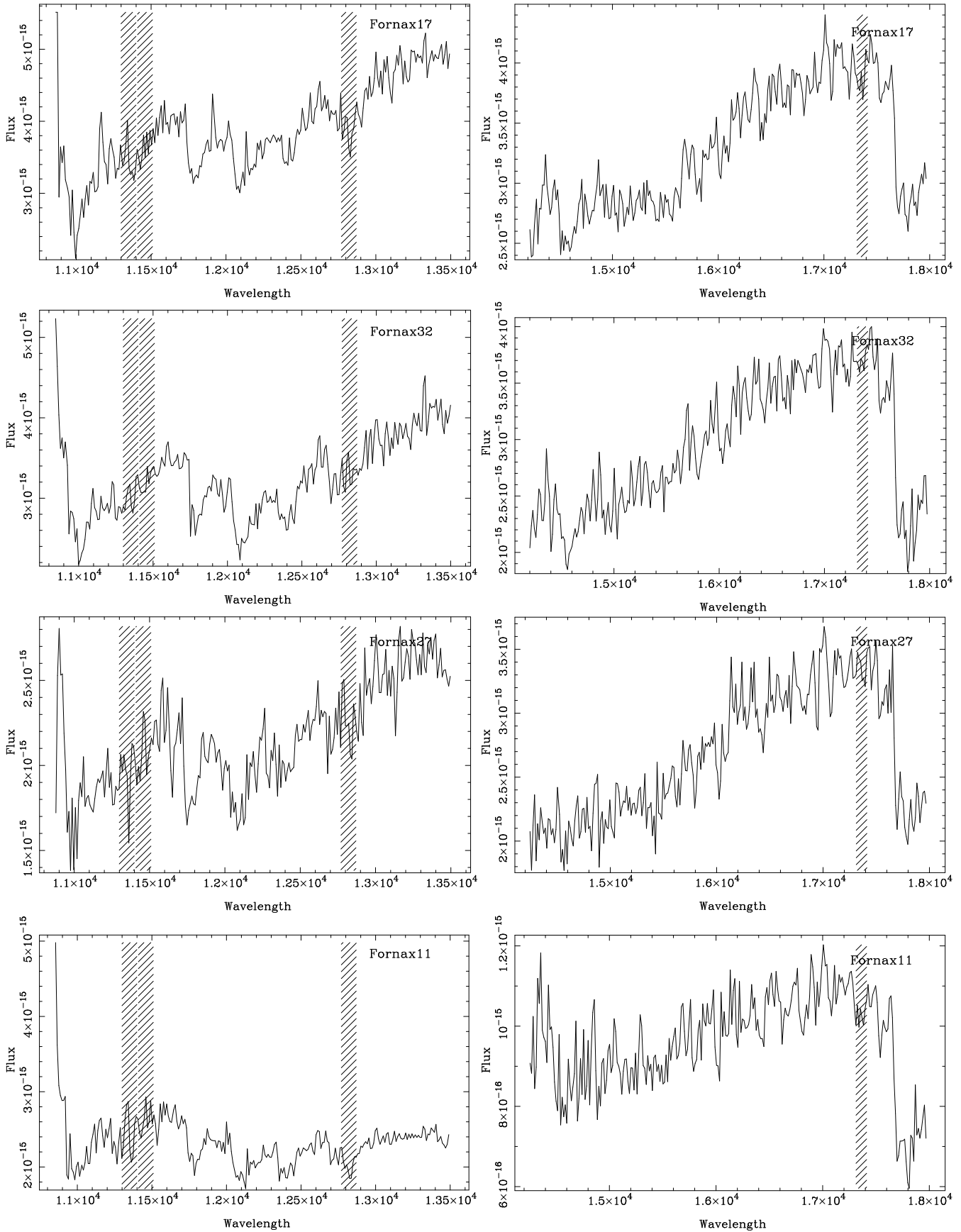


\begin{minipage}{0.48\textwidth}
\resizebox{\hsize}{!}{\includegraphics{FIN_Fornax17_2005-09-25T03.48.29.049_J.ps}} 
\end{minipage}
\begin{minipage}{0.48\textwidth}
\resizebox{\hsize}{!}{\includegraphics{FIN_Fornax17_2005-09-25T03.38.50.139_H.ps}} 
\end{minipage}
 
\begin{minipage}{0.48\textwidth}
\resizebox{\hsize}{!}{\includegraphics{FIN_Fornax32_2005-09-27T03.35.51.394_J.ps}} 
\end{minipage}
\begin{minipage}{0.48\textwidth}
\resizebox{\hsize}{!}{\includegraphics{FIN_Fornax32_2005-09-27T03.26.16.082_H.ps}} 
\end{minipage}
 
\begin{minipage}{0.48\textwidth}
\resizebox{\hsize}{!}{\includegraphics{FIN_Fornax27_2005-09-24T02.27.44.965_J.ps}} 
\end{minipage}
\begin{minipage}{0.48\textwidth}
\resizebox{\hsize}{!}{\includegraphics{FIN_Fornax27_2005-09-24T02.18.07.417_H.ps}} 
\end{minipage}
 
\begin{minipage}{0.48\textwidth}
\resizebox{\hsize}{!}{\includegraphics{FIN_Fornax11_2005-09-24T03.10.14.349_J.ps}} 
\end{minipage}
\begin{minipage}{0.48\textwidth}
\resizebox{\hsize}{!}{\includegraphics{FIN_Fornax11_2005-09-24T03.00.30.986_H.ps}} 
\end{minipage}

\caption[]{
$J$ and $H$-band spectra obtained with ISAAC. 
Wavelengths are in \AA\, fluxes in W/m$^2/\mu$m.
Shaded areas indicate wavelength regions where the cancellation of 
the atmospheric features and/or of features in the standard stars 
may be imperfect. 
The carbon stars in Fornax are presented first,
ordered by increasing $(J-K)$ colour, followed by the one red carbon star 
in Sculptor. Subsequently the O-stars are plotted by galaxy, and by increasing
$(J-K)$ colour. Finally, for completeness,
we append the spectra of previously known
sources (observed for reference), and those of outliers (source 16
in Fornax for which our $H$-band spectrum is odd, as well as the confirmed 
or likely low-redshift galaxies).
 
The complete figure is available in electronic form via http://www.edpsciences.org
}
\label{Fig-ISAAC}

\end{figure*}
}
 
\onlfig{6}{
\begin{figure*}[H]

\begin{minipage}{0.48\textwidth}
\resizebox{\hsize}{!}{\includegraphics{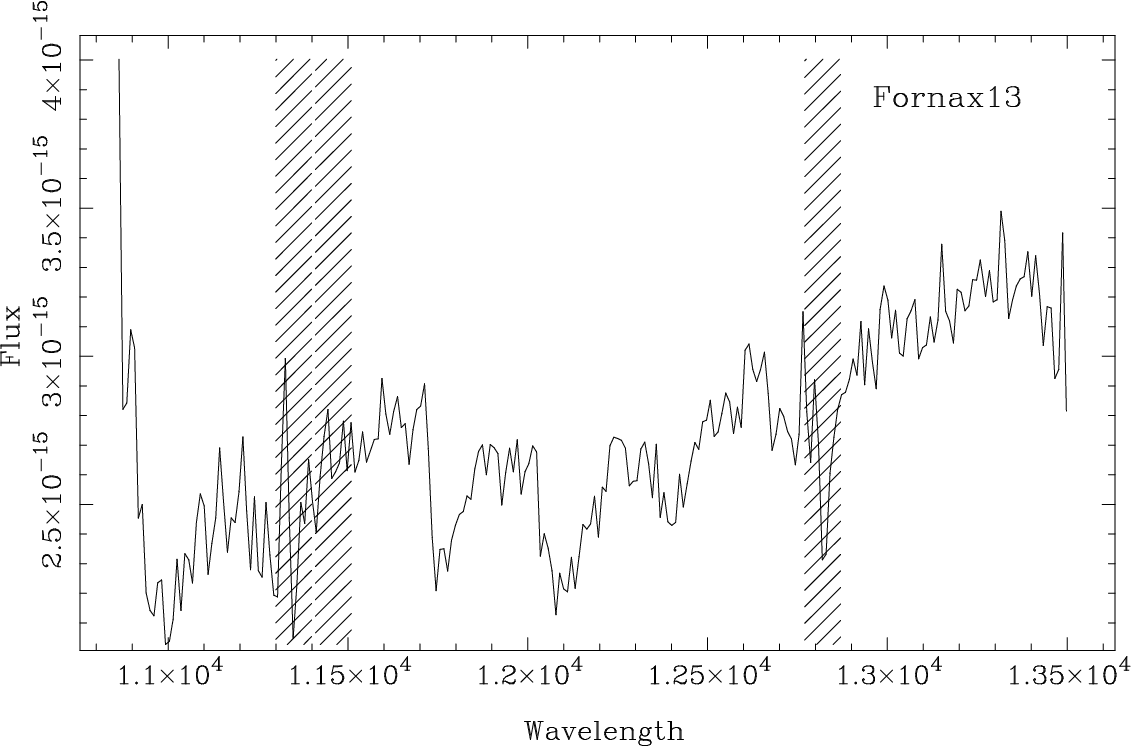}} 
\end{minipage}
\begin{minipage}{0.48\textwidth}
\resizebox{\hsize}{!}{\includegraphics{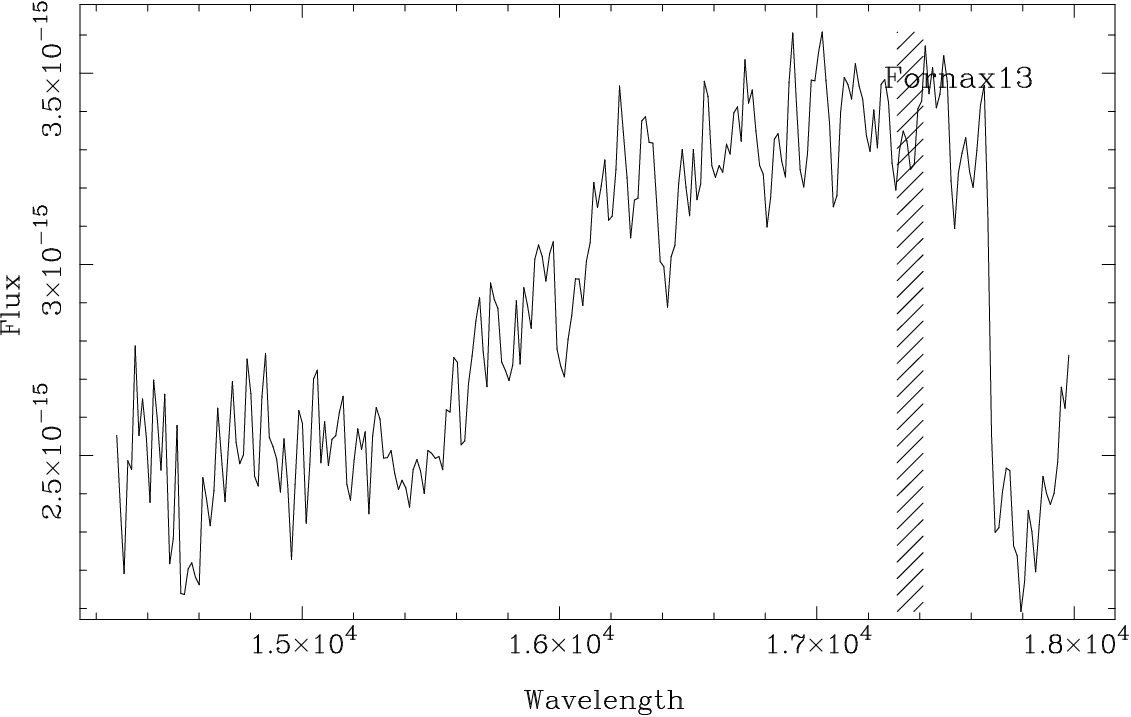}} 
\end{minipage}
 
\begin{minipage}{0.48\textwidth}
\resizebox{\hsize}{!}{\includegraphics{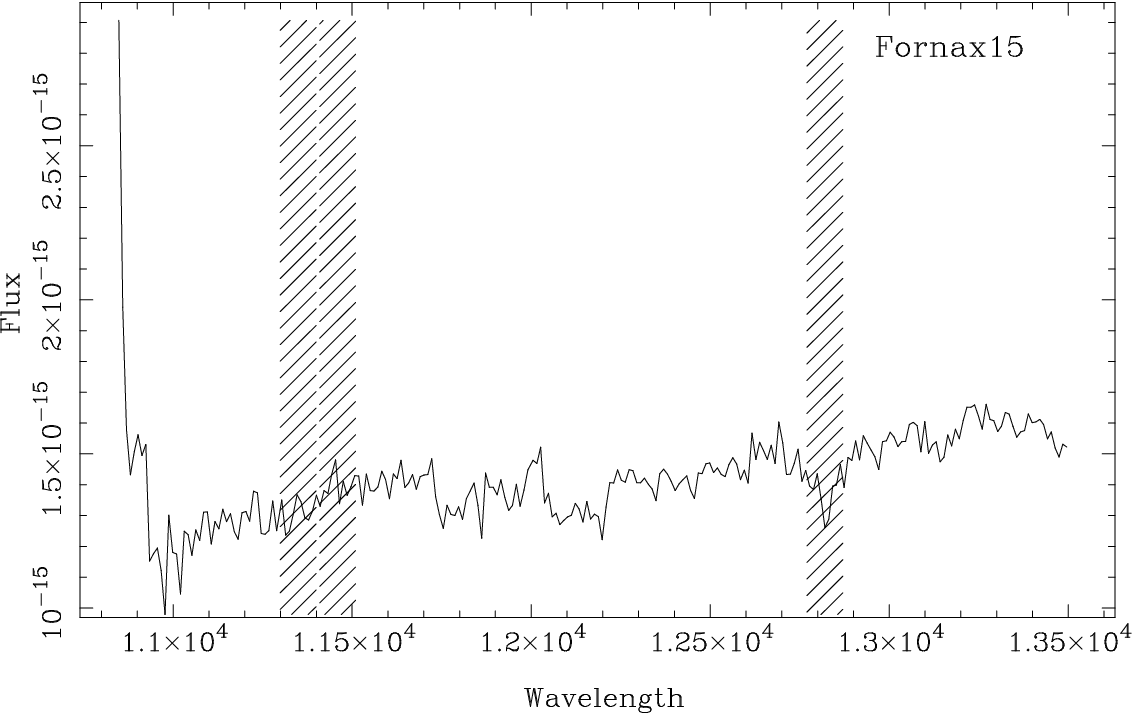}} 
\end{minipage}
\begin{minipage}{0.48\textwidth}
\resizebox{\hsize}{!}{\includegraphics{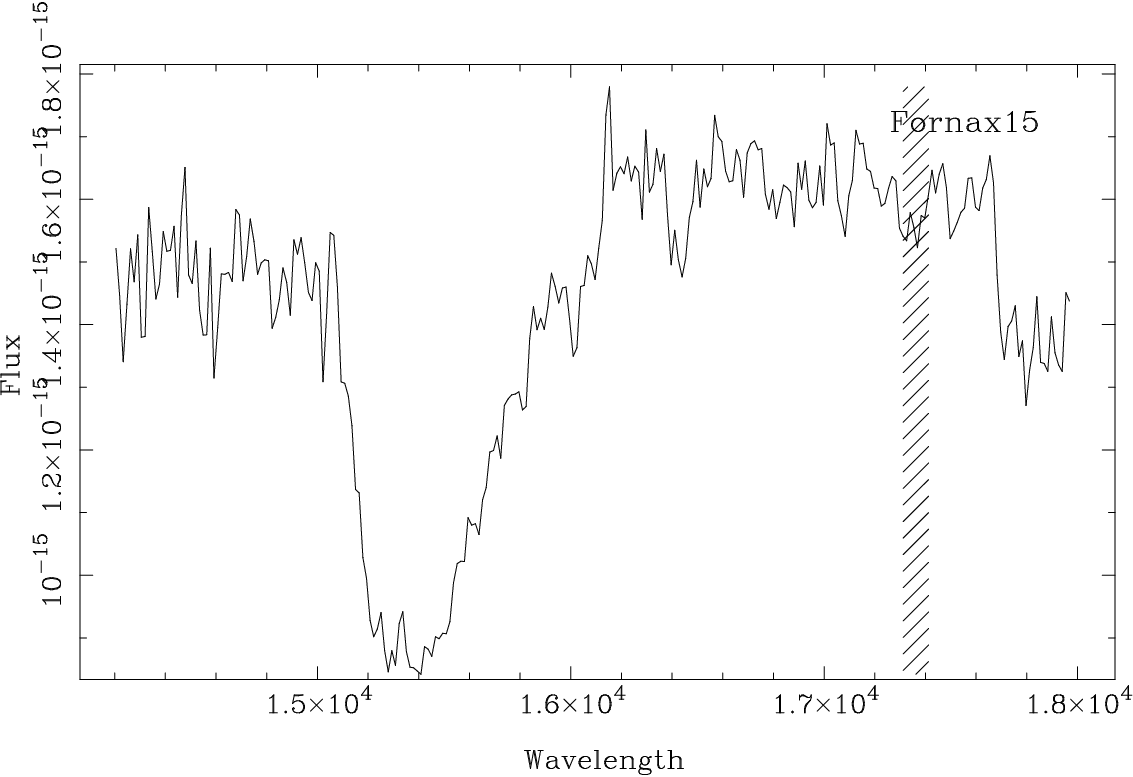}} 
\end{minipage}
 
\begin{minipage}{0.48\textwidth}
\resizebox{\hsize}{!}{\includegraphics{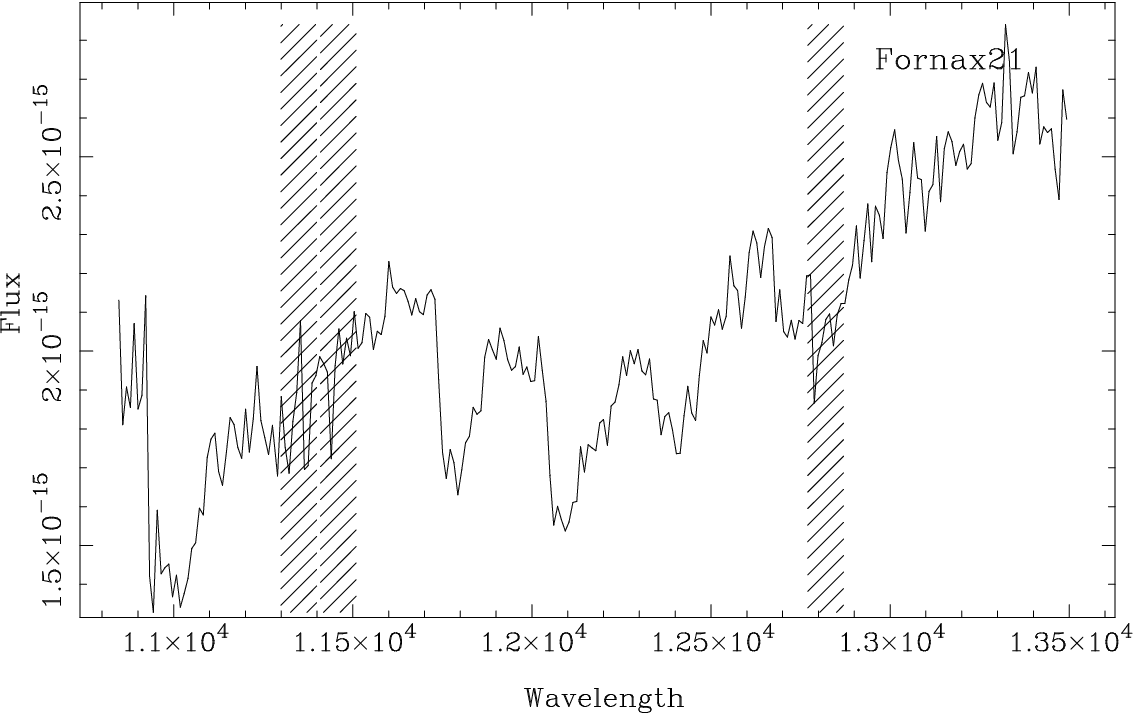}} 
\end{minipage}
\begin{minipage}{0.48\textwidth}
\resizebox{\hsize}{!}{\includegraphics{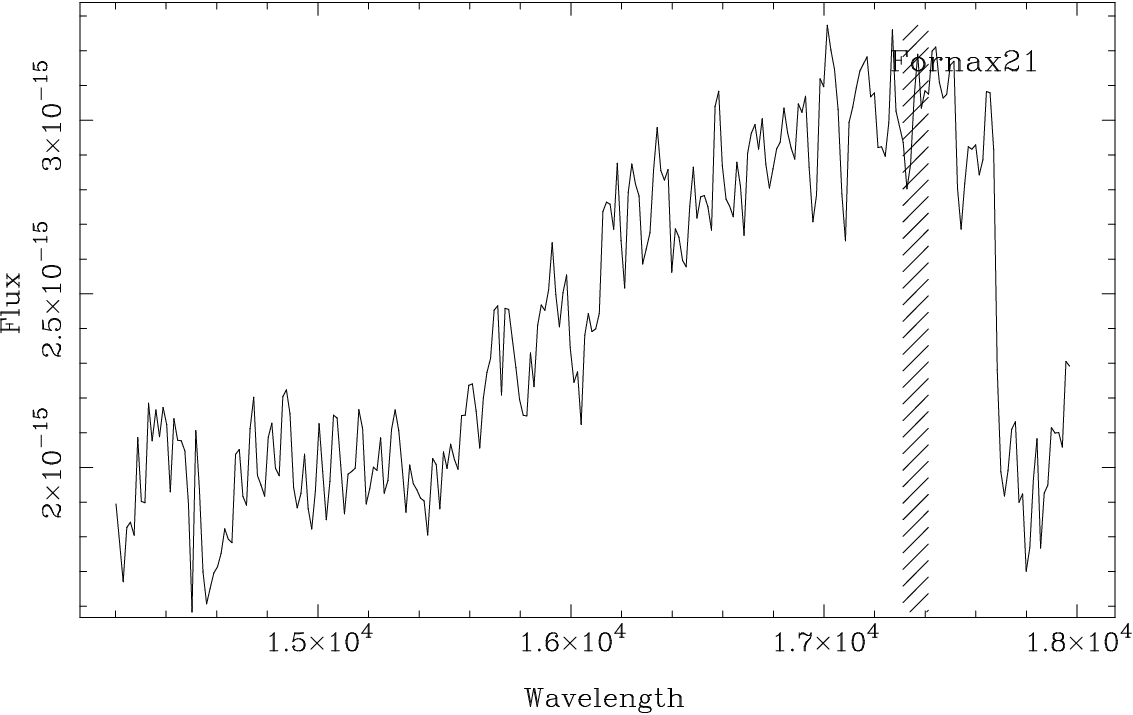}} 
\end{minipage}
 
\begin{minipage}{0.48\textwidth}
\resizebox{\hsize}{!}{\includegraphics{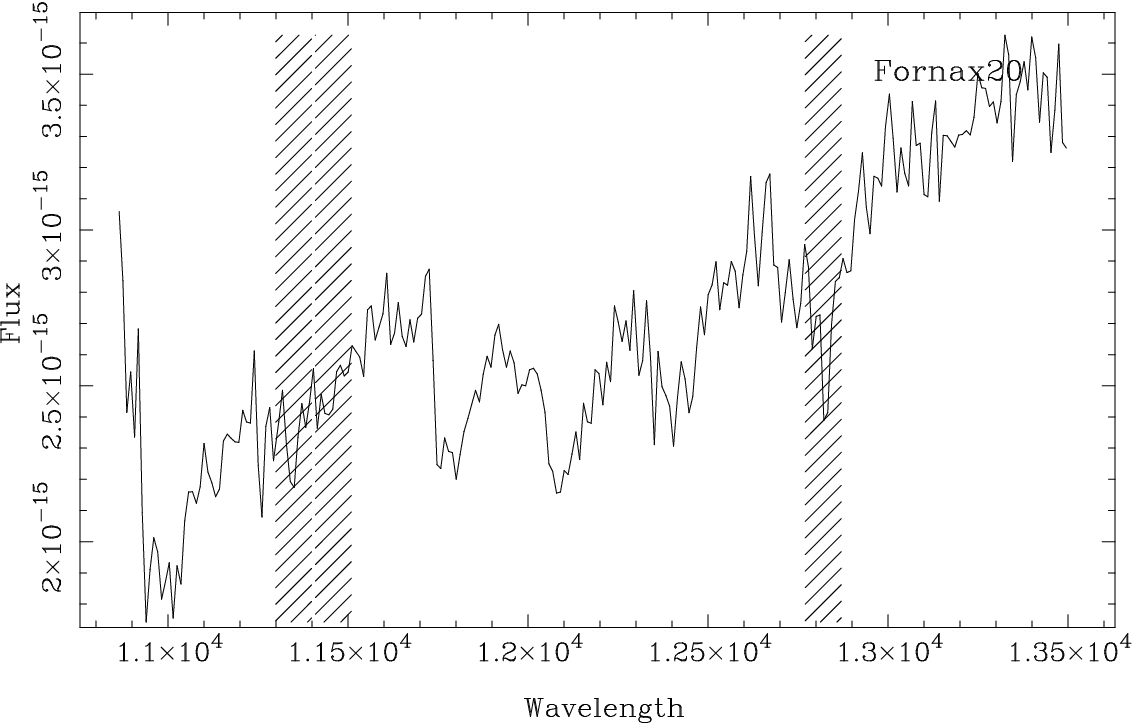}} 
\end{minipage}
\begin{minipage}{0.48\textwidth}
\resizebox{\hsize}{!}{\includegraphics{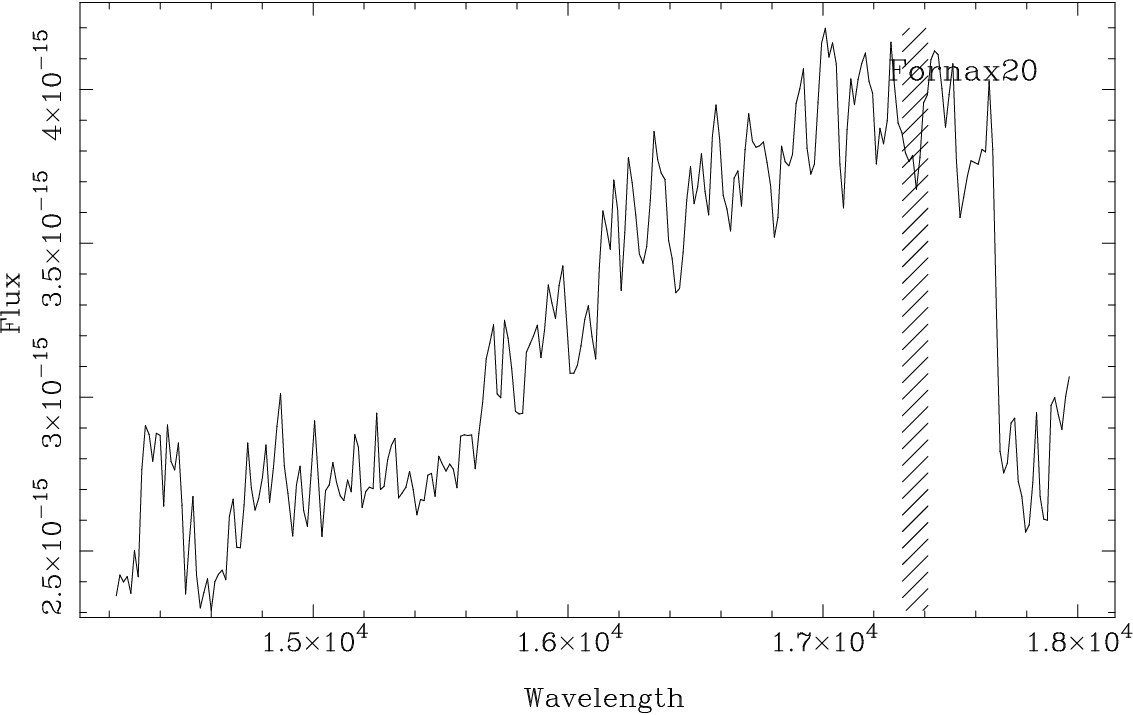}} 
\end{minipage}
 
\end{figure*}
}

\onlfig{6}{
\begin{figure*}[H]

\begin{minipage}{0.48\textwidth}
\resizebox{\hsize}{!}{\includegraphics{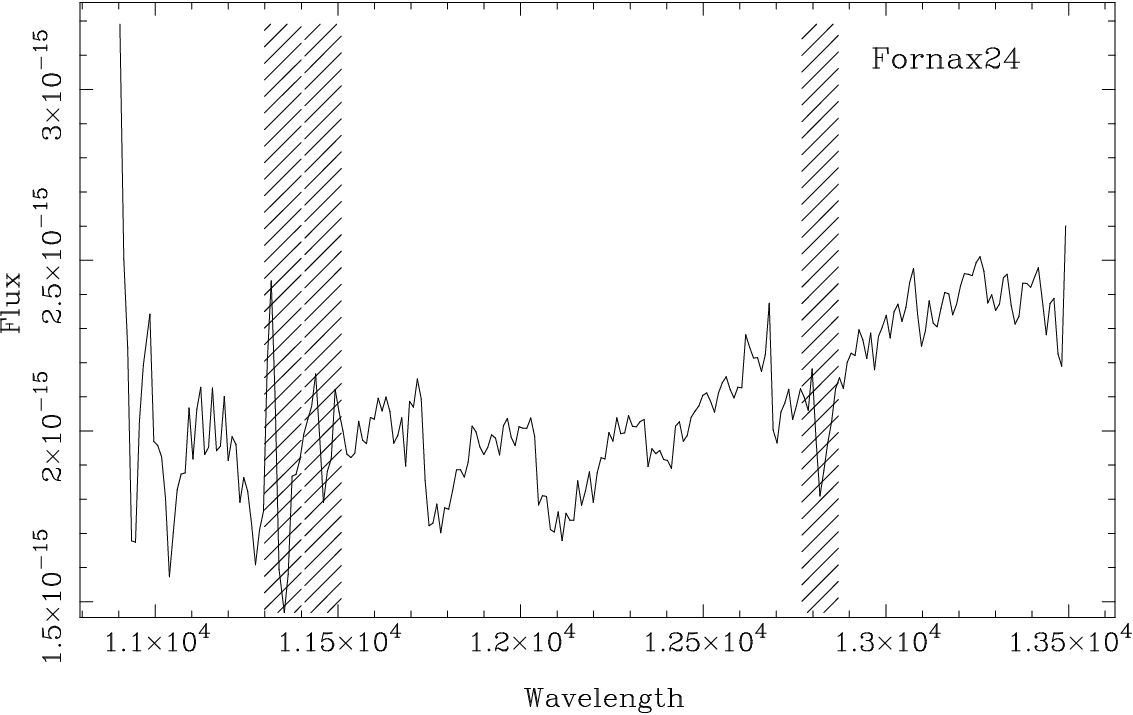}} 
\end{minipage}
\begin{minipage}{0.48\textwidth}
\resizebox{\hsize}{!}{\includegraphics{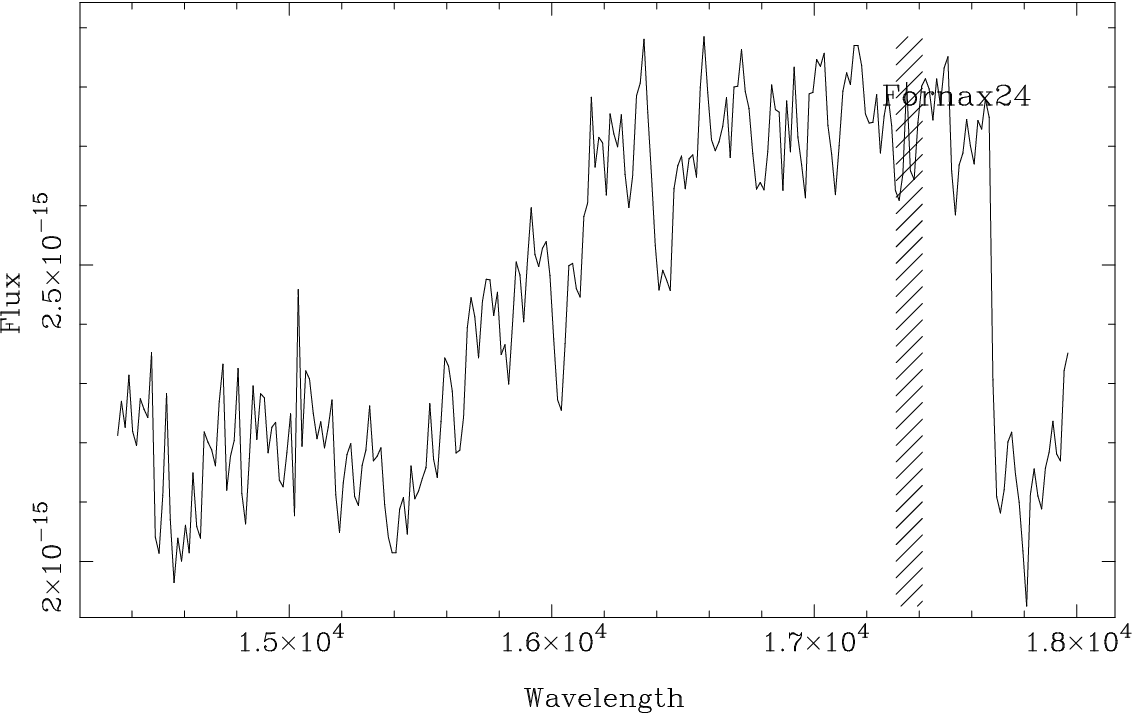}} 
\end{minipage}
 
\begin{minipage}{0.48\textwidth}
\resizebox{\hsize}{!}{\includegraphics{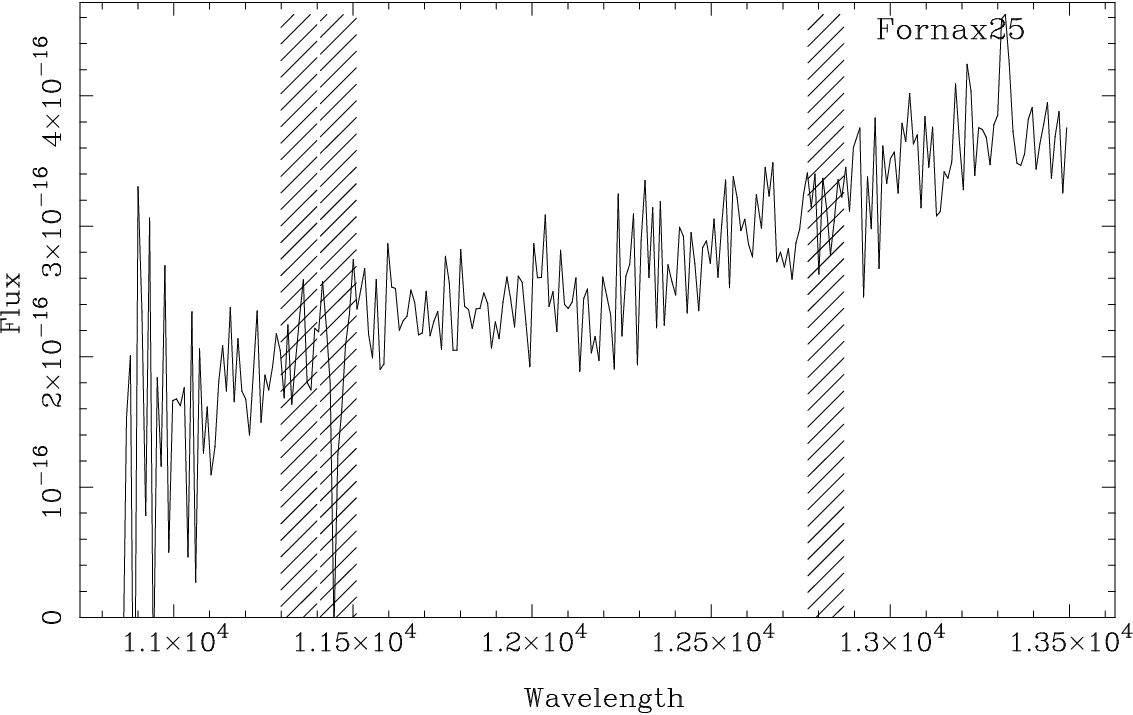}} 
\end{minipage}
\begin{minipage}{0.48\textwidth}
\resizebox{\hsize}{!}{\includegraphics{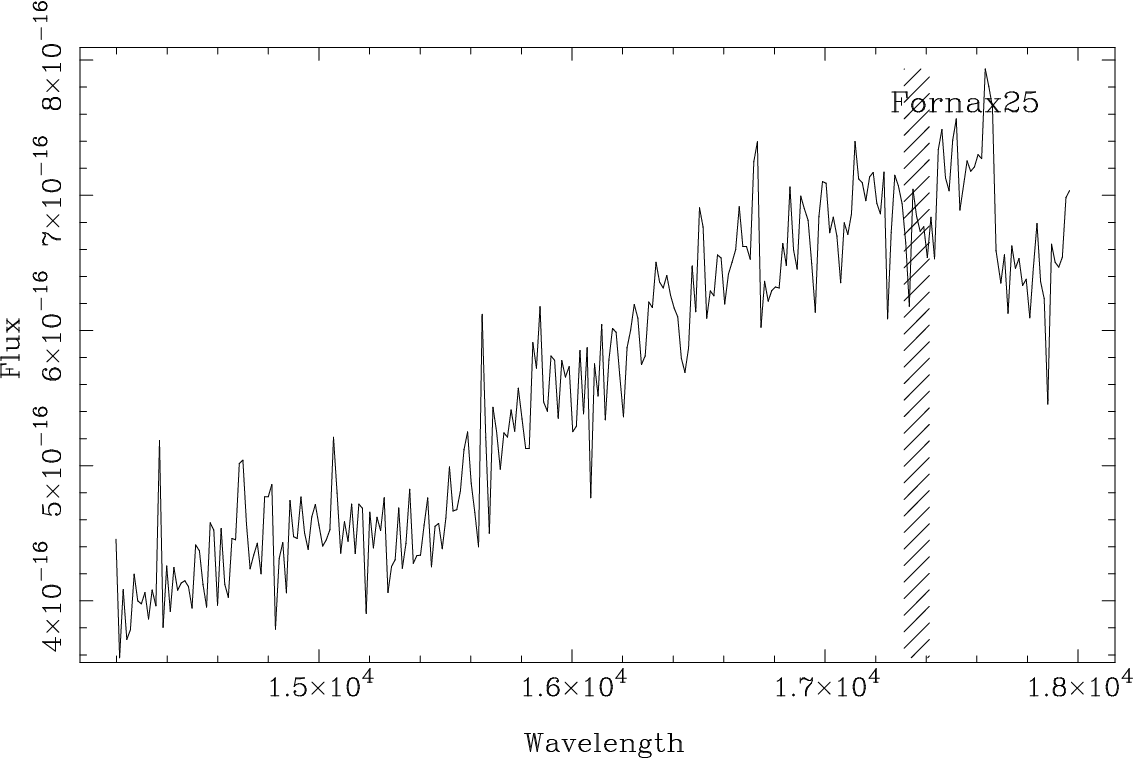}} 
\end{minipage}
  
\begin{minipage}{0.48\textwidth}
\resizebox{\hsize}{!}{\includegraphics{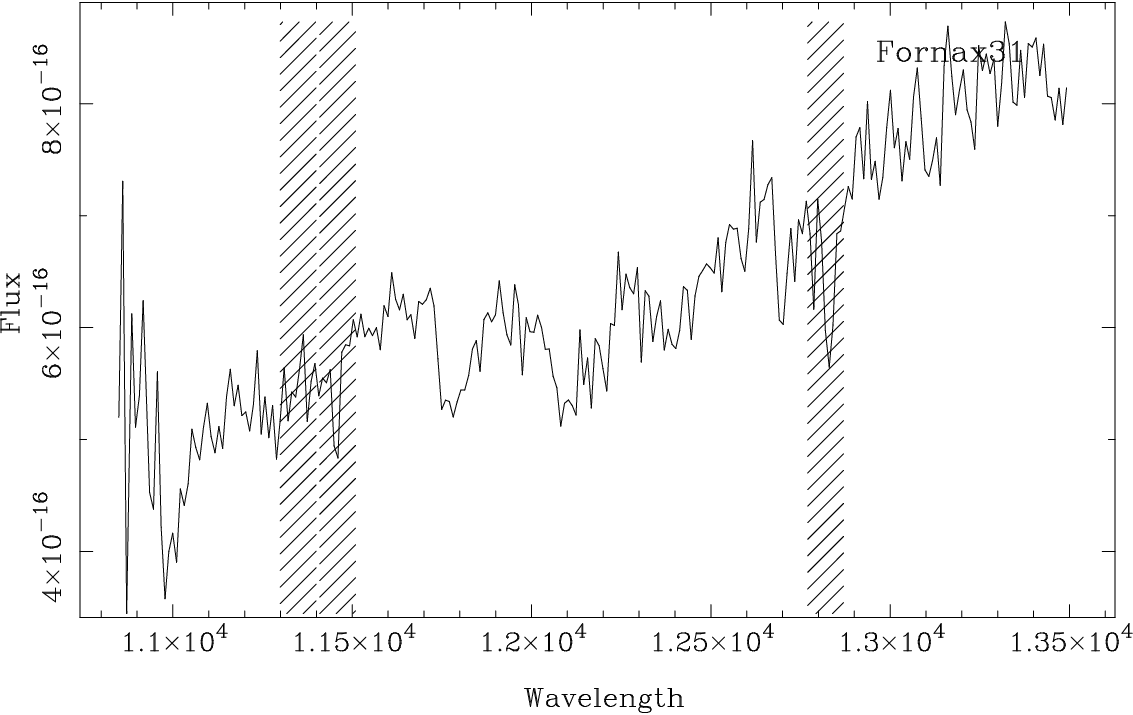}} 
\end{minipage}
\begin{minipage}{0.48\textwidth}
\resizebox{\hsize}{!}{\includegraphics{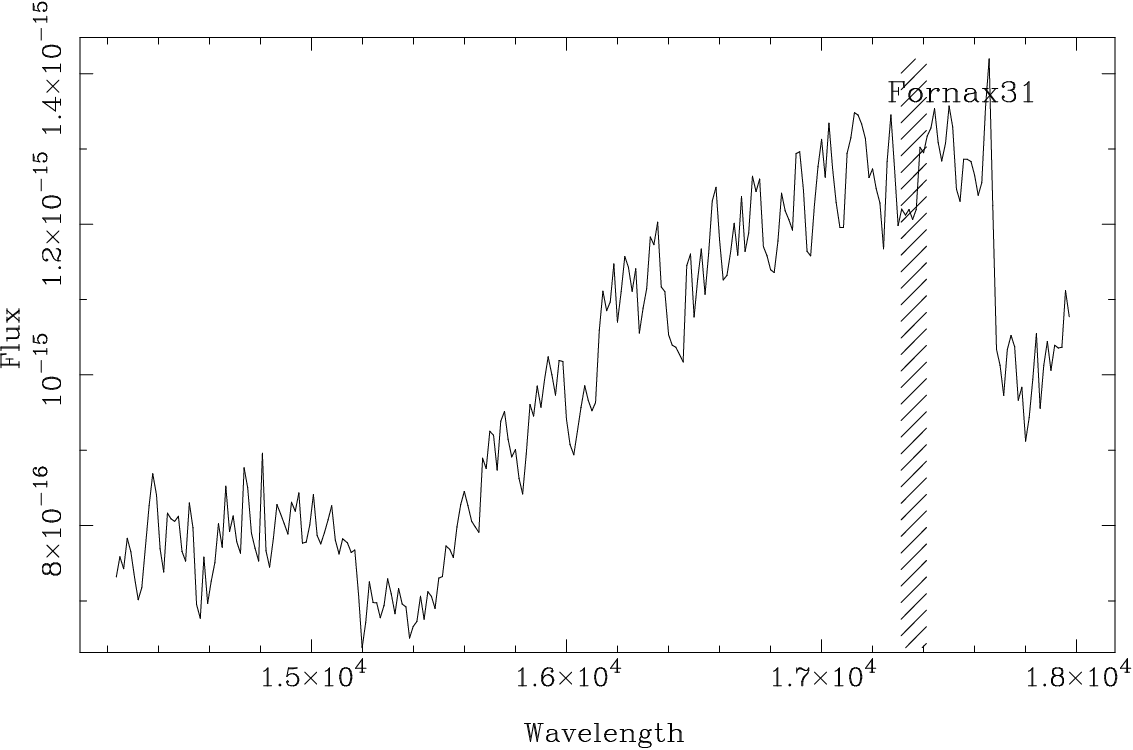}} 
\end{minipage}
 
\begin{minipage}{0.48\textwidth}
\resizebox{\hsize}{!}{\includegraphics{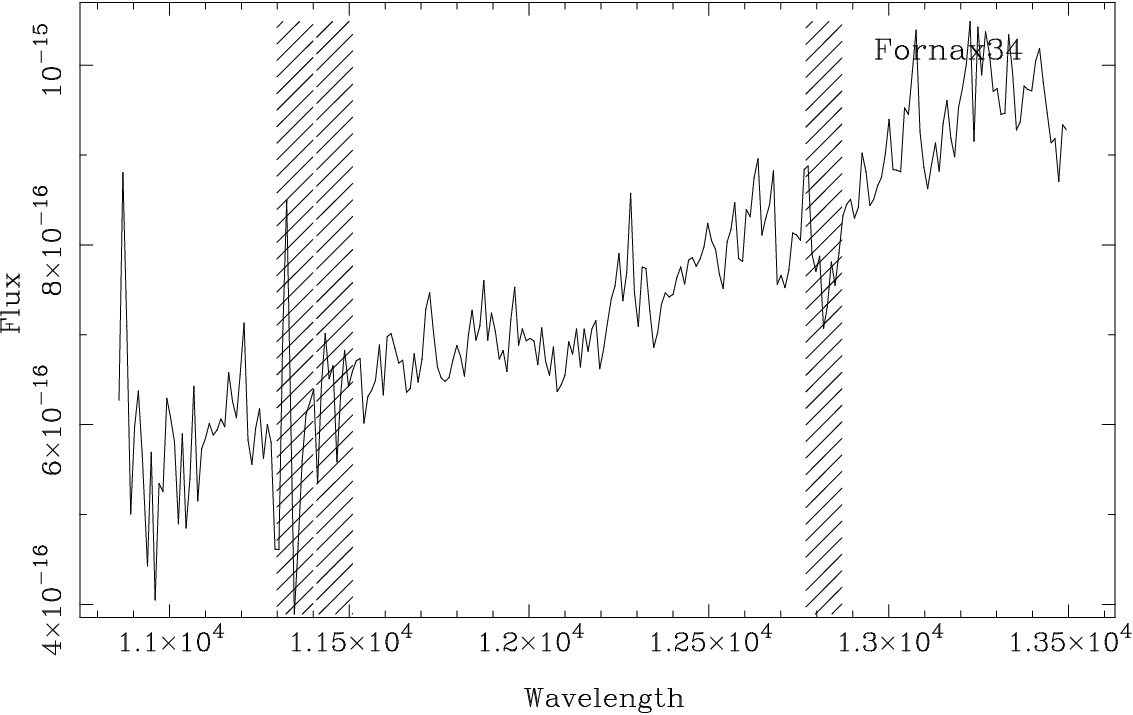}} 
\end{minipage}
\begin{minipage}{0.48\textwidth}
\resizebox{\hsize}{!}{\includegraphics{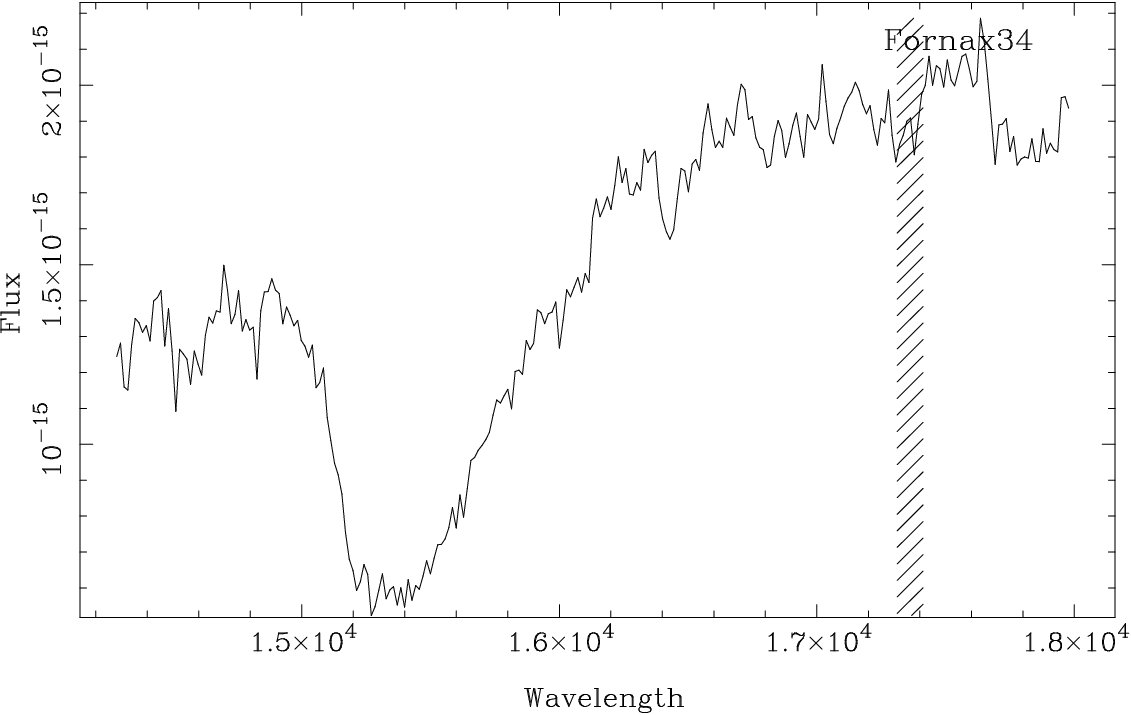}} 
\end{minipage}

\end{figure*}
}

\onlfig{6}{
\begin{figure*}[H]

\begin{minipage}{0.48\textwidth}
\resizebox{\hsize}{!}{\includegraphics{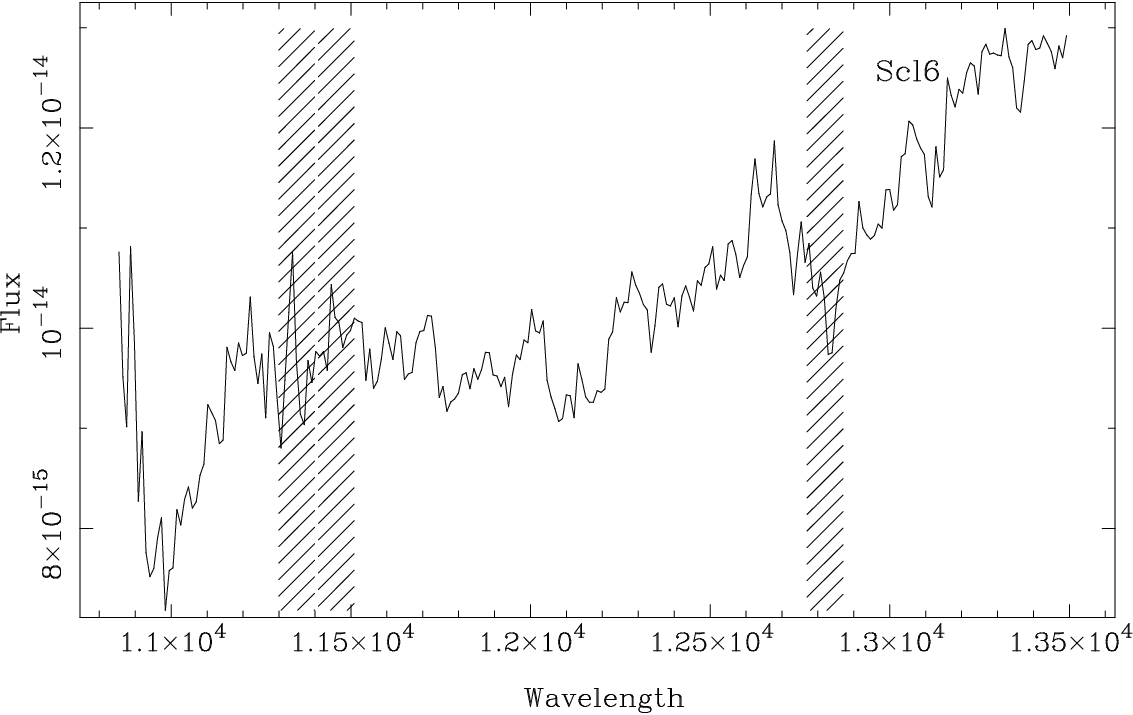}}
\end{minipage}
\begin{minipage}{0.48\textwidth}
\resizebox{\hsize}{!}{\includegraphics{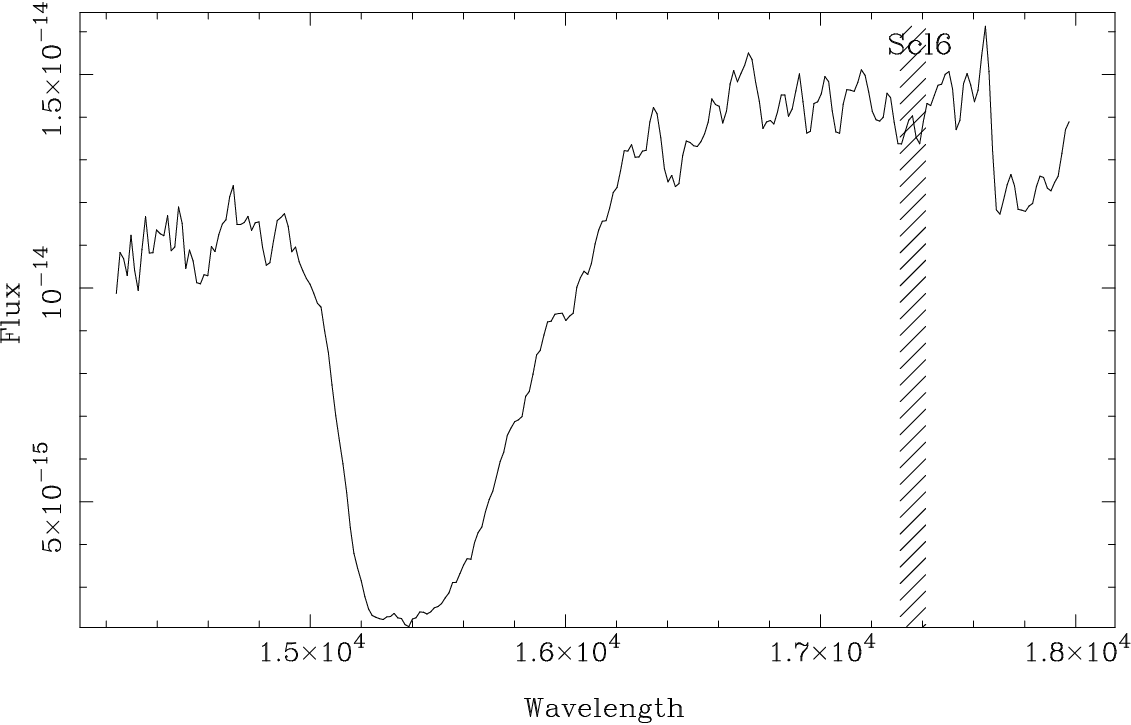}}
\end{minipage}
 
\begin{minipage}{0.48\textwidth}
\resizebox{\hsize}{!}{\includegraphics{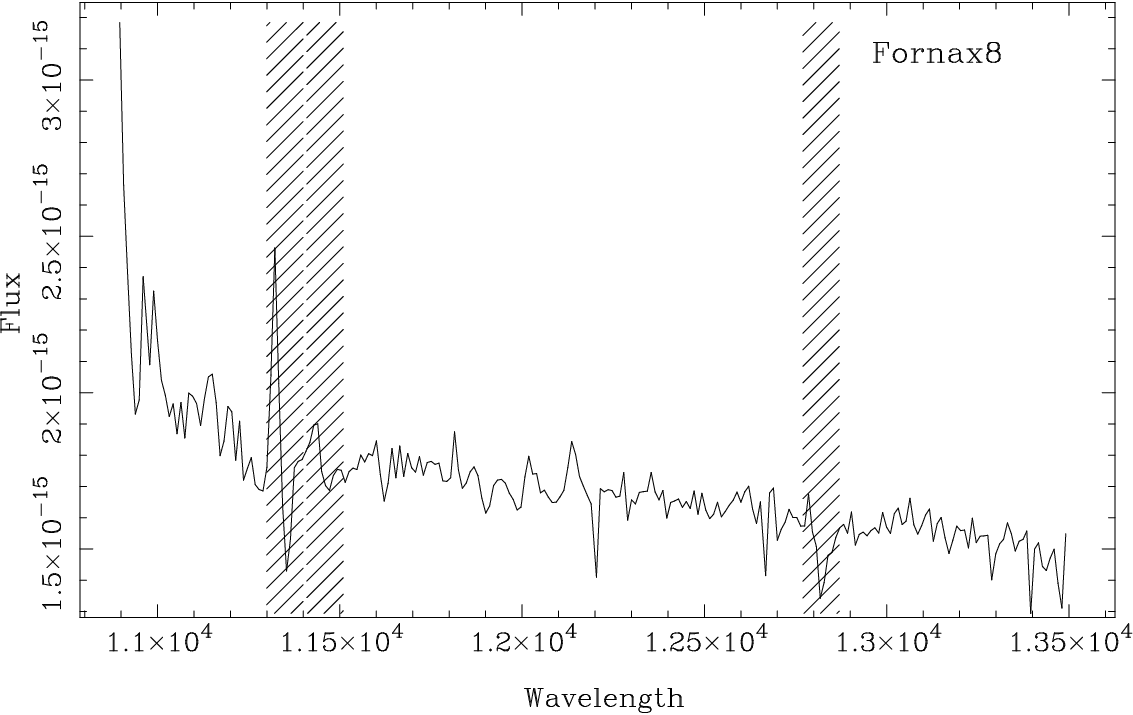}} 
\end{minipage}
\begin{minipage}{0.48\textwidth}
\resizebox{\hsize}{!}{\includegraphics{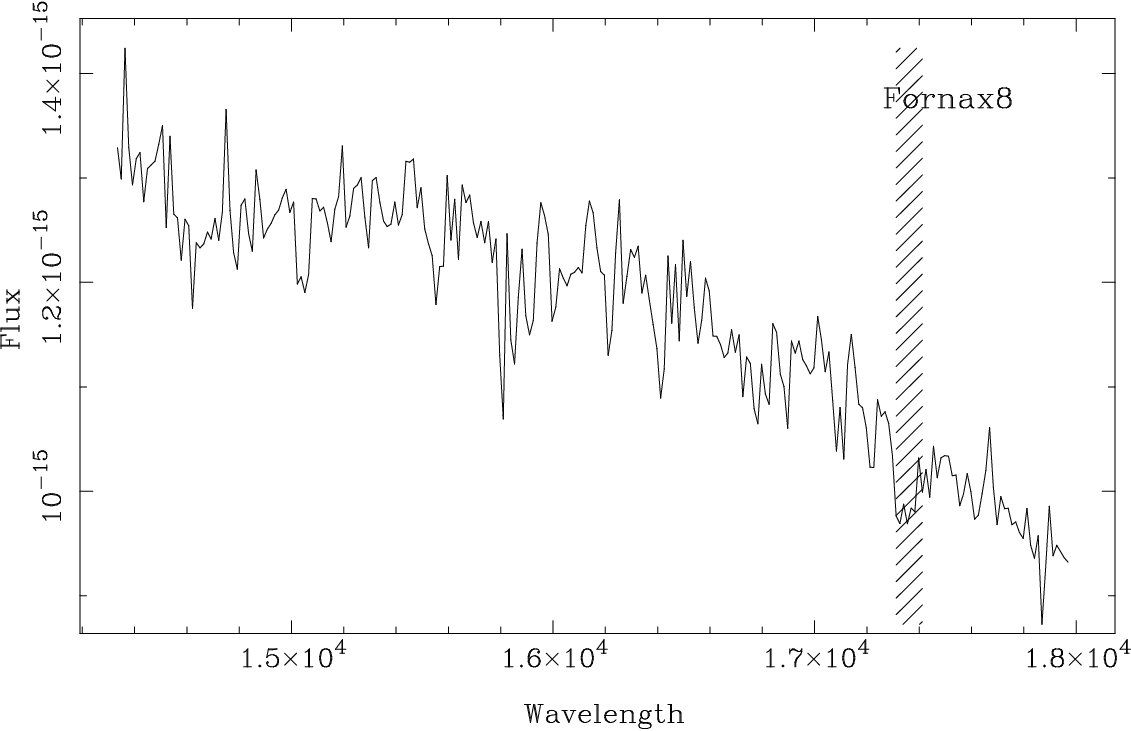}} 
\end{minipage}

\begin{minipage}{0.48\textwidth}
\resizebox{\hsize}{!}{\includegraphics{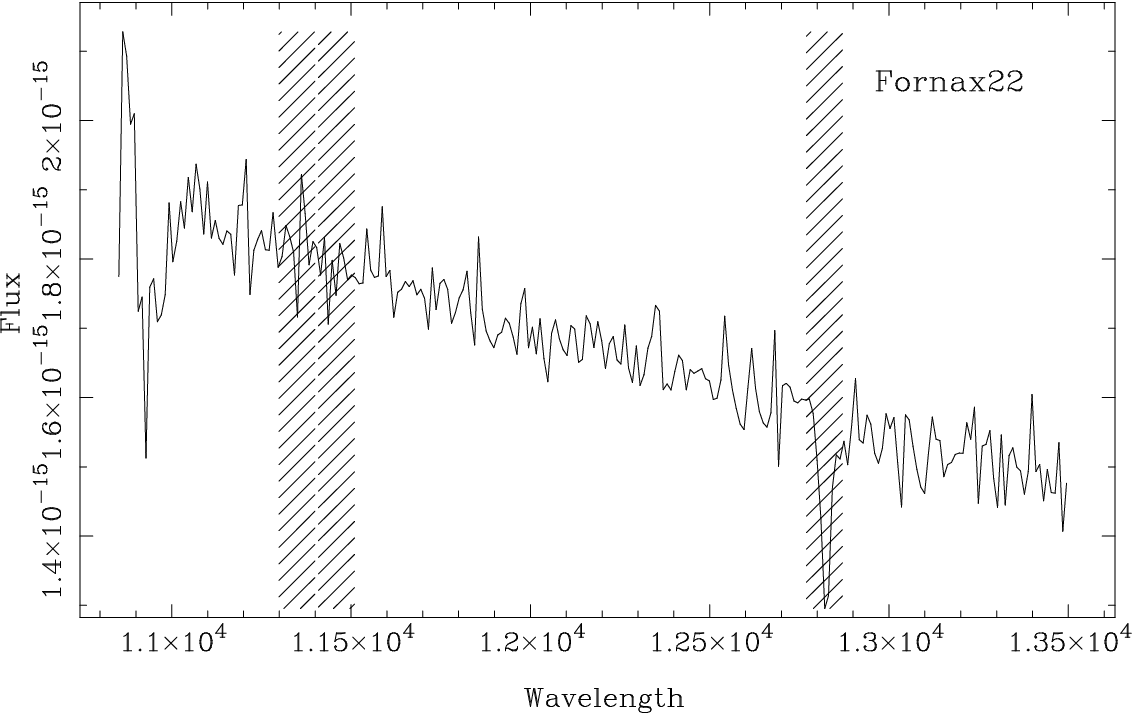}} 
\end{minipage}
\begin{minipage}{0.48\textwidth}
\resizebox{\hsize}{!}{\includegraphics{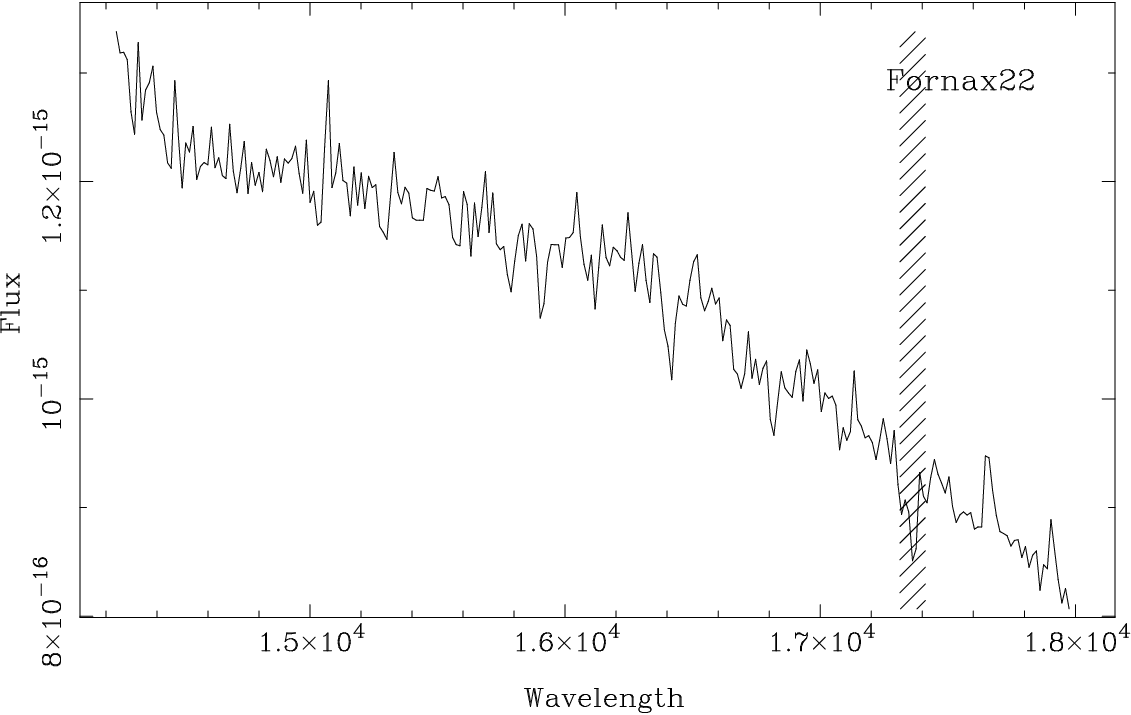}} 
\end{minipage}
 
\begin{minipage}{0.48\textwidth}
\resizebox{\hsize}{!}{\includegraphics{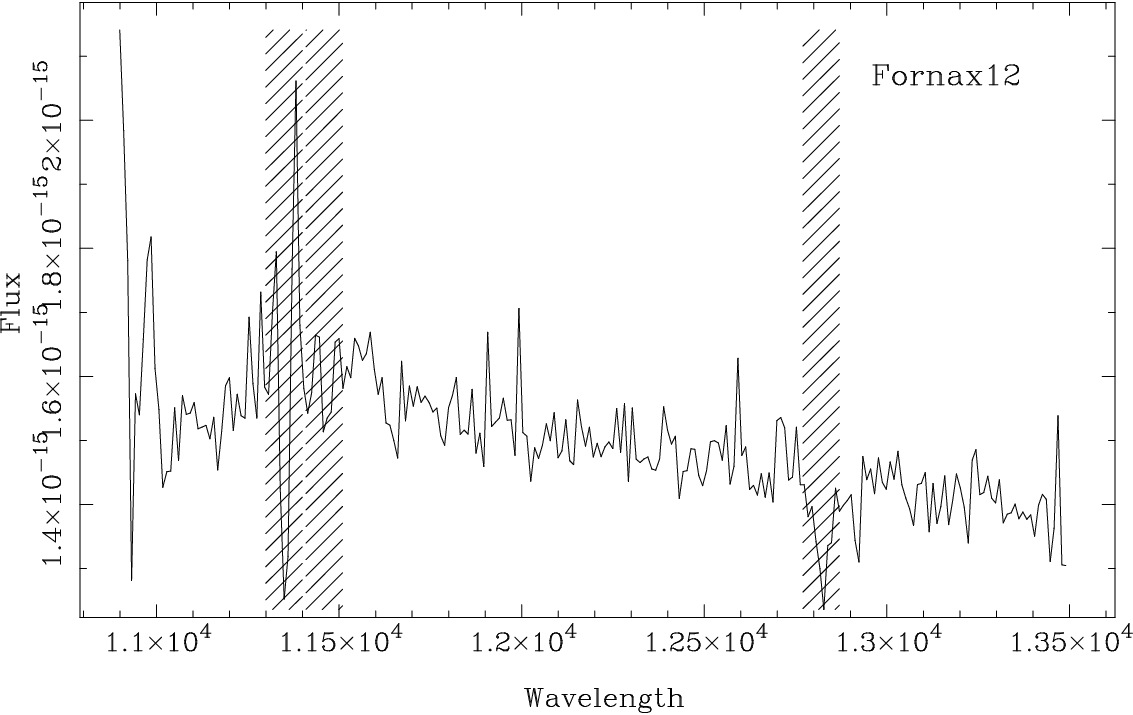}} 
\end{minipage}
\begin{minipage}{0.48\textwidth}
\resizebox{\hsize}{!}{\includegraphics{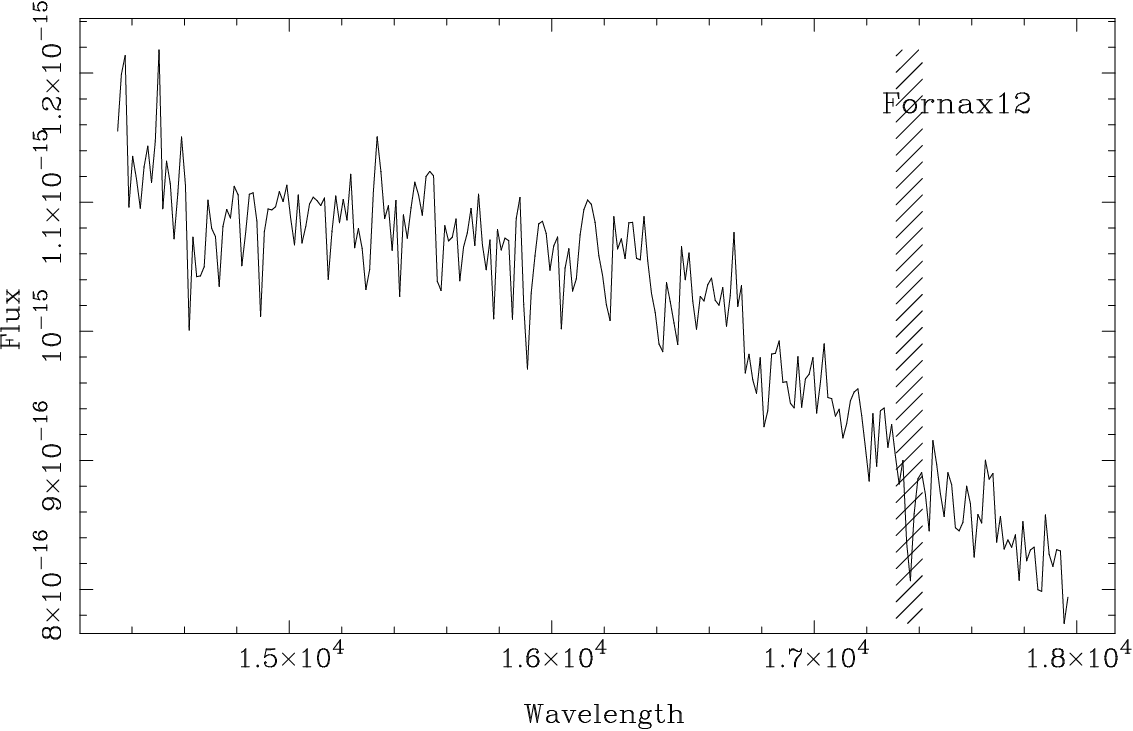}} 
\end{minipage}

\end{figure*}
}

\onlfig{6}{
\begin{figure*}[H]

\begin{minipage}{0.48\textwidth}
\resizebox{\hsize}{!}{\includegraphics{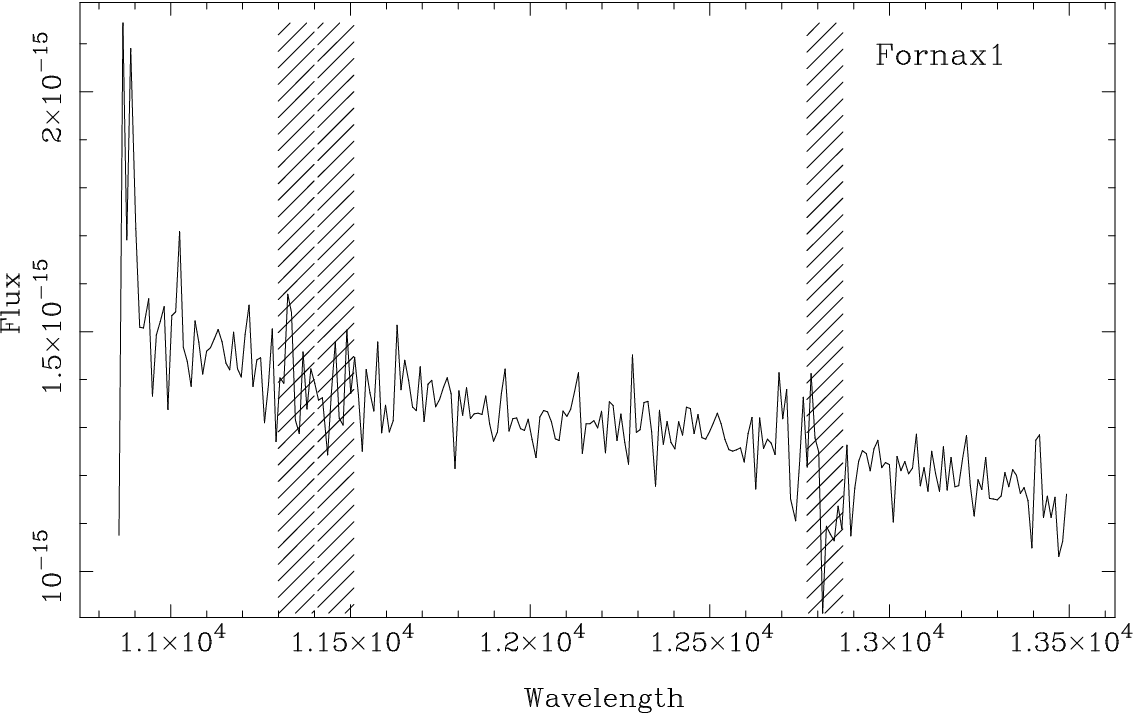}} 
\end{minipage}
\begin{minipage}{0.48\textwidth}
\resizebox{\hsize}{!}{\includegraphics{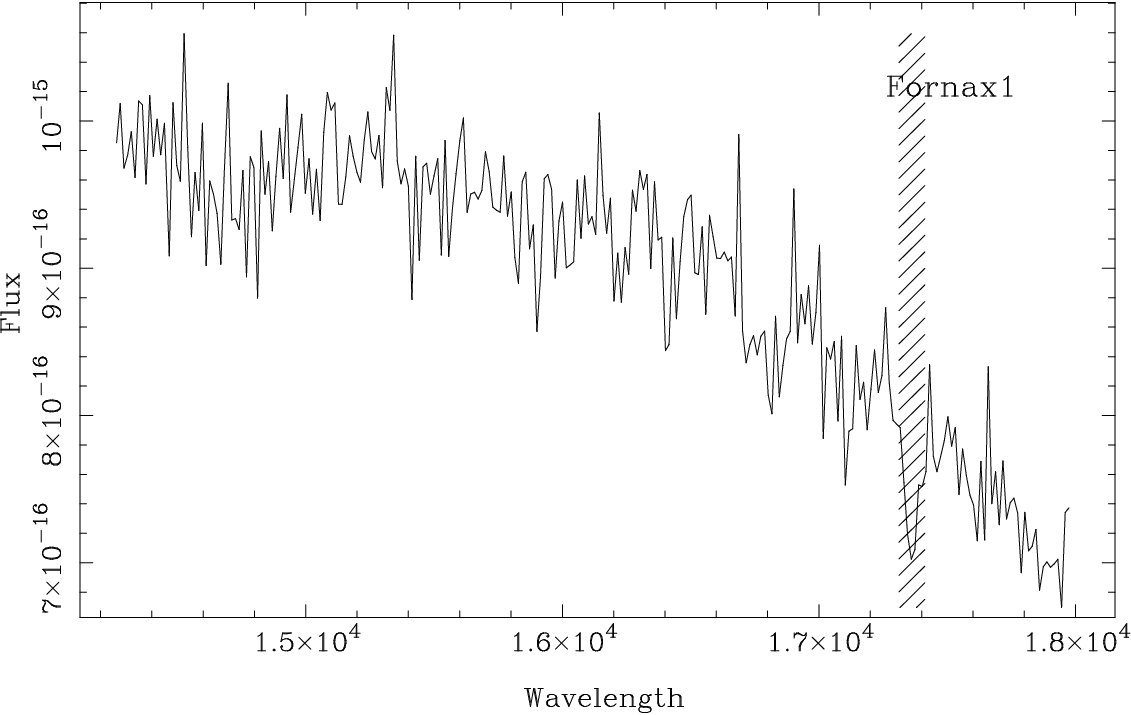}} 
\end{minipage}
 
\begin{minipage}{0.48\textwidth}
\resizebox{\hsize}{!}{\includegraphics{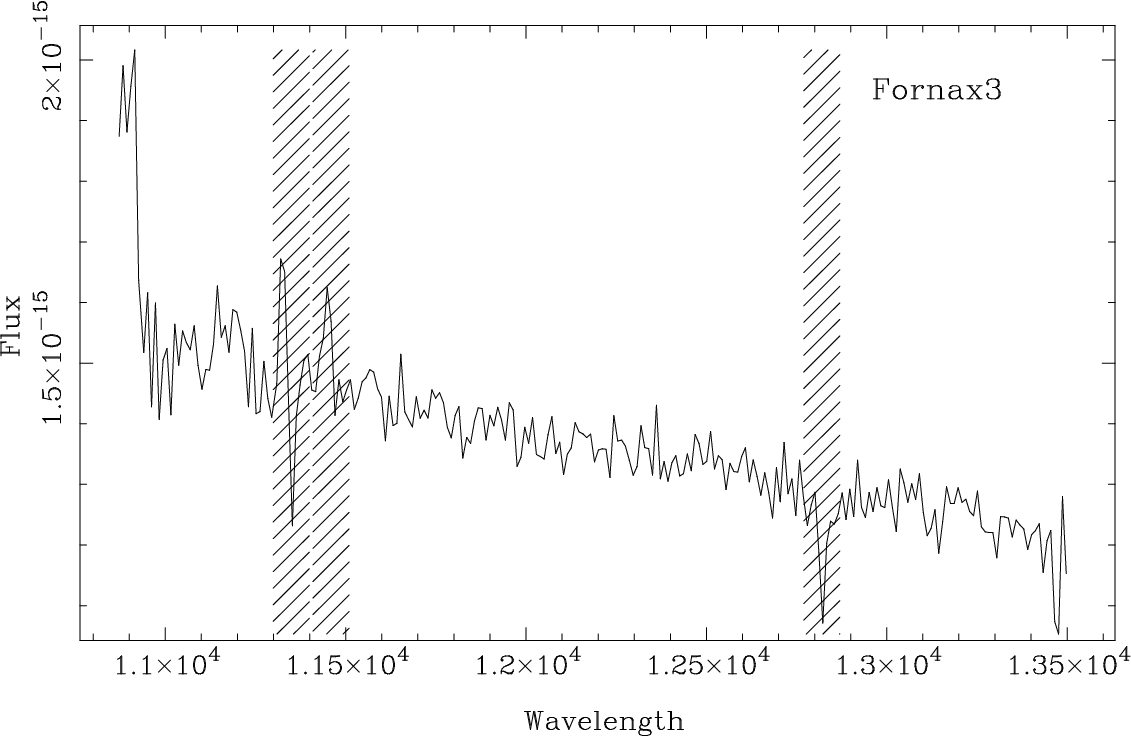}} 
\end{minipage}
\begin{minipage}{0.48\textwidth}
\resizebox{\hsize}{!}{\includegraphics{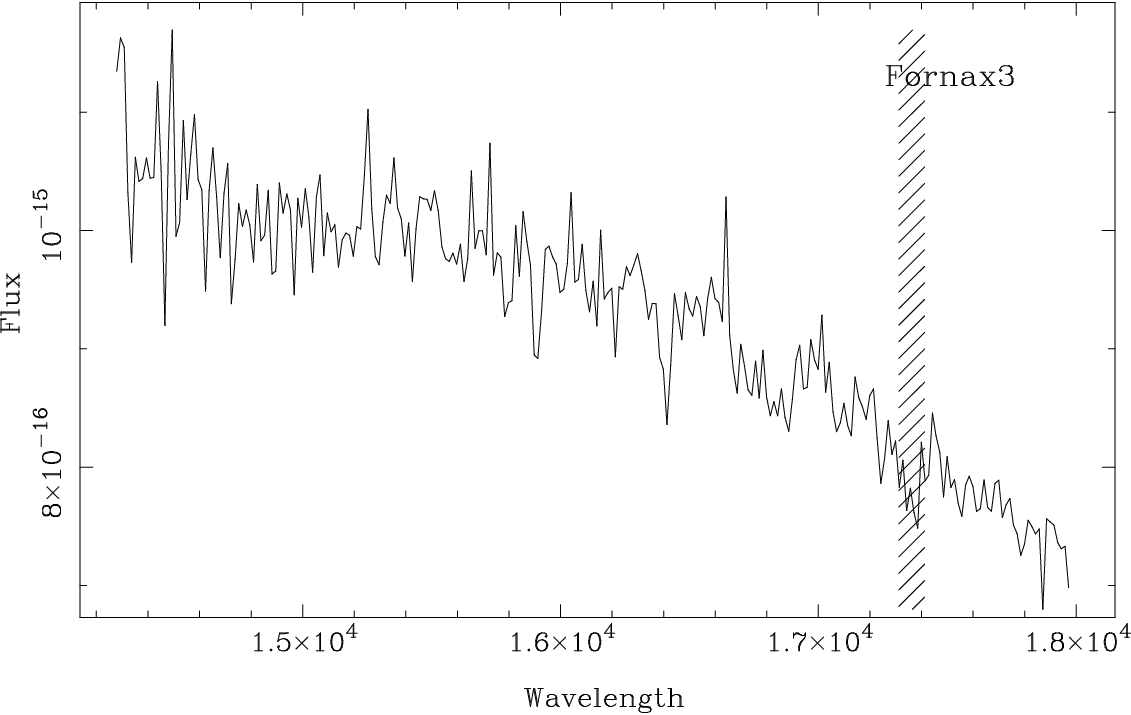}} 
\end{minipage}
 
\begin{minipage}{0.48\textwidth}
\resizebox{\hsize}{!}{\includegraphics{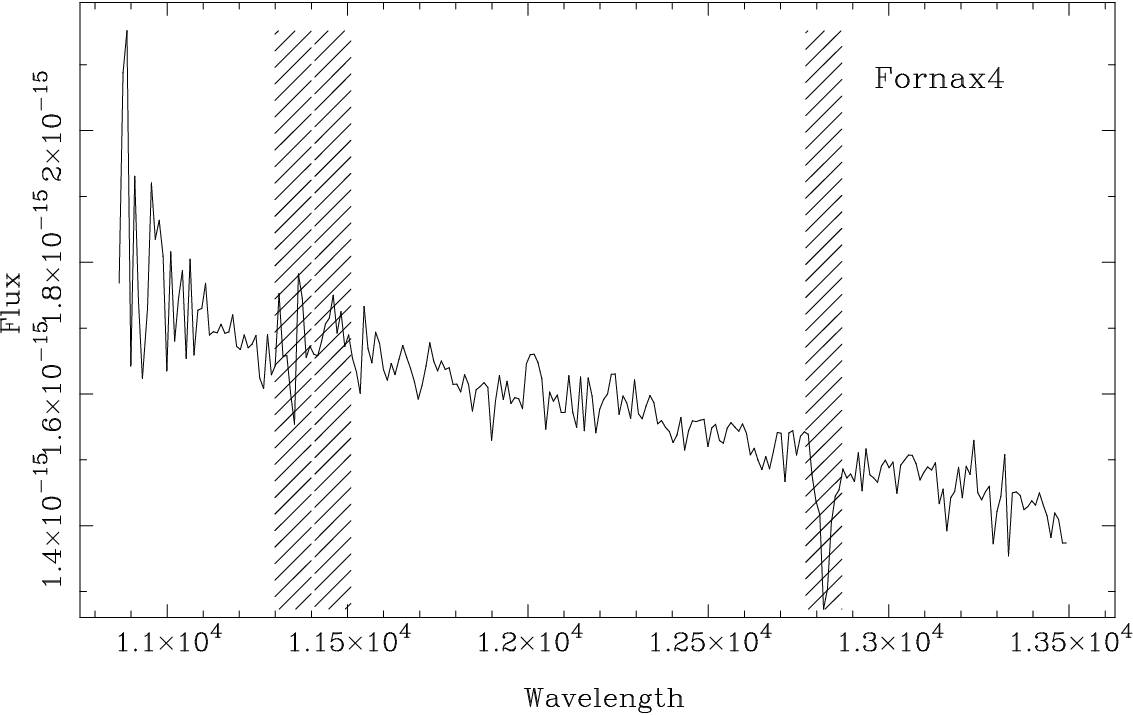}} 
\end{minipage}
\begin{minipage}{0.48\textwidth}
\resizebox{\hsize}{!}{\includegraphics{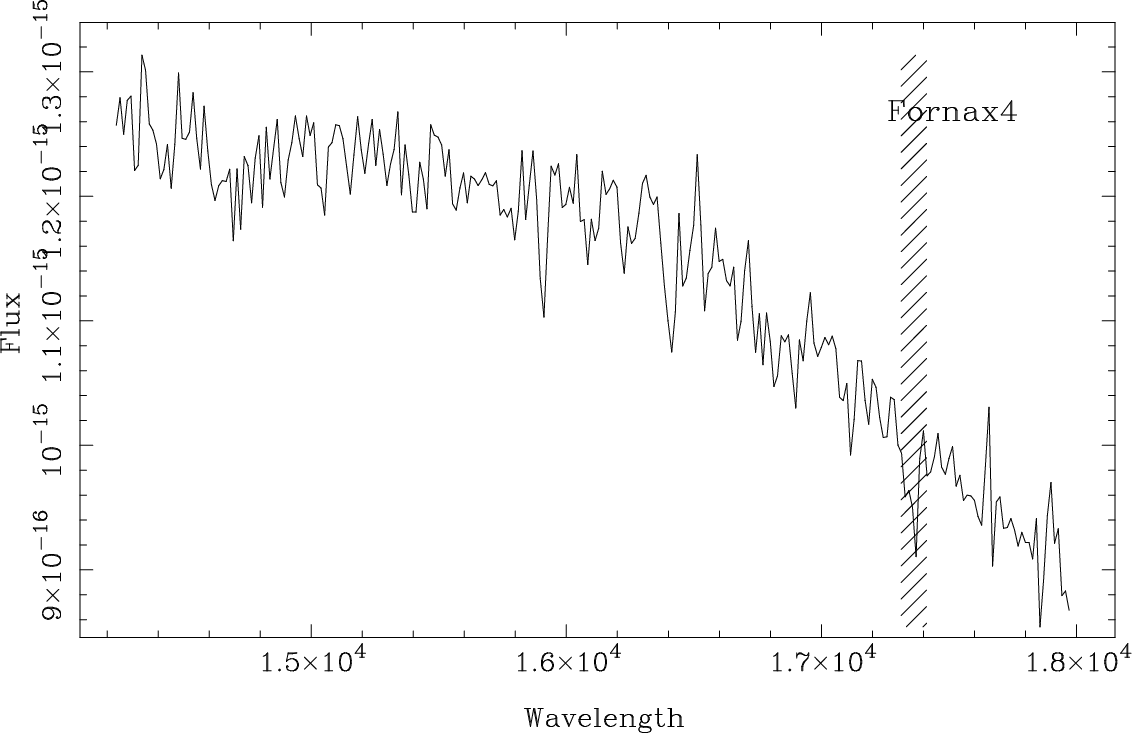}} 
\end{minipage}

\begin{minipage}{0.48\textwidth}
\resizebox{\hsize}{!}{\includegraphics{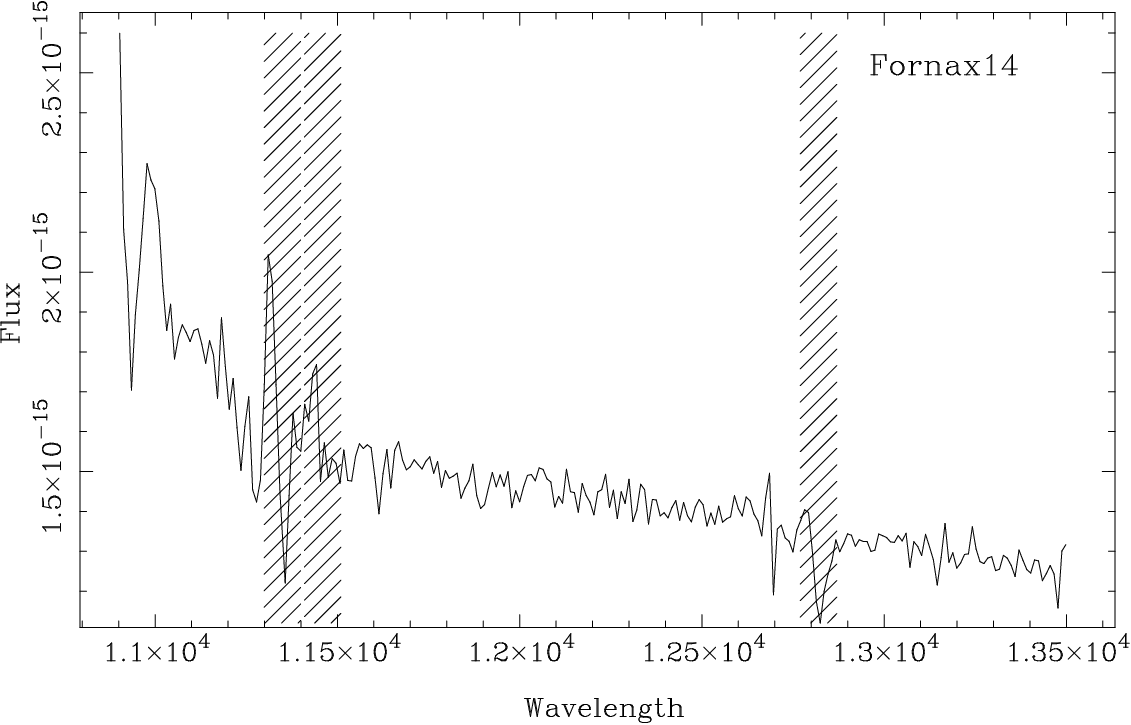}} 
\end{minipage}
\begin{minipage}{0.48\textwidth}
\resizebox{\hsize}{!}{\includegraphics{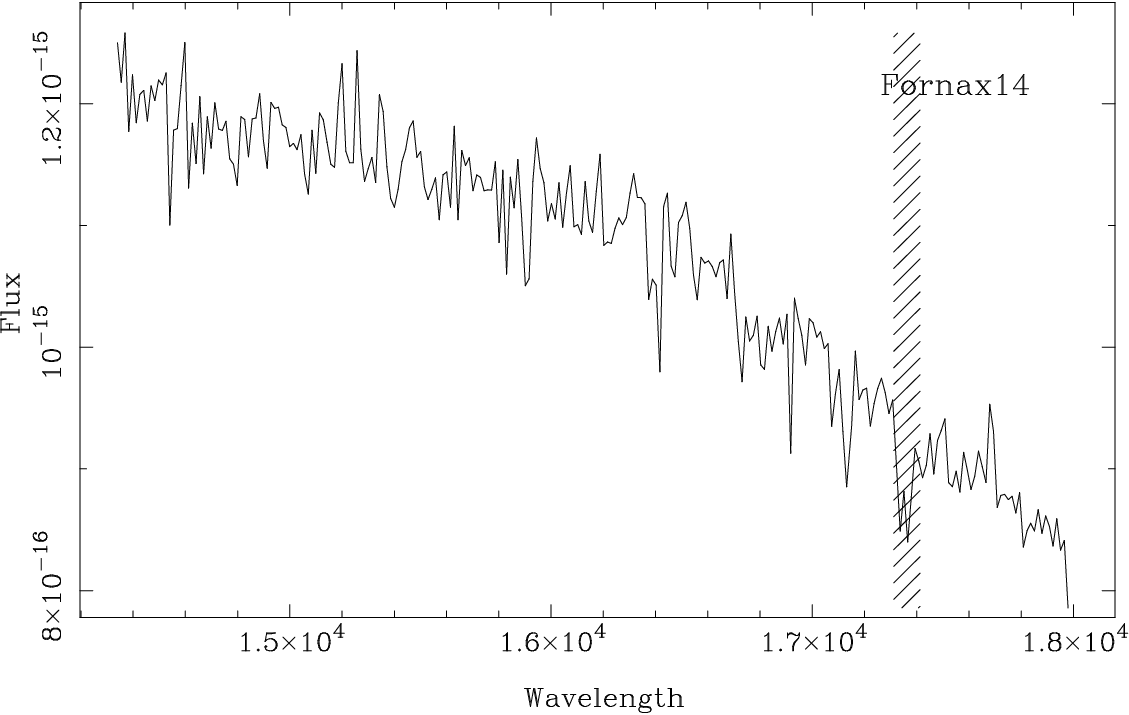}} 
\end{minipage}

\end{figure*}
}

\onlfig{6}{
\begin{figure*}[H]
 
\begin{minipage}{0.48\textwidth}
\resizebox{\hsize}{!}{\includegraphics{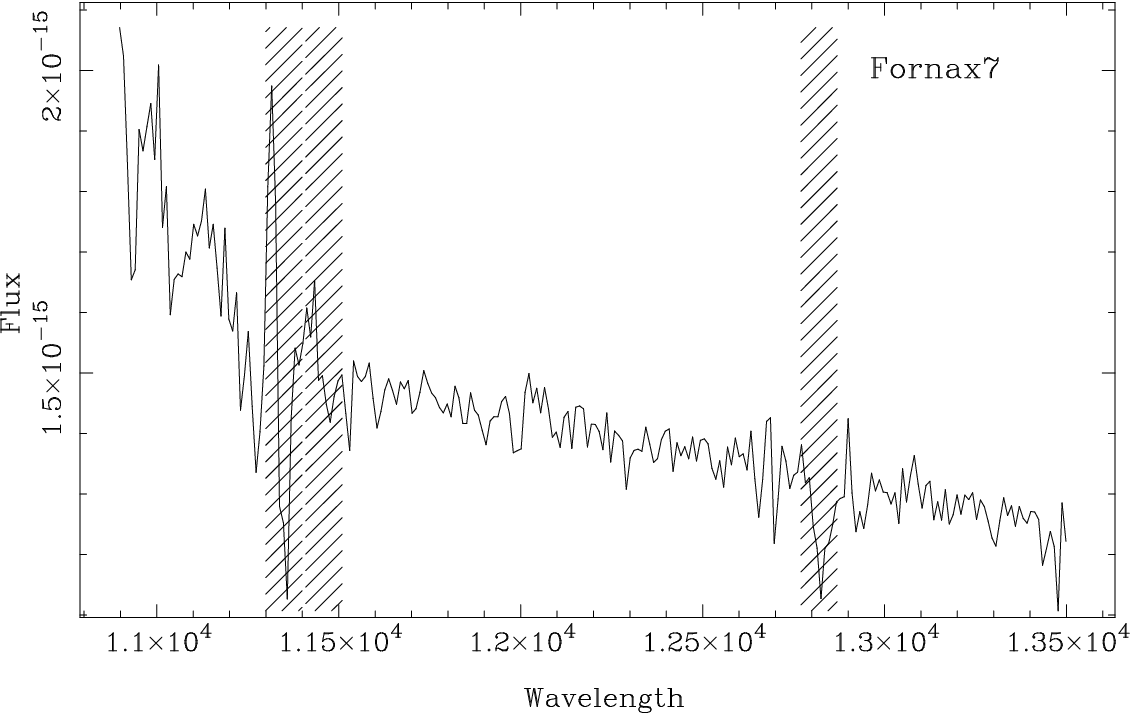}} 
\end{minipage}
\begin{minipage}{0.48\textwidth}
\resizebox{\hsize}{!}{\includegraphics{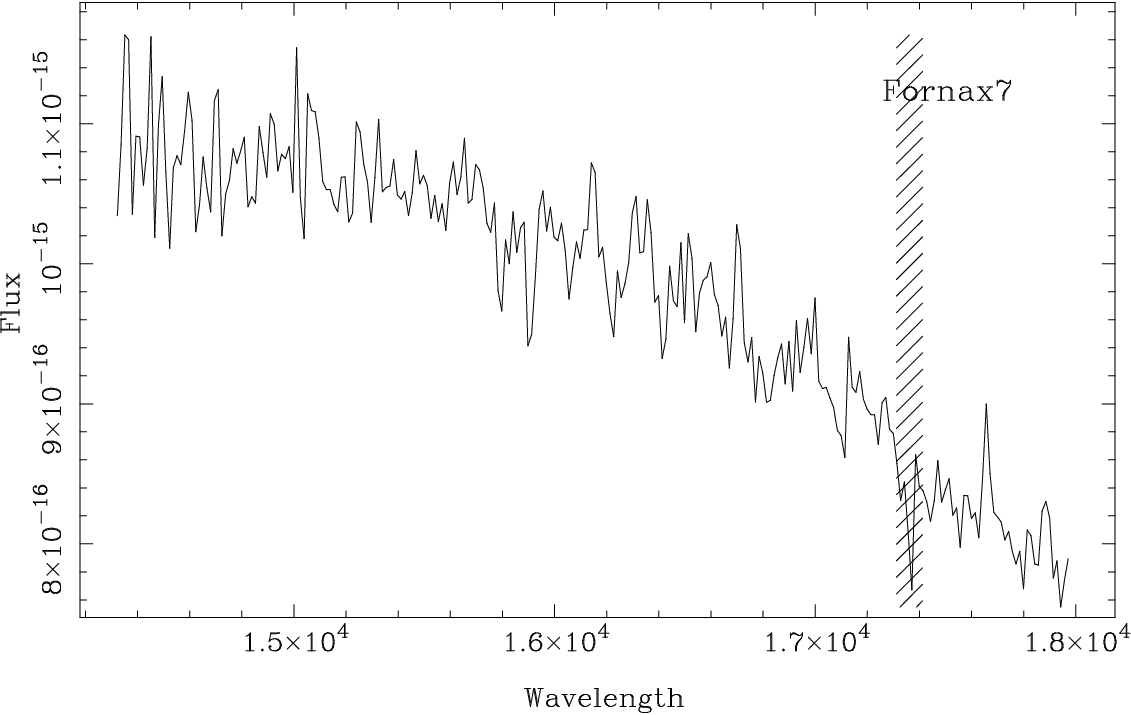}} 
\end{minipage}
 
\begin{minipage}{0.48\textwidth}
\resizebox{\hsize}{!}{\includegraphics{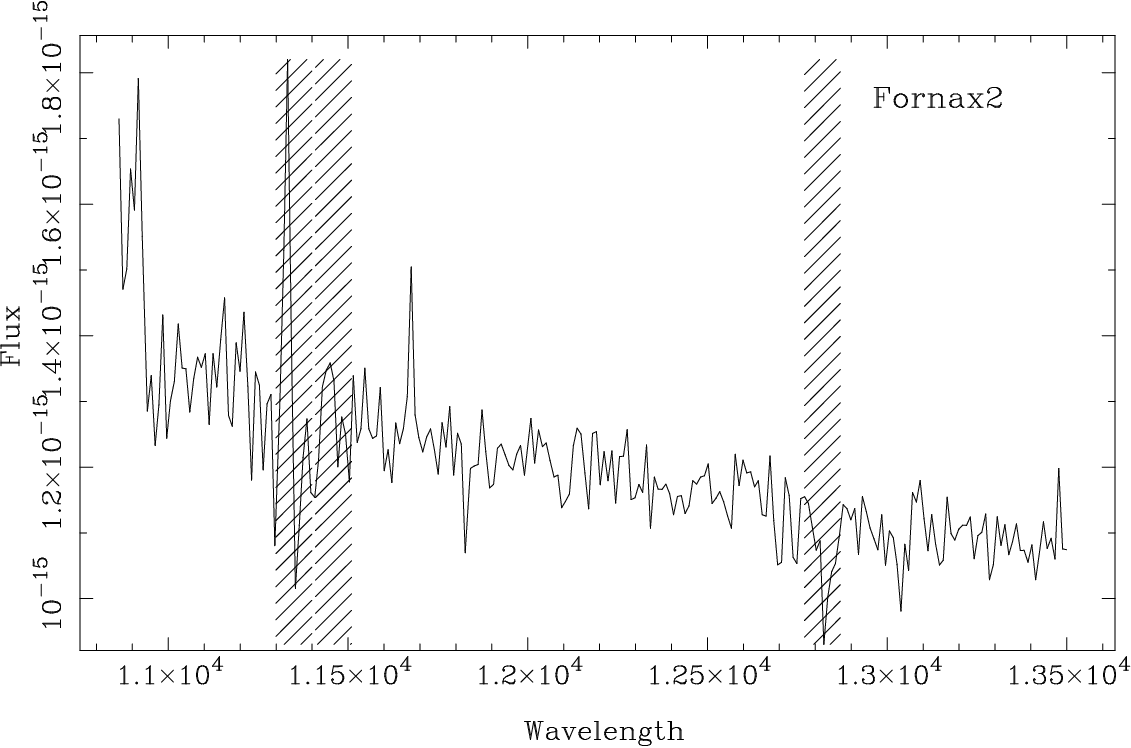}} 
\end{minipage}
\begin{minipage}{0.48\textwidth}
\resizebox{\hsize}{!}{\includegraphics{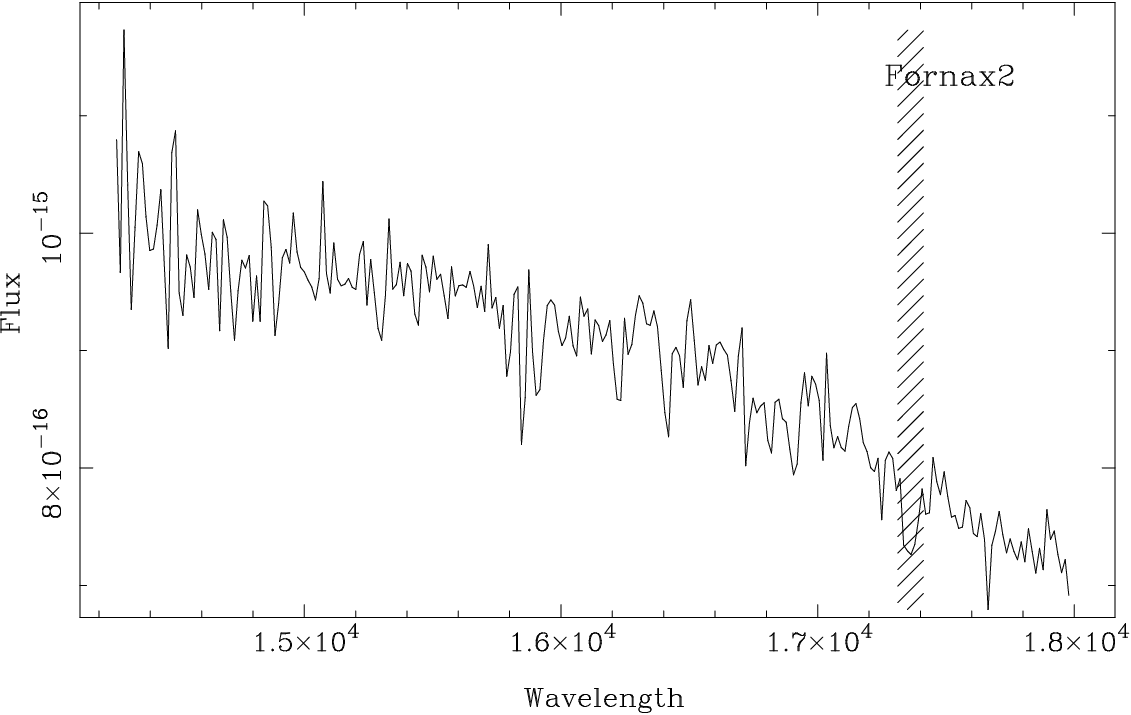}} 
\end{minipage}
 
\begin{minipage}{0.48\textwidth}
\resizebox{\hsize}{!}{\includegraphics{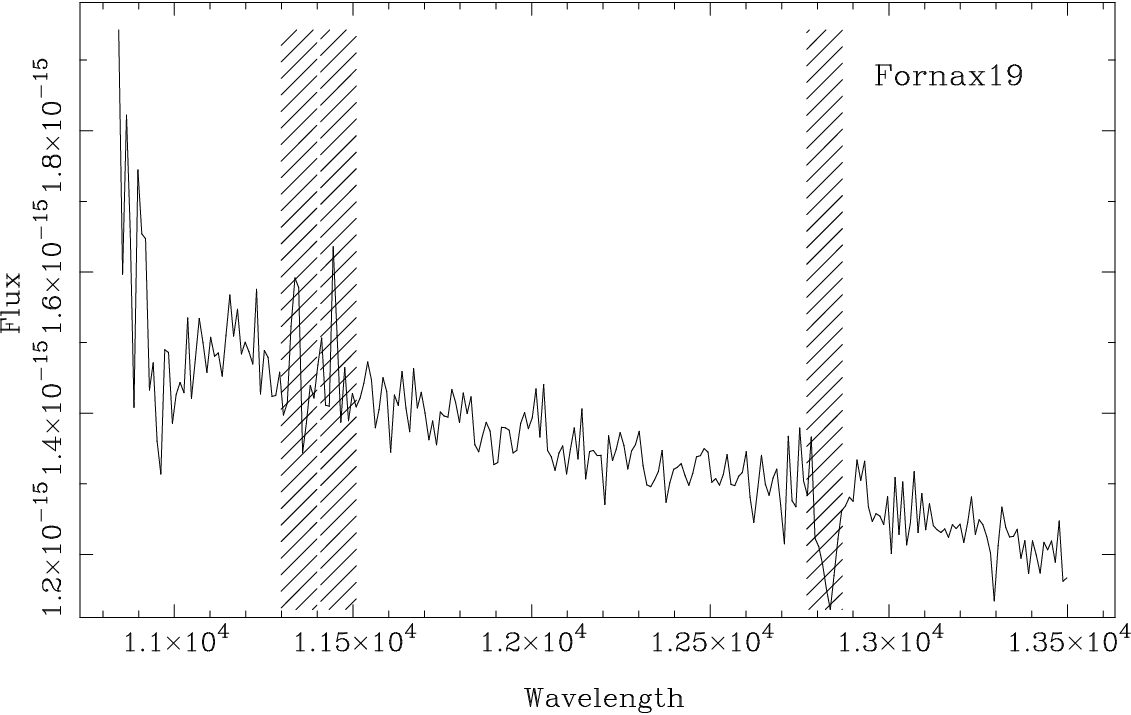}} 
\end{minipage}
\begin{minipage}{0.48\textwidth}
\resizebox{\hsize}{!}{\includegraphics{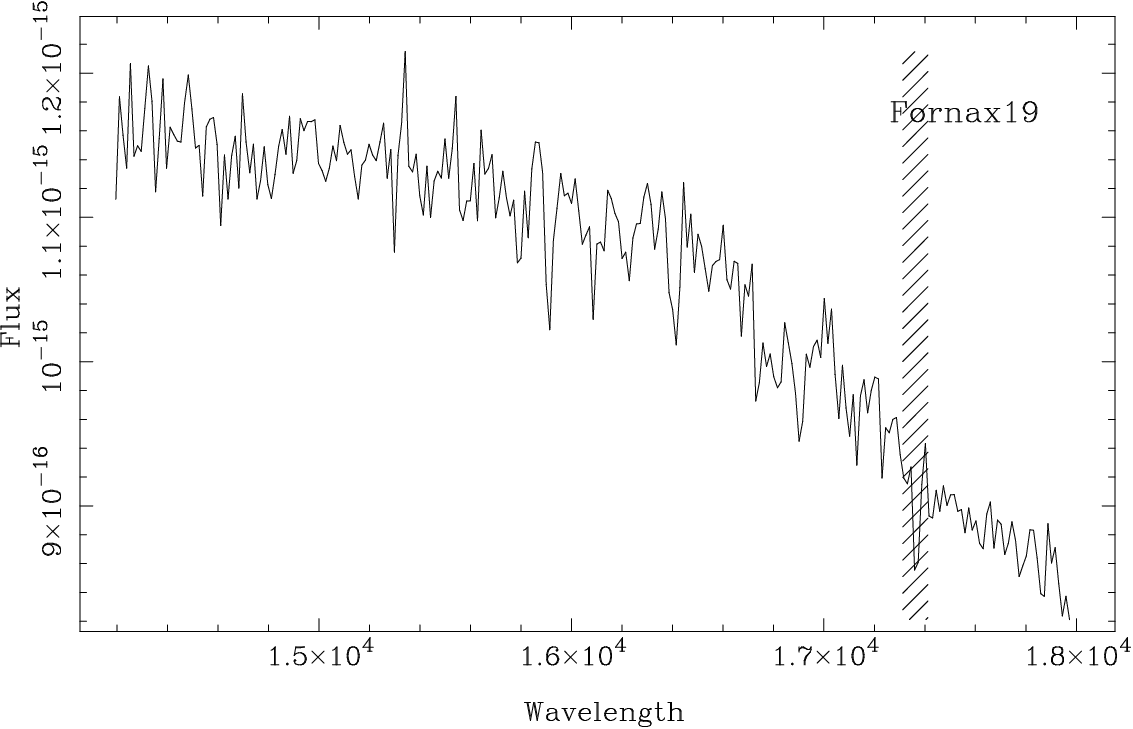}} 
\end{minipage}
 
\begin{minipage}{0.48\textwidth}
\resizebox{\hsize}{!}{\includegraphics{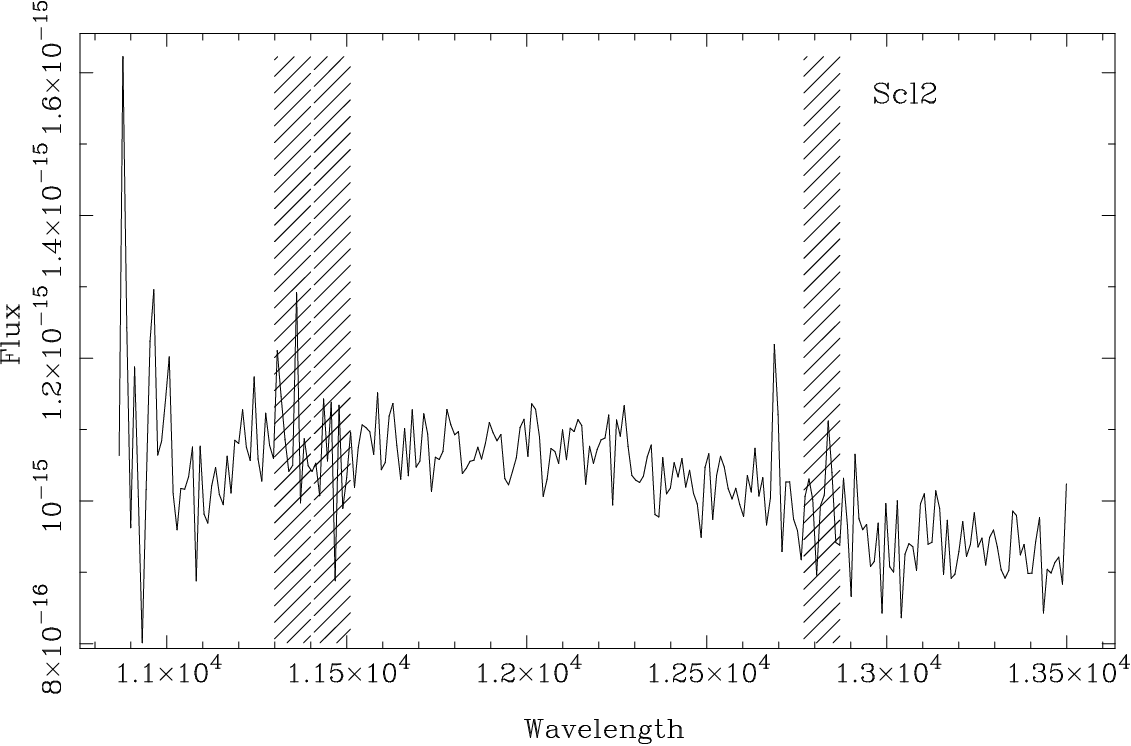}} 
\end{minipage}
\begin{minipage}{0.48\textwidth}
\resizebox{\hsize}{!}{\includegraphics{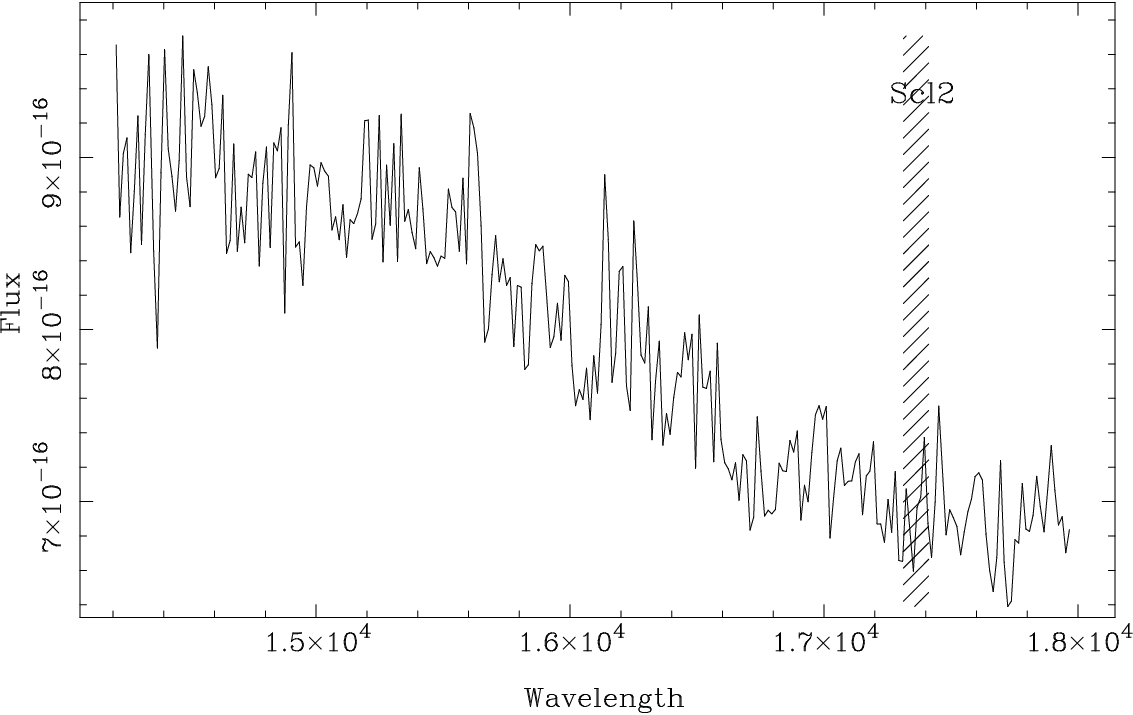}} 
\end{minipage}

\end{figure*}
}

\onlfig{6}{
\begin{figure*}[H]

\begin{minipage}{0.48\textwidth}
\resizebox{\hsize}{!}{\includegraphics{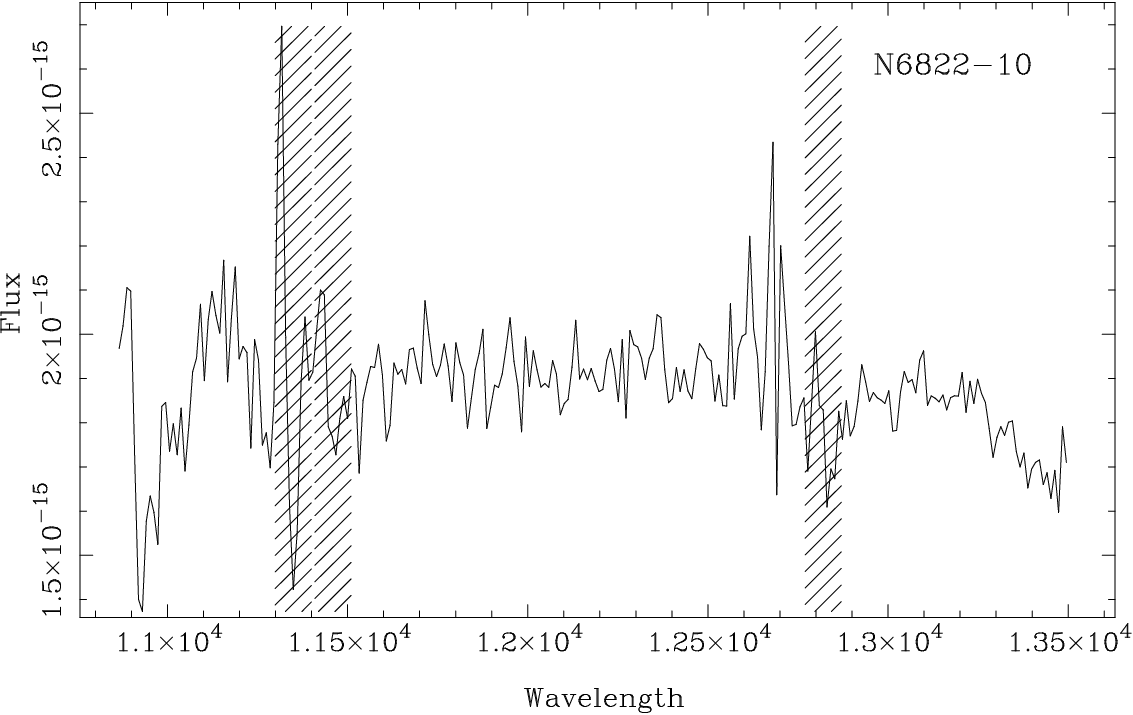}} 
\end{minipage}
\begin{minipage}{0.48\textwidth}
\resizebox{\hsize}{!}{\includegraphics{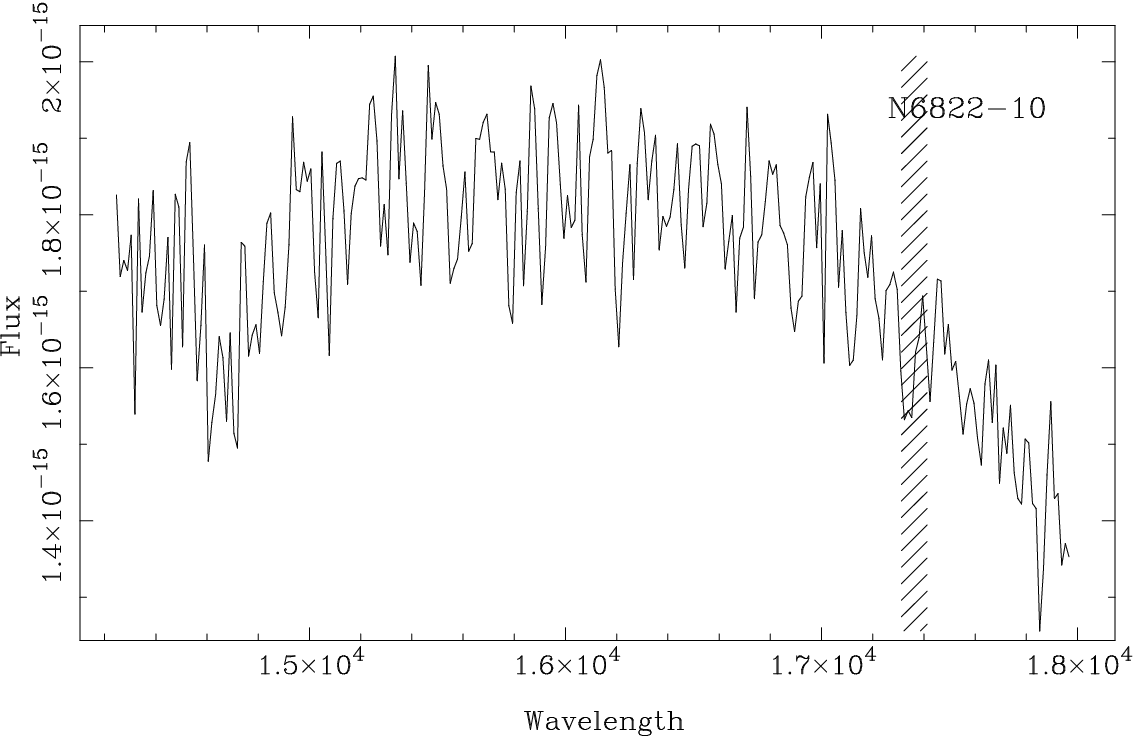}} 
\end{minipage}

\begin{minipage}{0.48\textwidth}
\resizebox{\hsize}{!}{\includegraphics{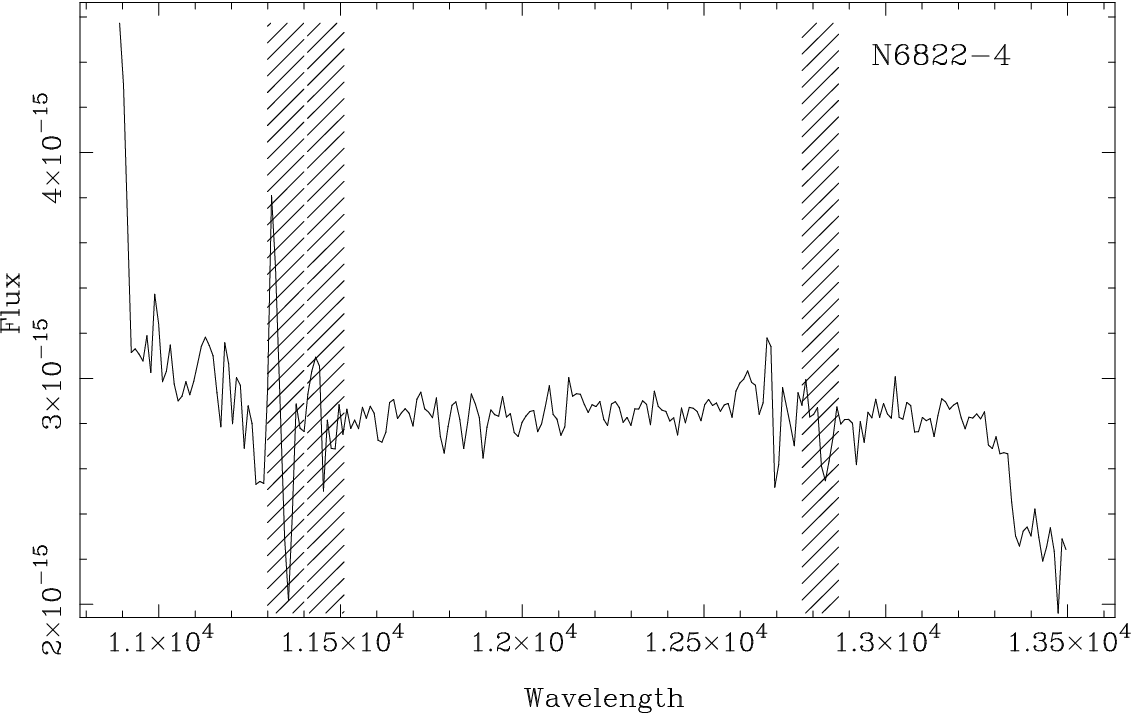}} 
\end{minipage}
\begin{minipage}{0.48\textwidth}
\resizebox{\hsize}{!}{\includegraphics{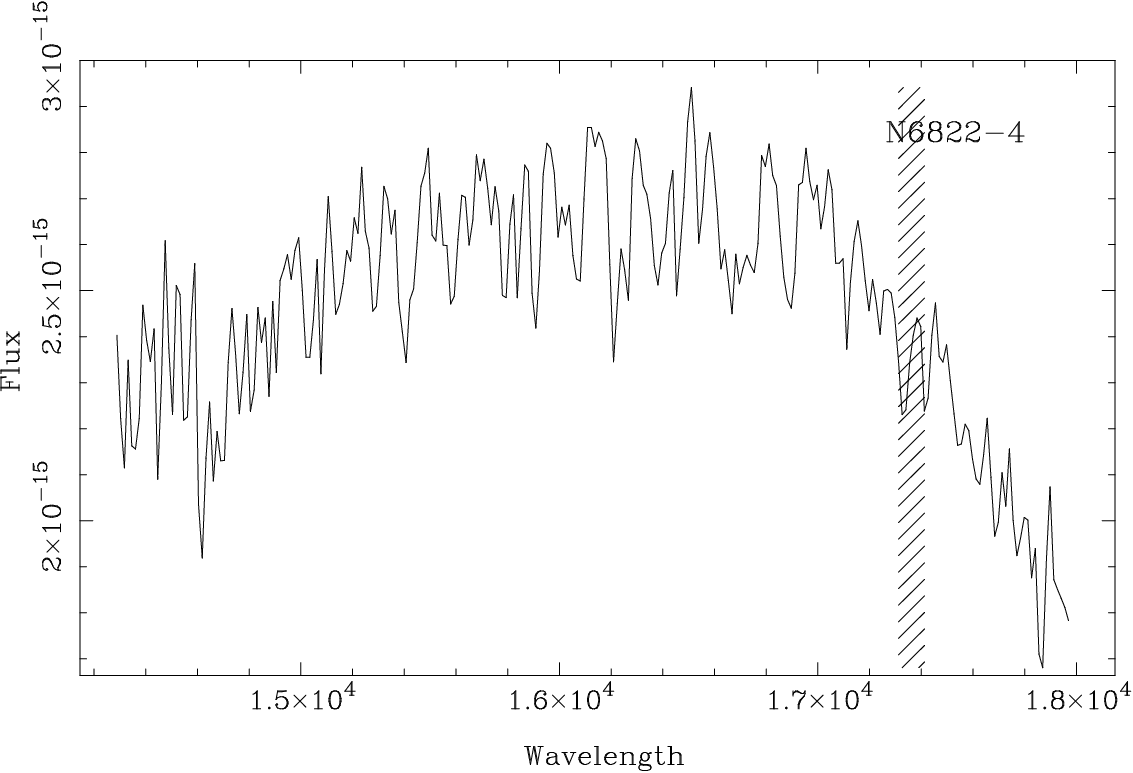}} 
\end{minipage}

\begin{minipage}{0.48\textwidth}
\resizebox{\hsize}{!}{\includegraphics{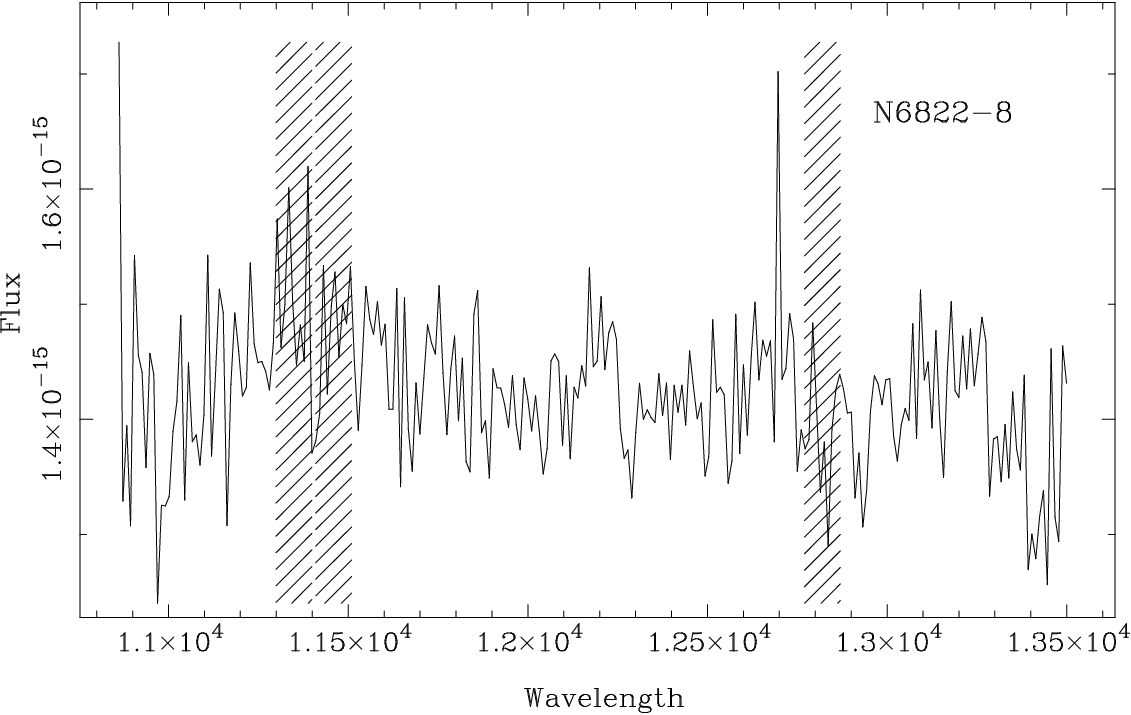}} 
\end{minipage}
\begin{minipage}{0.48\textwidth}
\resizebox{\hsize}{!}{\includegraphics{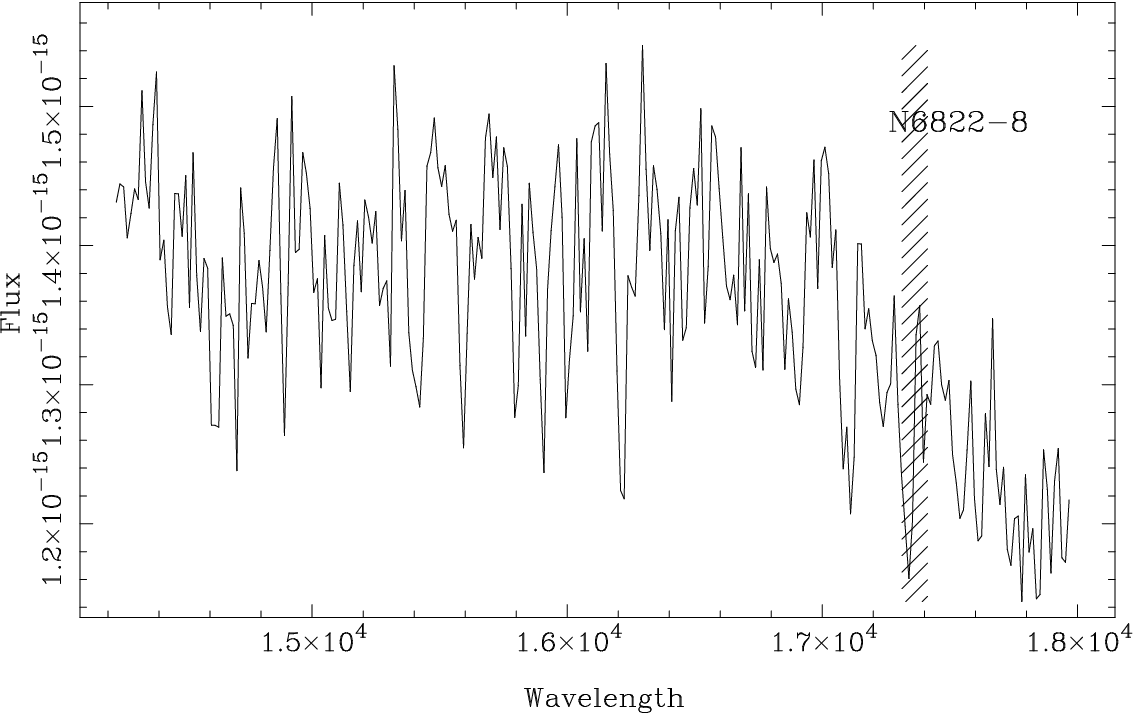}} 
\end{minipage}

\begin{minipage}{0.48\textwidth}
\resizebox{\hsize}{!}{\includegraphics{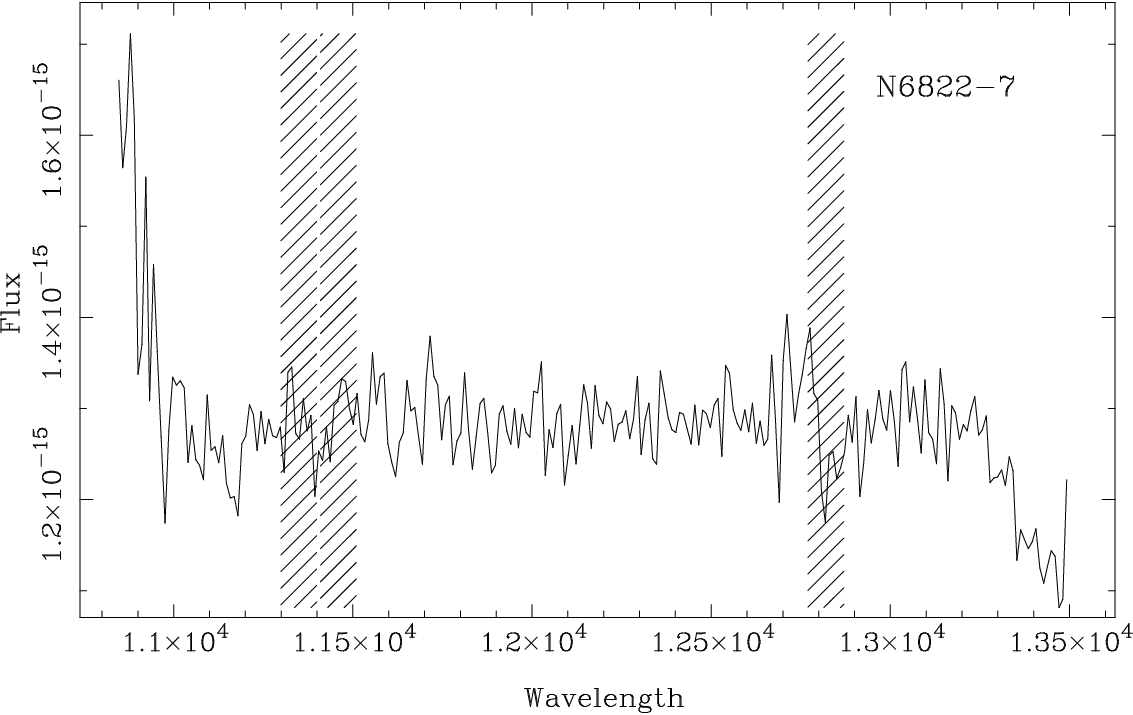}} 
\end{minipage}
\begin{minipage}{0.48\textwidth}
\resizebox{\hsize}{!}{\includegraphics{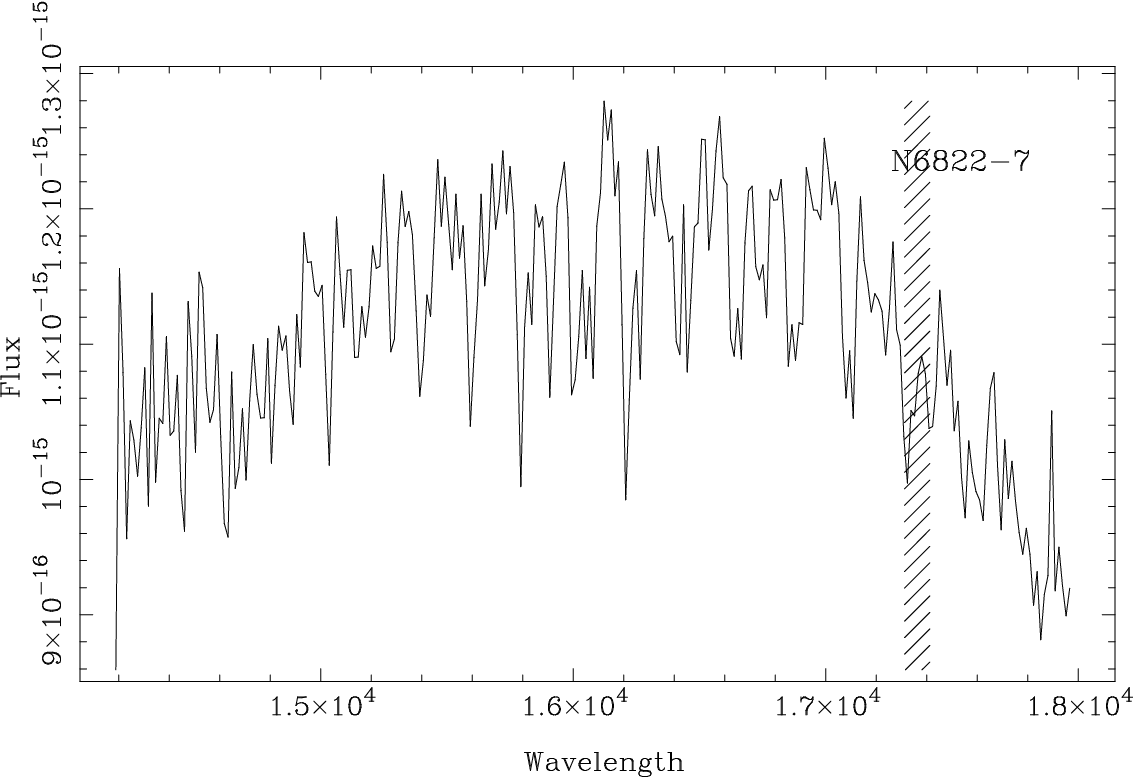}} 
\end{minipage}

\end{figure*}
}

\onlfig{6}{
\begin{figure*}[H]
 
\begin{minipage}{0.48\textwidth}
\resizebox{\hsize}{!}{\includegraphics{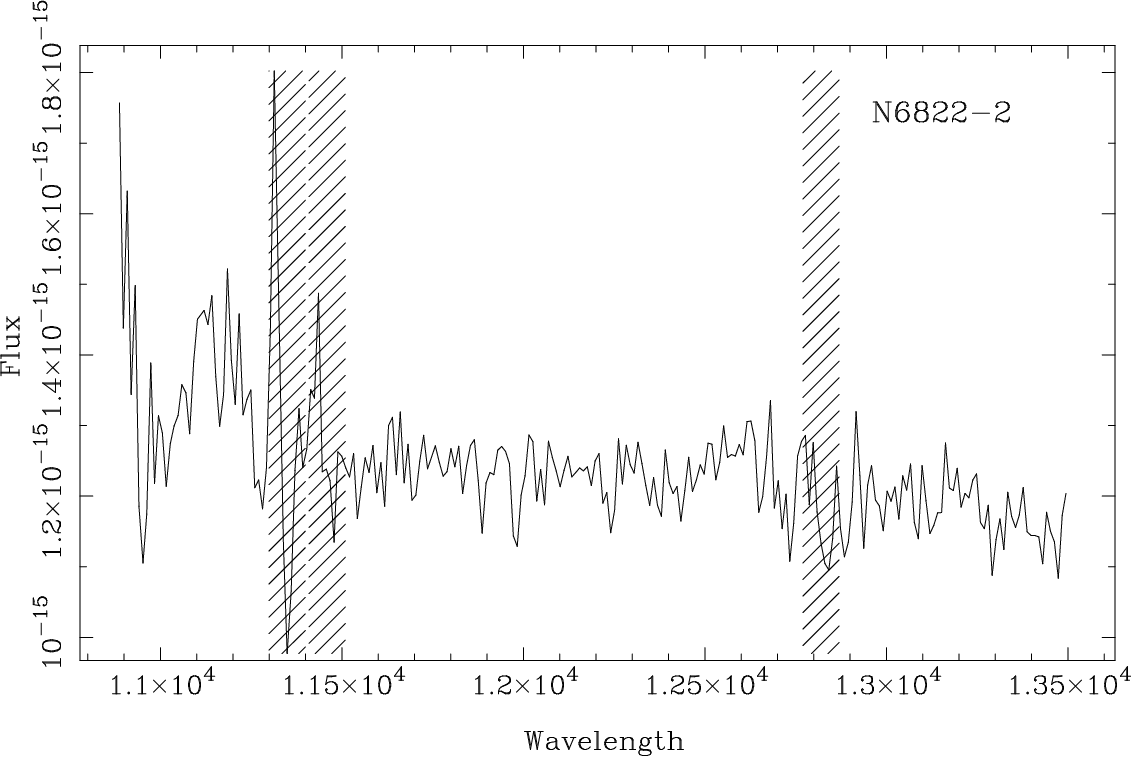}} 
\end{minipage}
\begin{minipage}{0.48\textwidth}
\resizebox{\hsize}{!}{\includegraphics{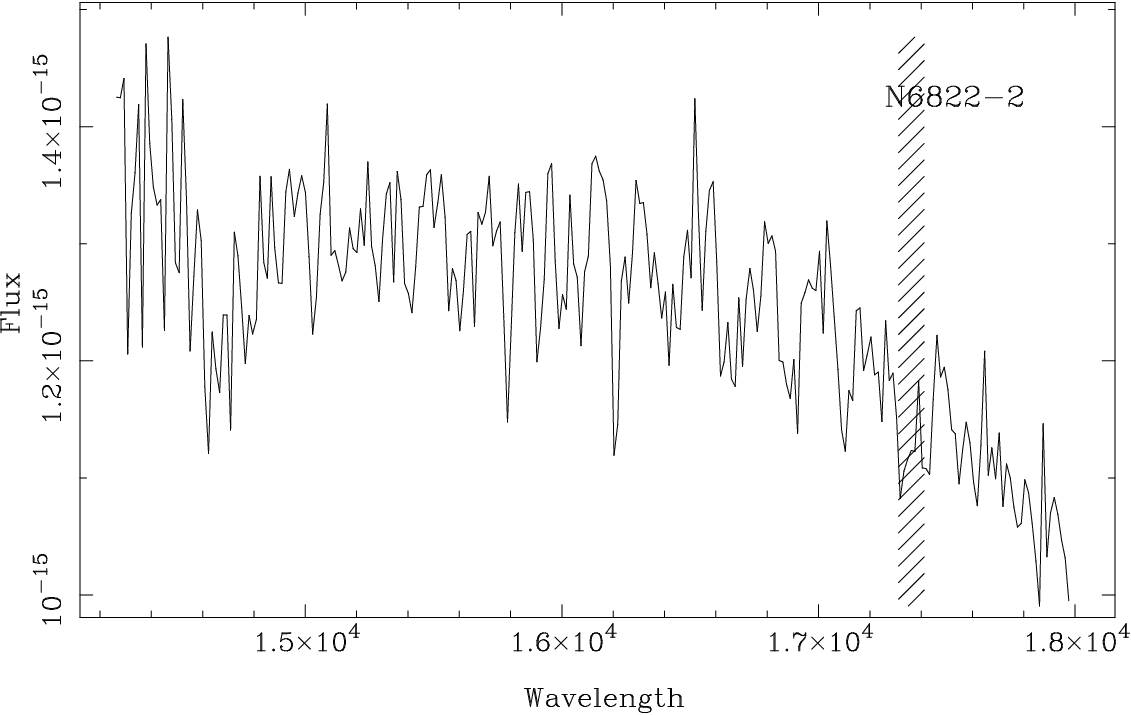}} 
\end{minipage}

\begin{minipage}{0.48\textwidth}
\resizebox{\hsize}{!}{\includegraphics{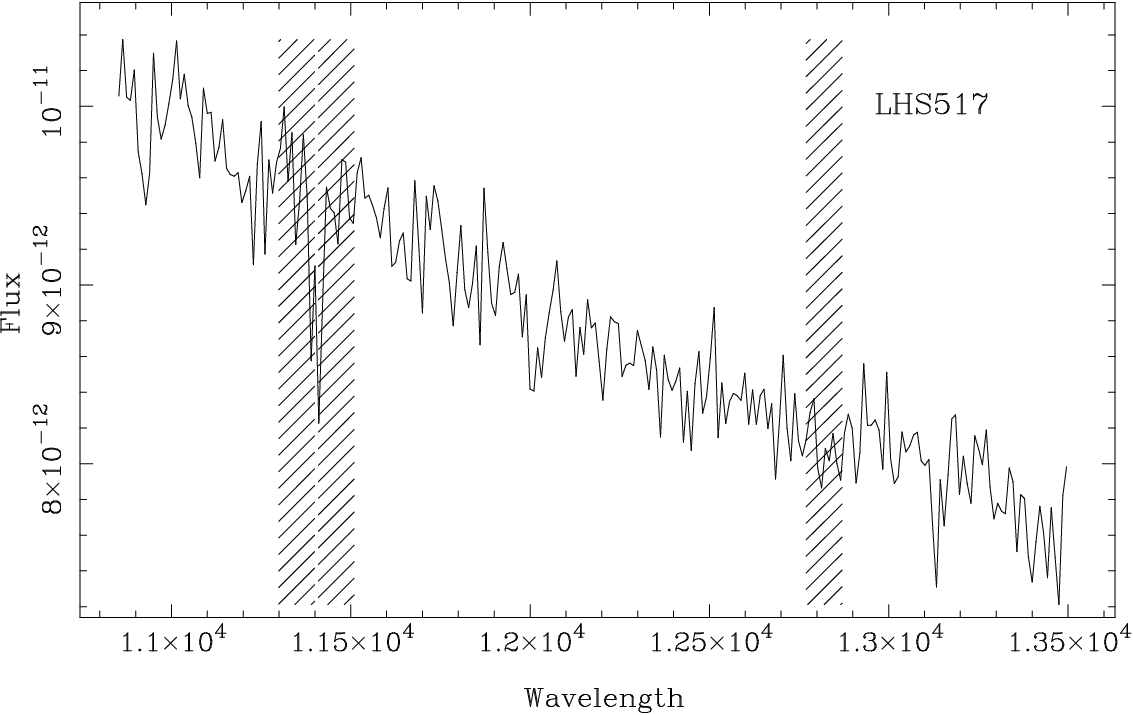}} 
\end{minipage}
\begin{minipage}{0.48\textwidth}
\resizebox{\hsize}{!}{\includegraphics{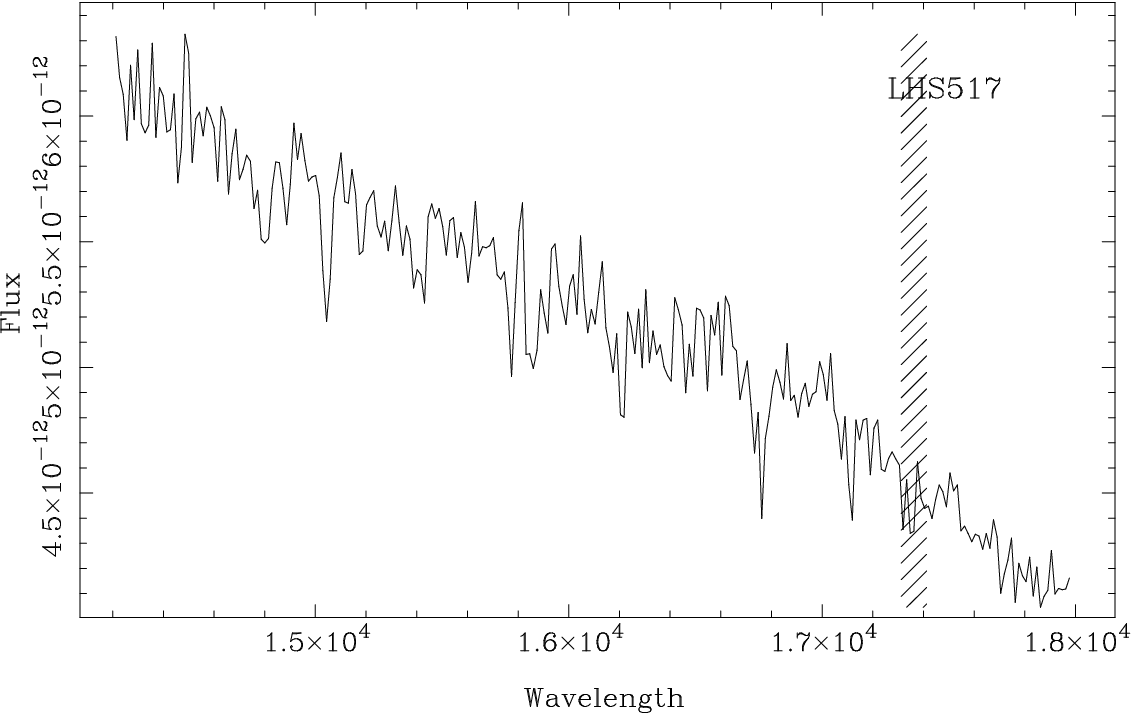}} 
\end{minipage}

\begin{minipage}{0.48\textwidth}
\resizebox{\hsize}{!}{\includegraphics{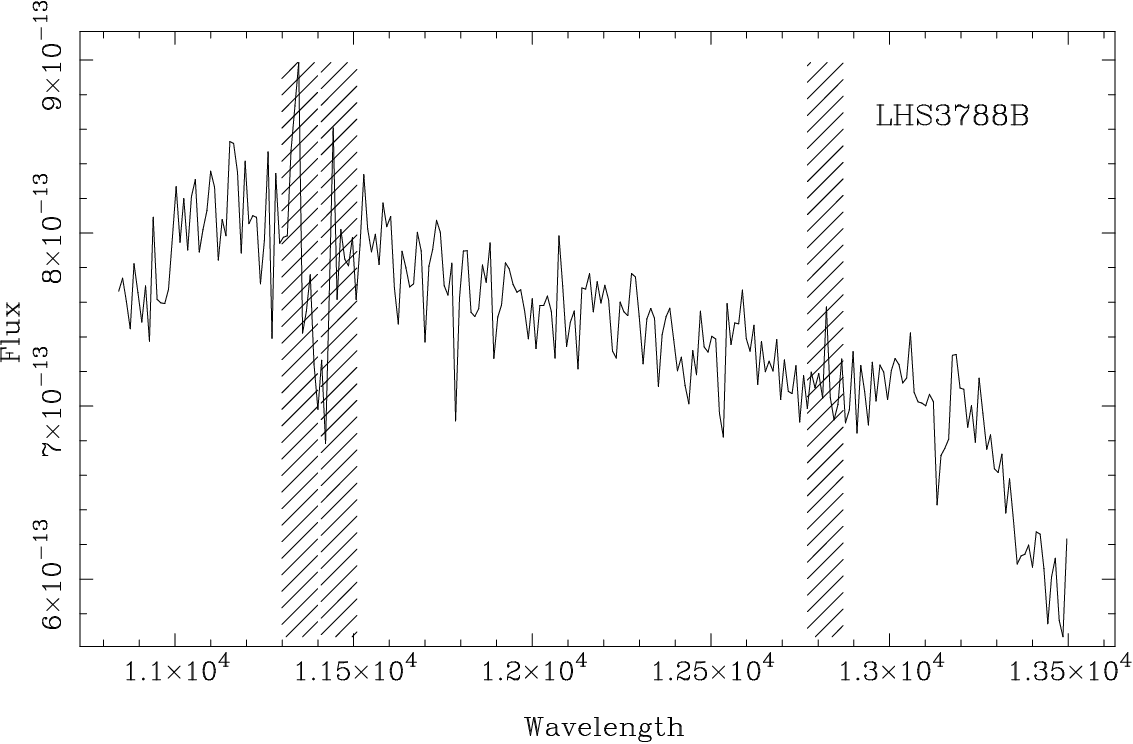}} 
\end{minipage}
\begin{minipage}{0.48\textwidth}
\resizebox{\hsize}{!}{\includegraphics{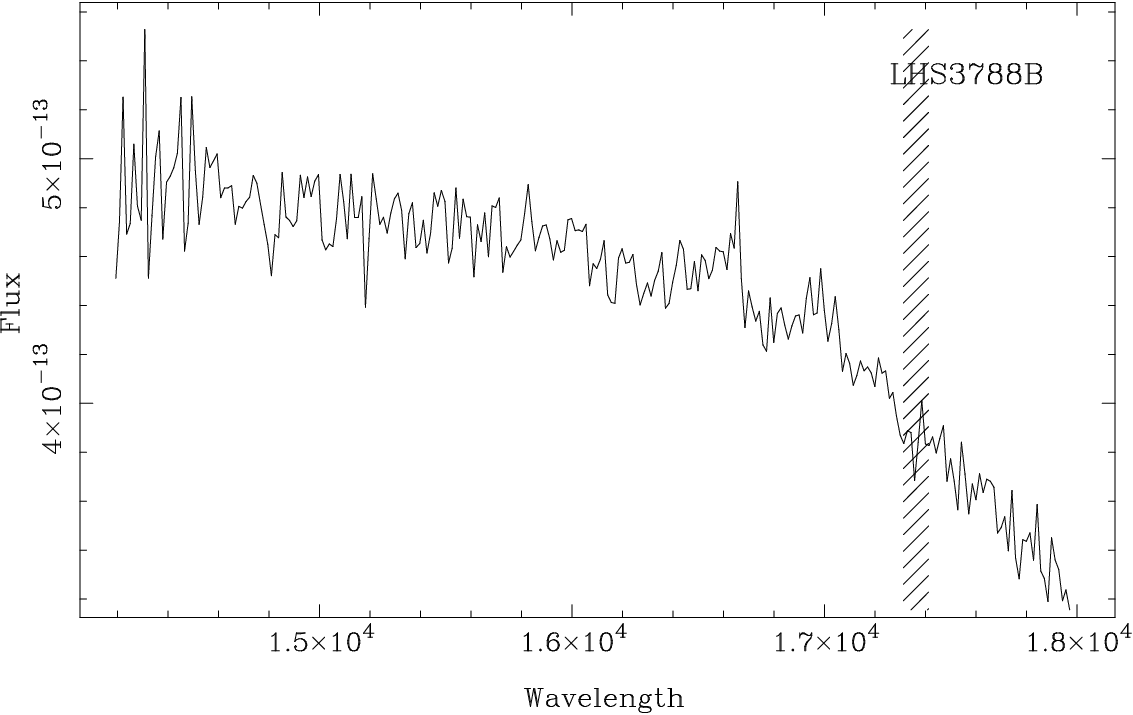}} 
\end{minipage}

\begin{minipage}{0.48\textwidth}
\resizebox{\hsize}{!}{\includegraphics{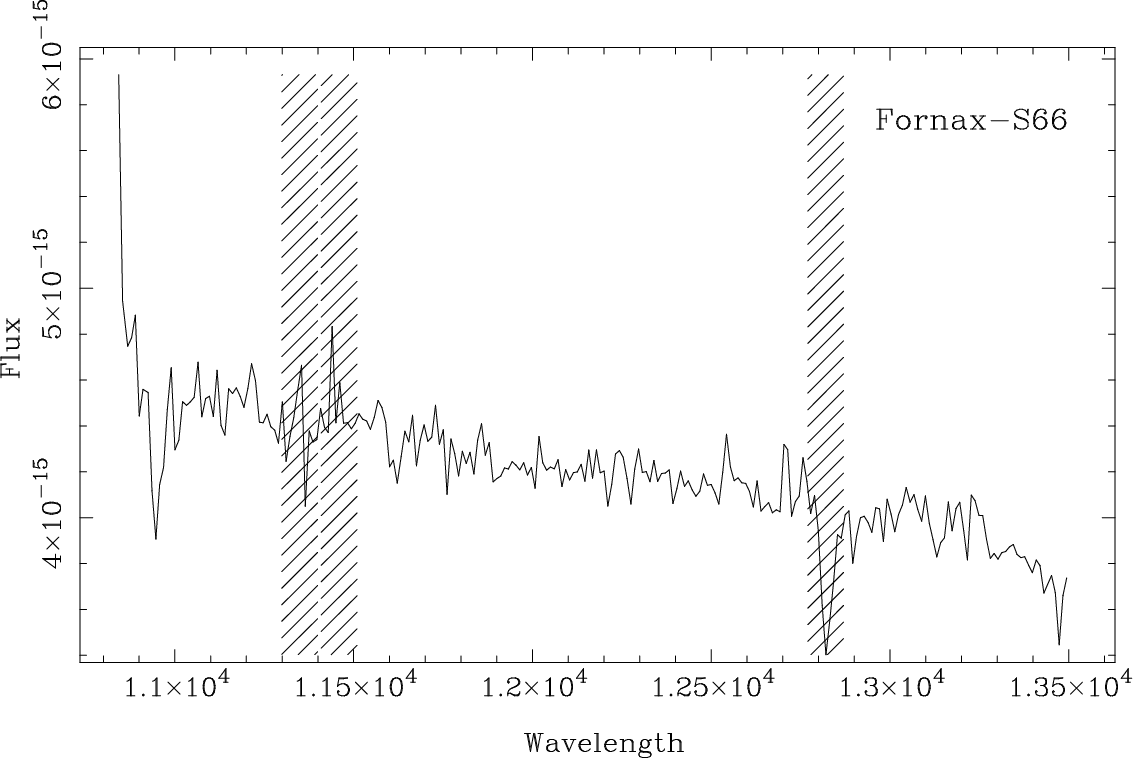}}
\end{minipage}
\begin{minipage}{0.48\textwidth}
\resizebox{\hsize}{!}{\includegraphics{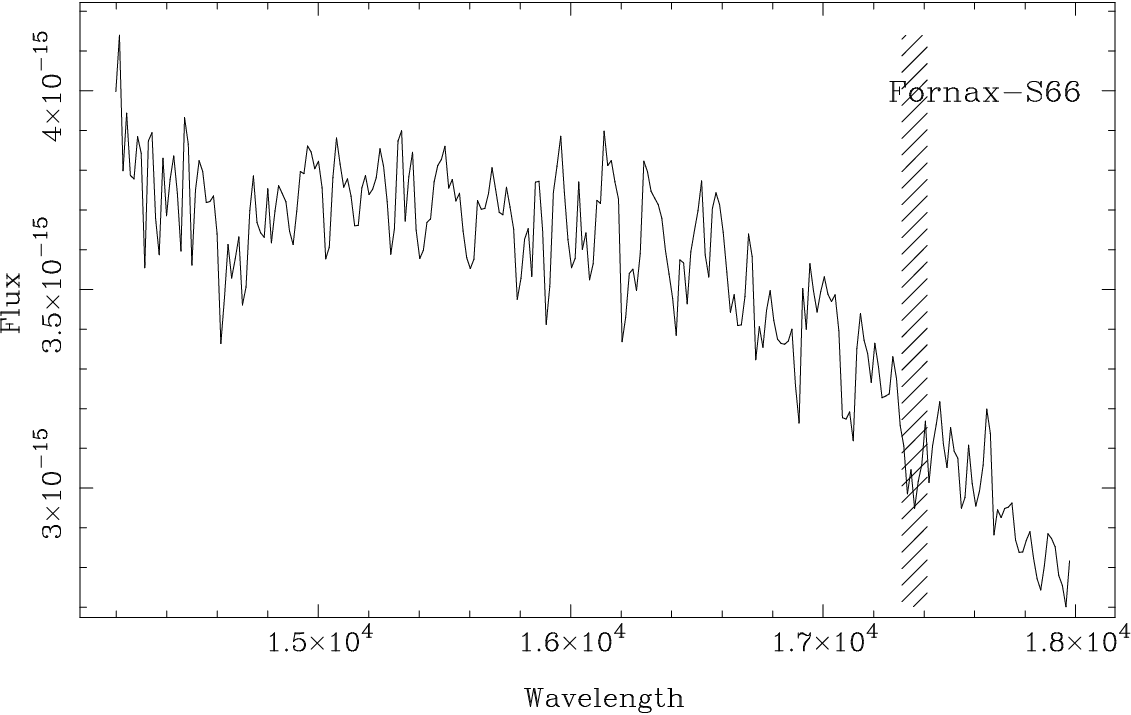}}
\end{minipage}

\end{figure*}
}

\onlfig{6}{
\begin{figure*}[H]

\begin{minipage}{0.48\textwidth}
\resizebox{\hsize}{!}{\includegraphics{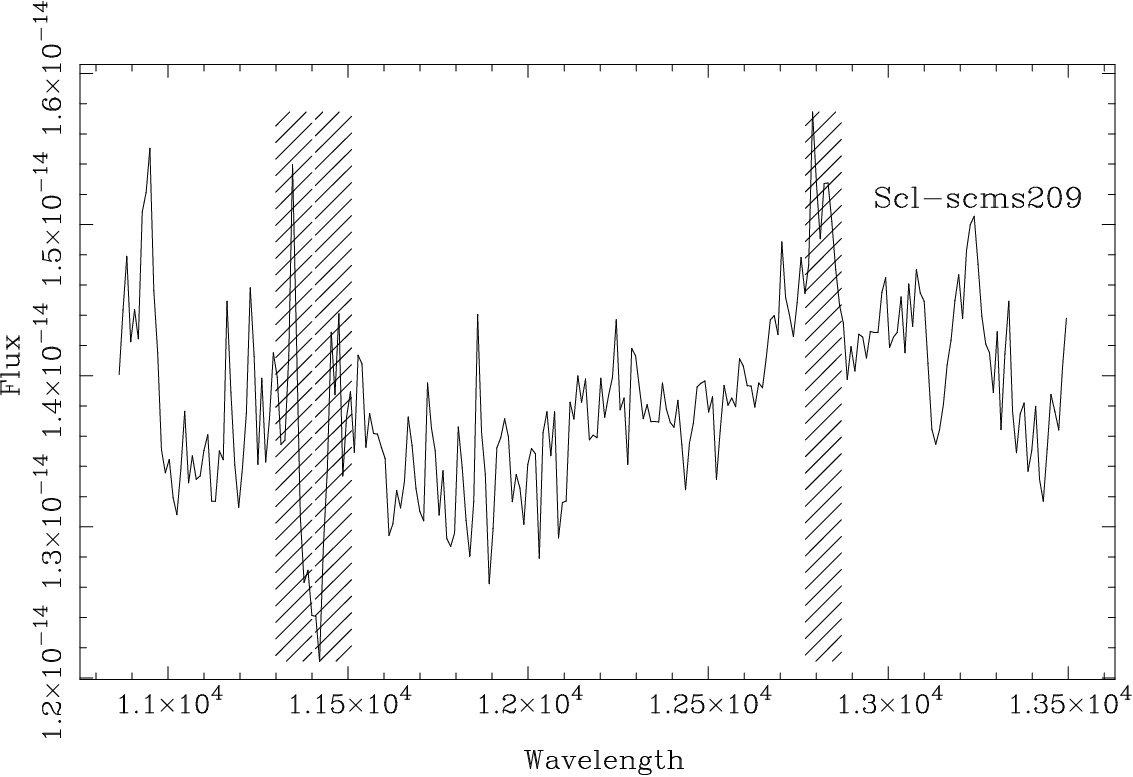}}
\end{minipage}
\begin{minipage}{0.48\textwidth}
\resizebox{\hsize}{!}{\includegraphics{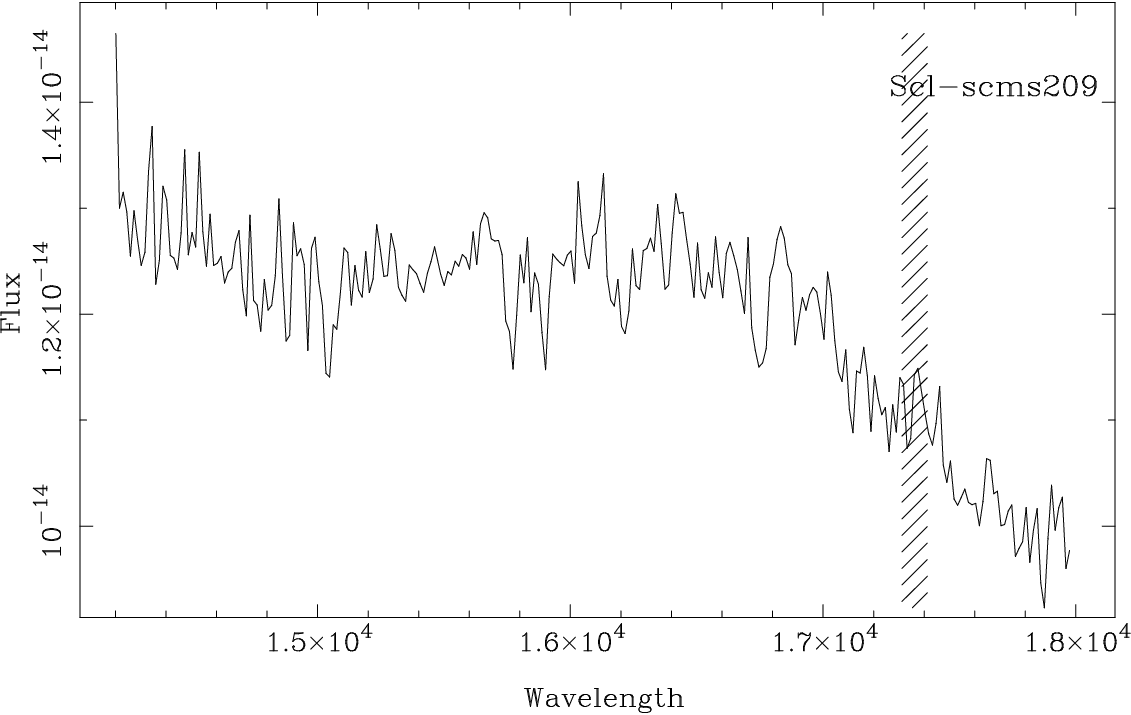}}
\end{minipage}

\begin{minipage}{0.48\textwidth}
\resizebox{\hsize}{!}{\includegraphics{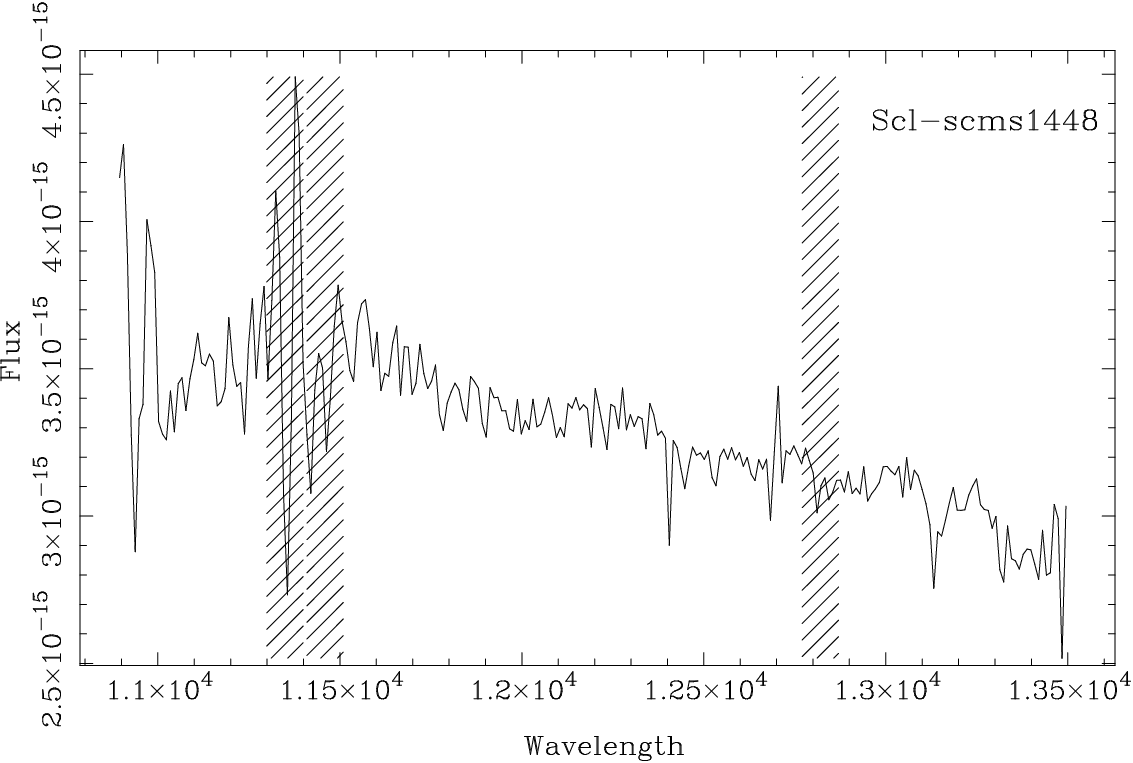}}
\end{minipage}
\begin{minipage}{0.48\textwidth}
\resizebox{\hsize}{!}{\includegraphics{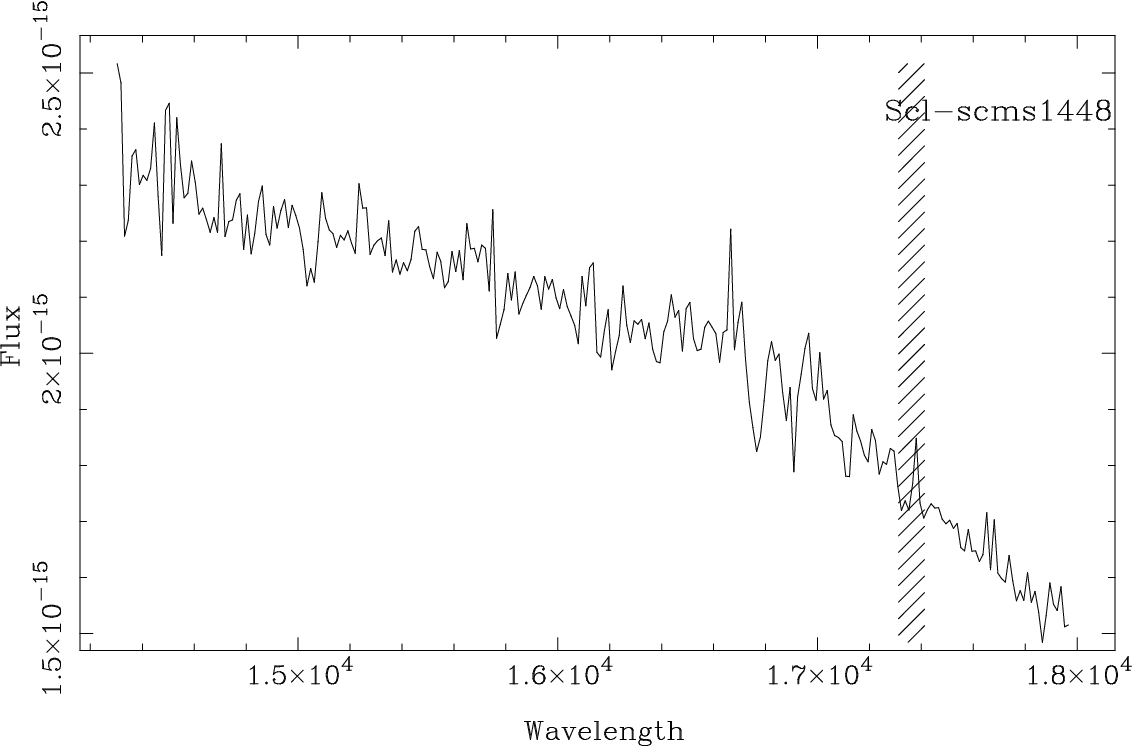}}
\end{minipage}
 
\begin{minipage}{0.48\textwidth}
\resizebox{\hsize}{!}{\includegraphics{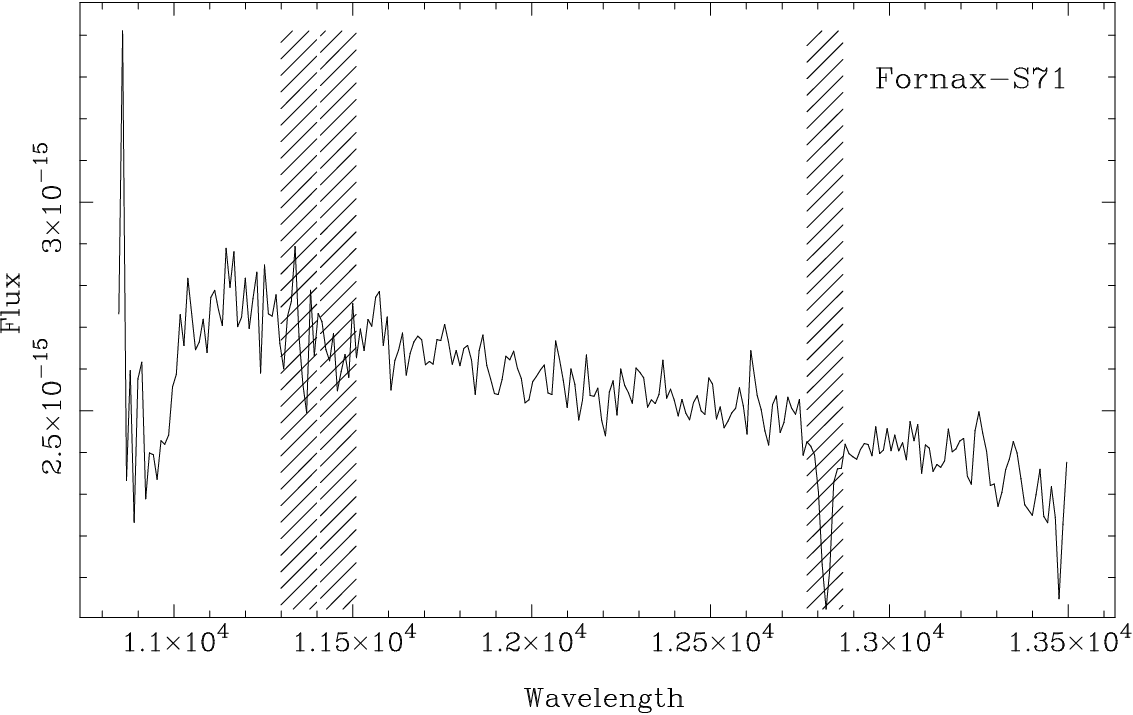}}
\end{minipage}
\begin{minipage}{0.48\textwidth}
\resizebox{\hsize}{!}{\includegraphics{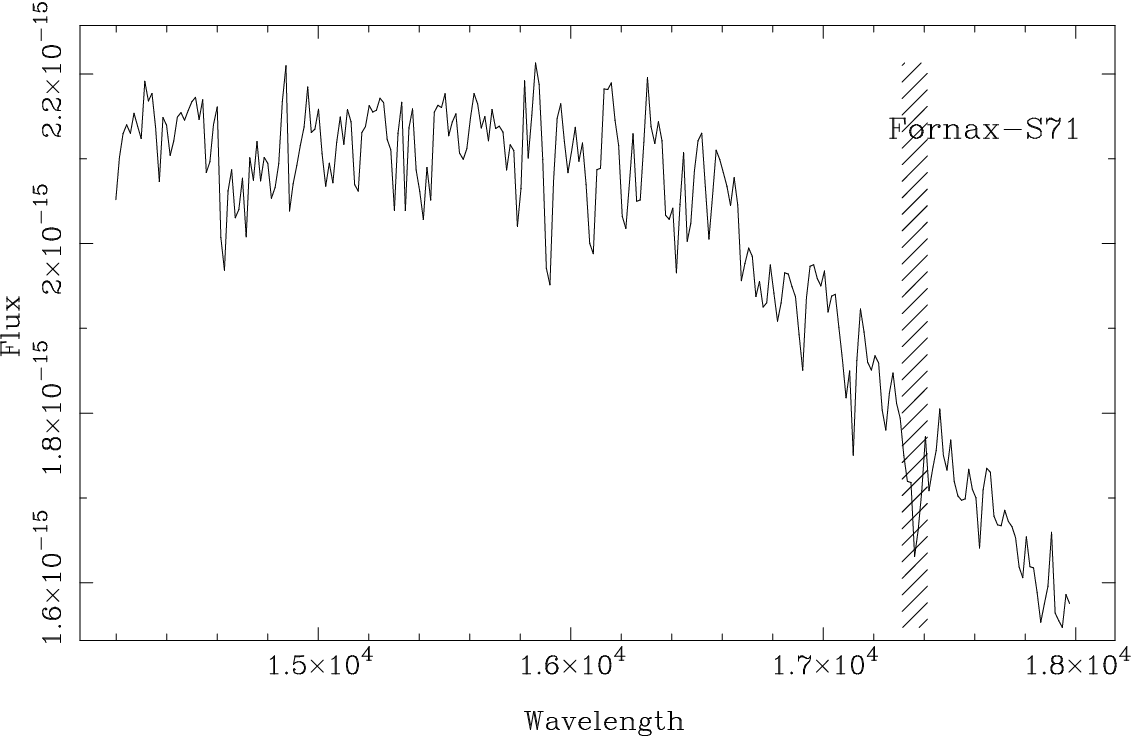}}
\end{minipage}

\begin{minipage}{0.48\textwidth}
\resizebox{\hsize}{!}{\includegraphics{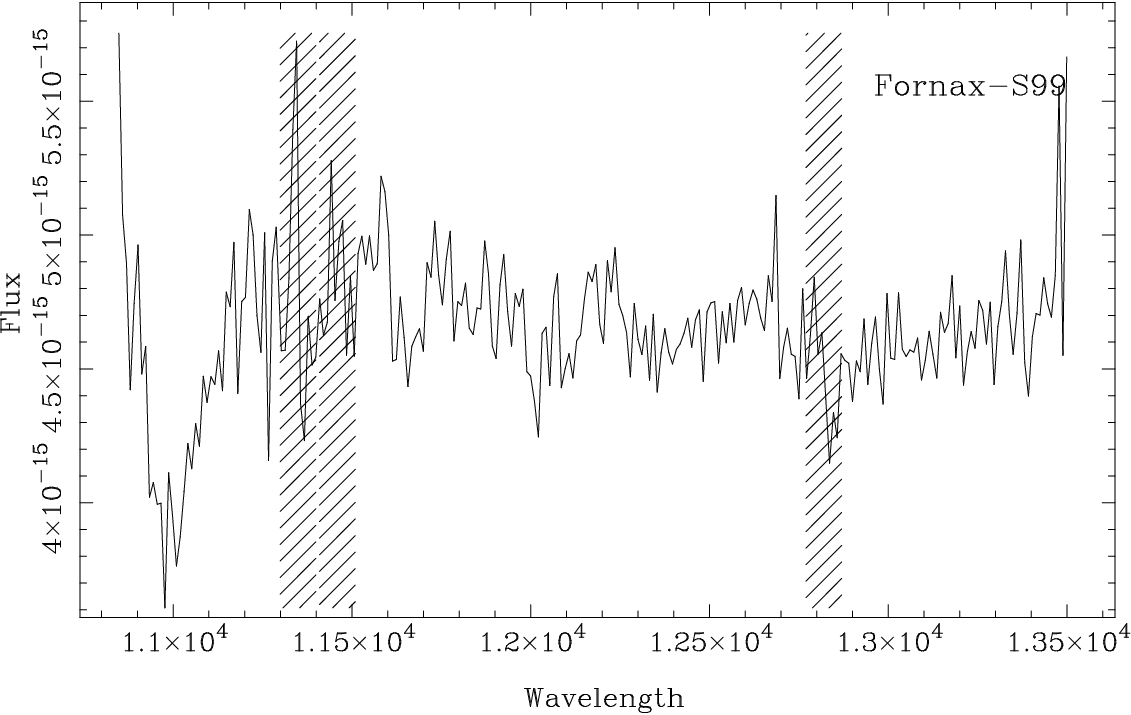}}
\end{minipage}
\begin{minipage}{0.48\textwidth}
\resizebox{\hsize}{!}{\includegraphics{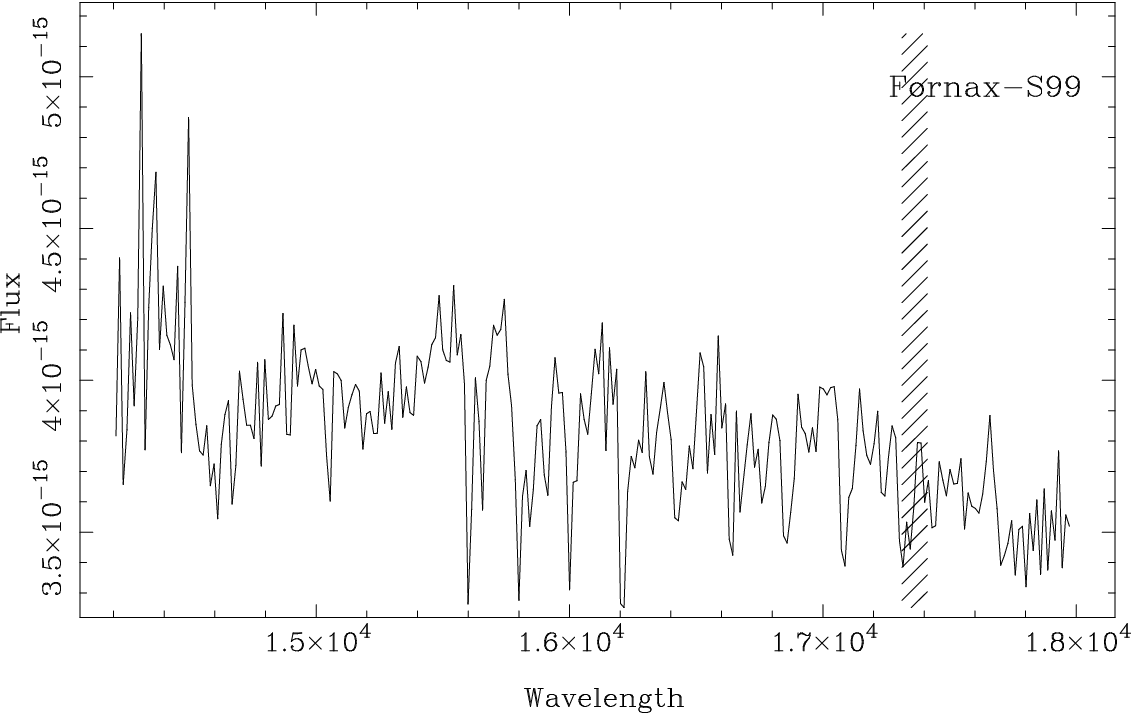}}
\end{minipage}

\end{figure*}
}

\onlfig{6}{
\begin{figure*}[H]

\begin{minipage}{0.48\textwidth}
\resizebox{\hsize}{!}{\includegraphics{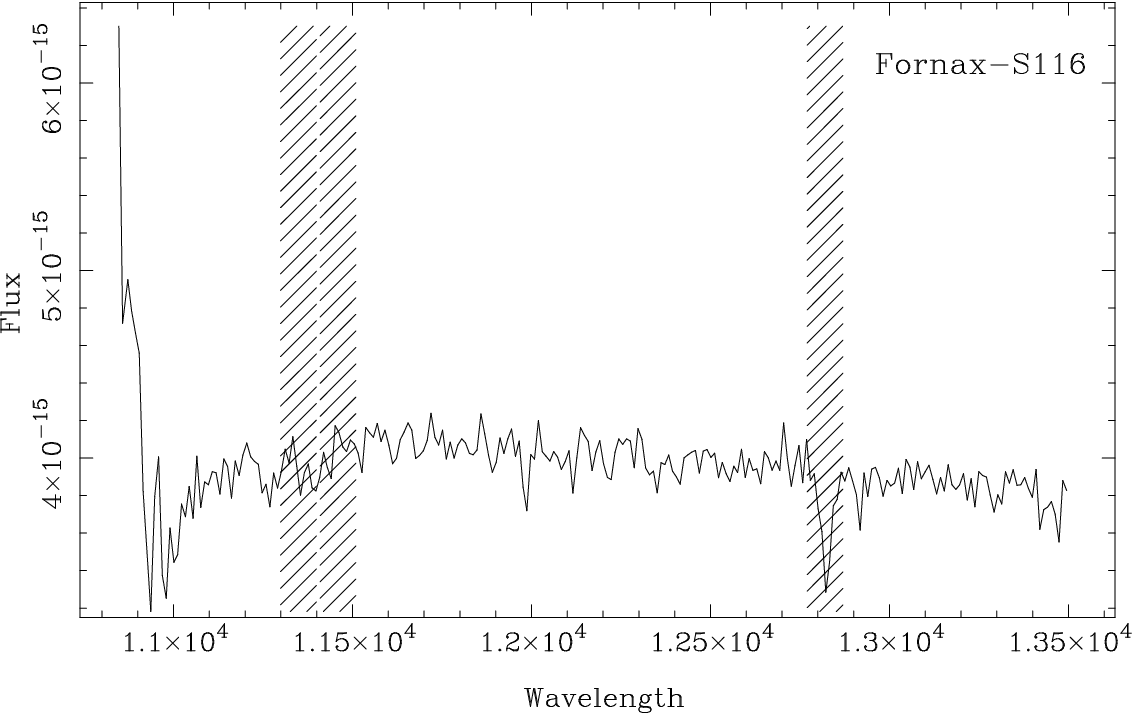}}
\end{minipage}
\begin{minipage}{0.48\textwidth}
\resizebox{\hsize}{!}{\includegraphics{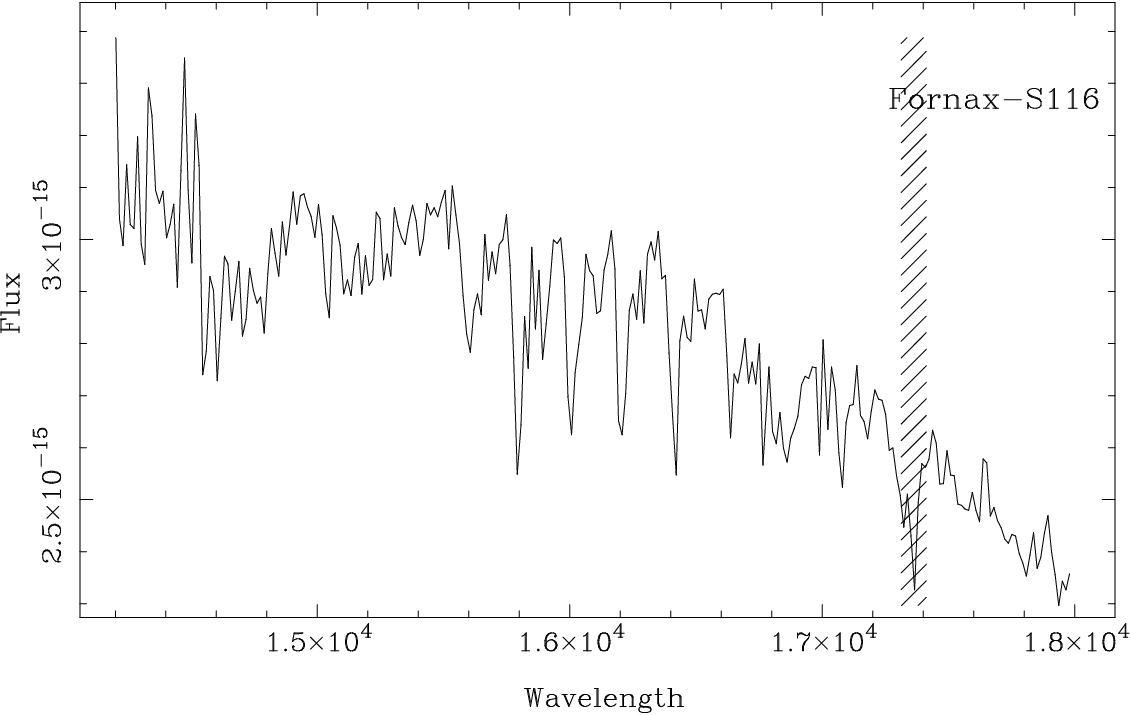}}
\end{minipage}

\begin{minipage}{0.48\textwidth}
\resizebox{\hsize}{!}{\includegraphics{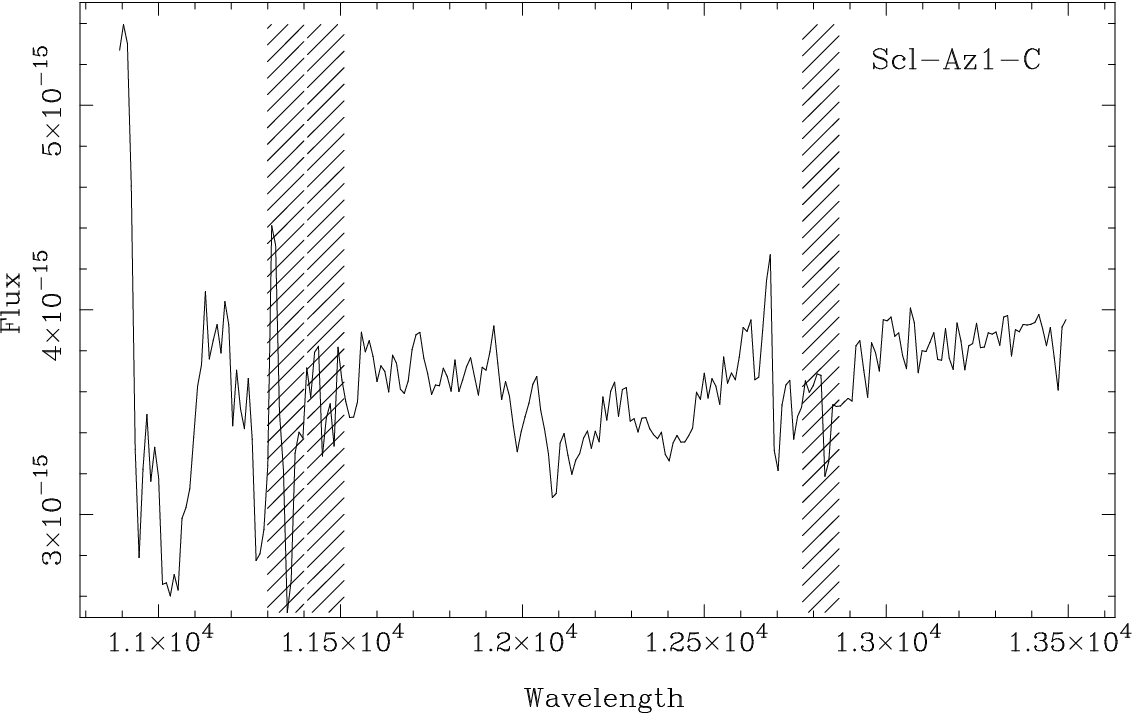}}
\end{minipage}
\begin{minipage}{0.48\textwidth}
\resizebox{\hsize}{!}{\includegraphics{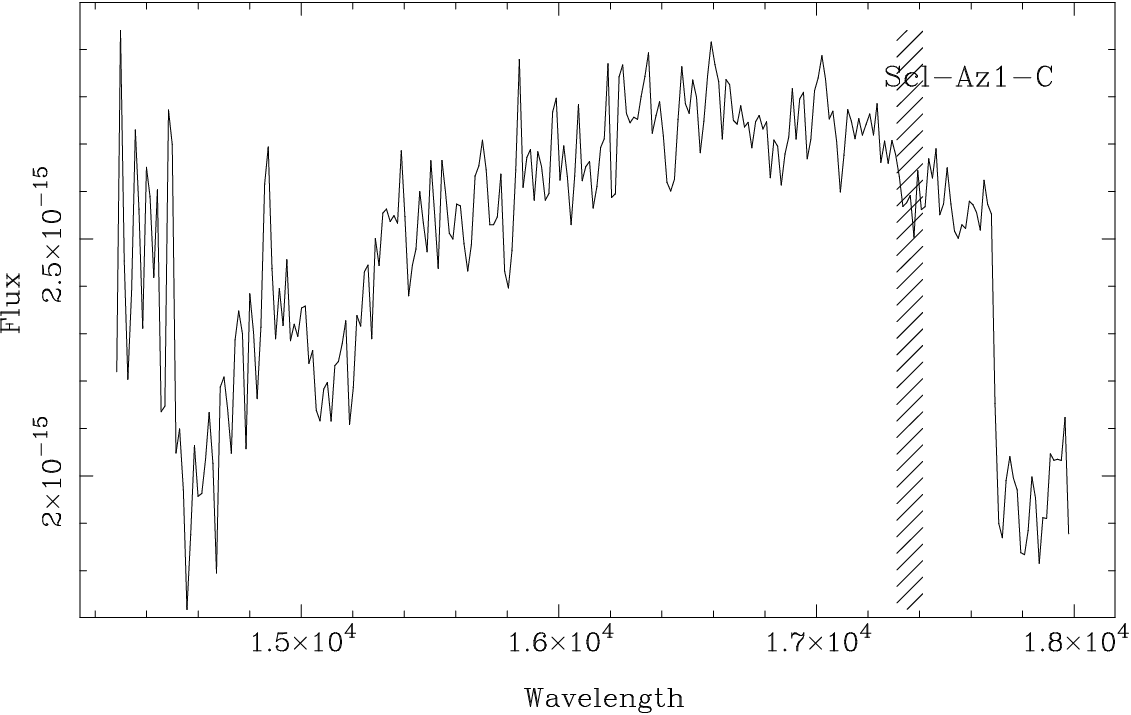}}
\end{minipage}

\begin{minipage}{0.48\textwidth}
\resizebox{\hsize}{!}{\includegraphics{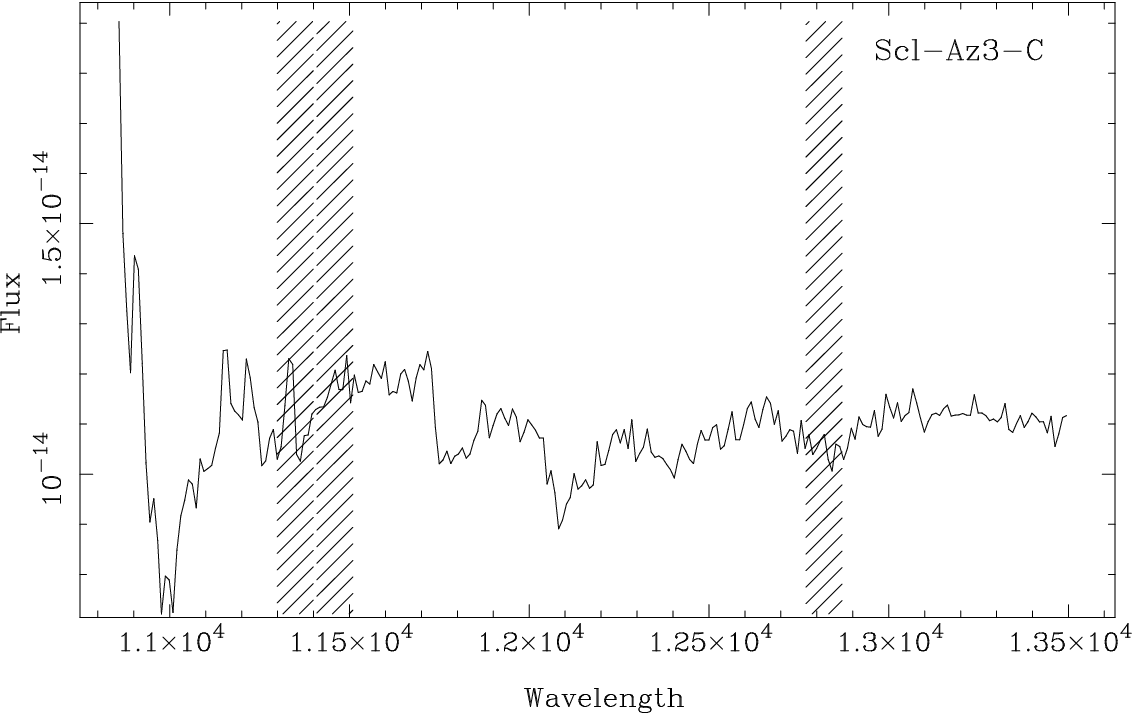}}
\end{minipage}
\begin{minipage}{0.48\textwidth}
\resizebox{\hsize}{!}{\includegraphics{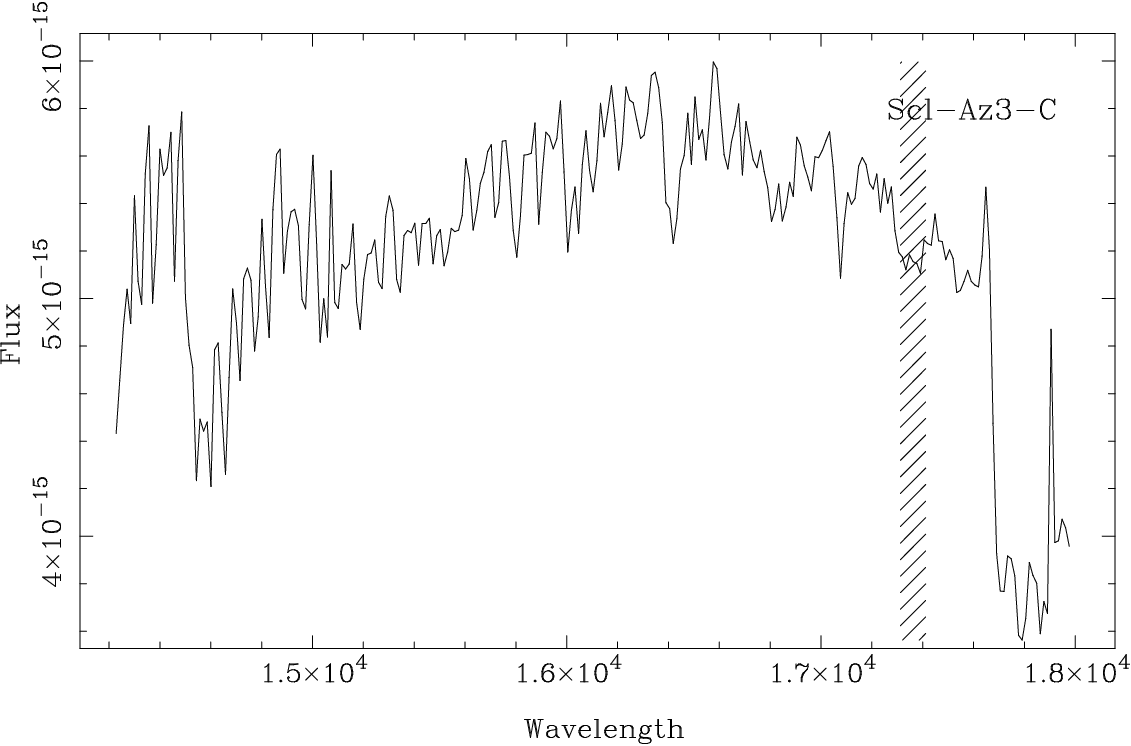}}
\end{minipage}

\begin{minipage}{0.48\textwidth}
\resizebox{\hsize}{!}{\includegraphics{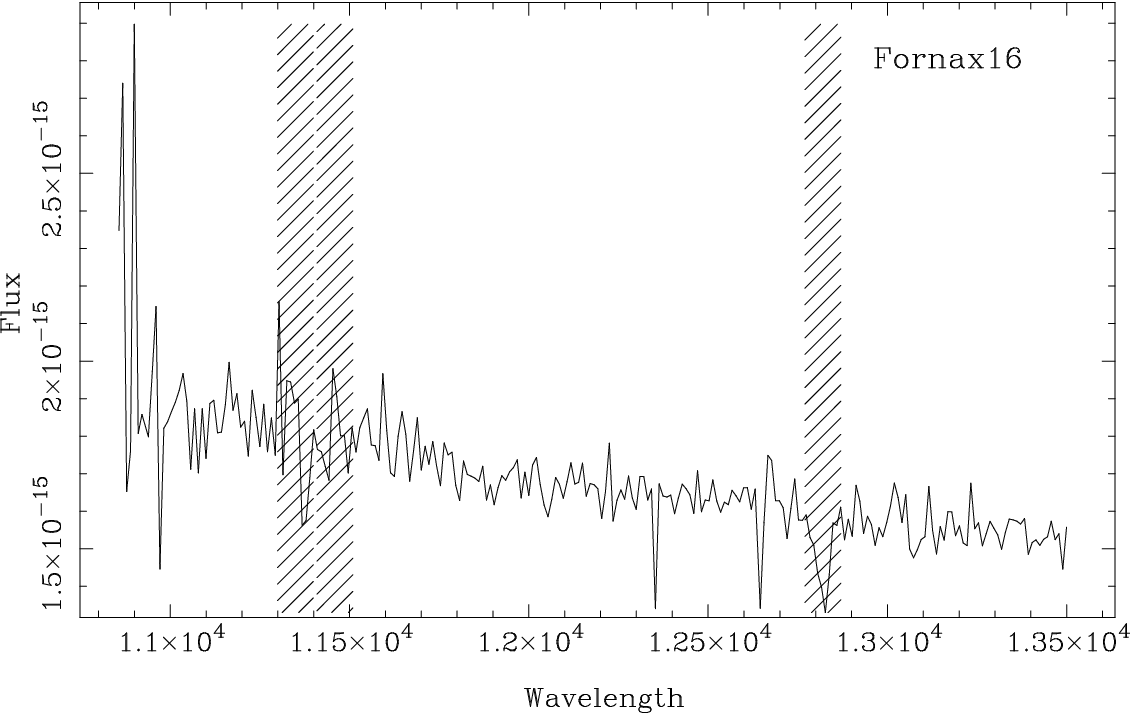}} 
\end{minipage}
\begin{minipage}{0.48\textwidth}
\resizebox{\hsize}{!}{\includegraphics{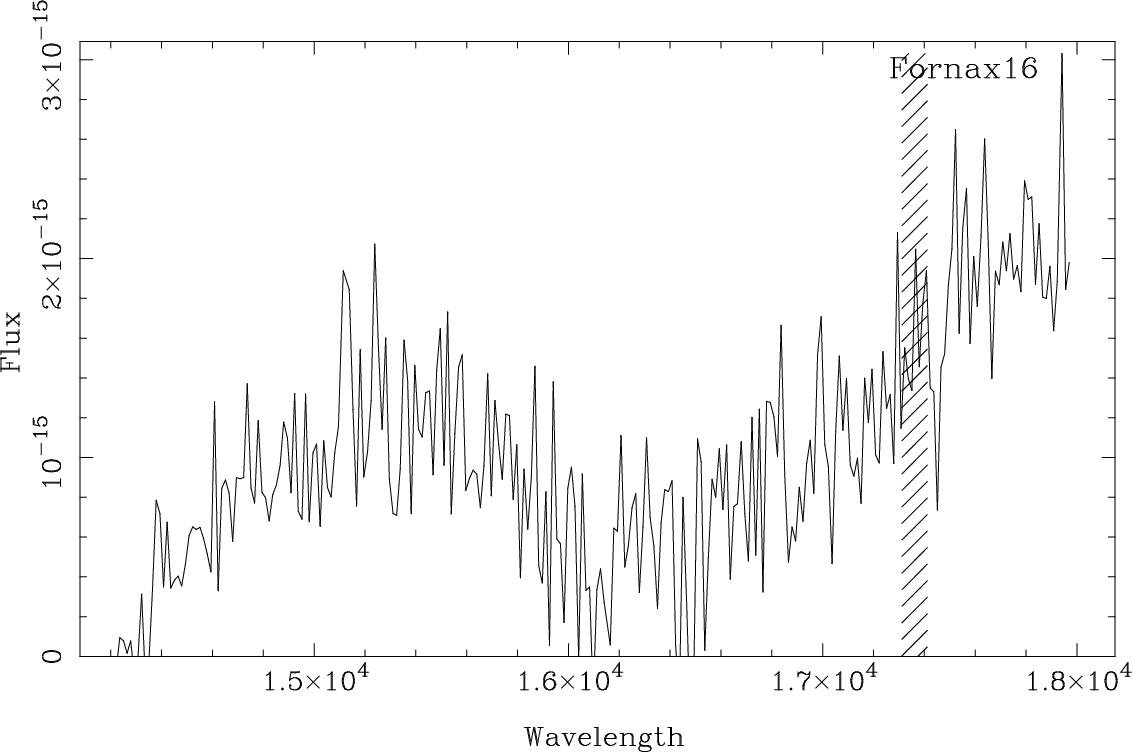}} 
\end{minipage}

\end{figure*}
}

\onlfig{6}{
\begin{figure*}[H]

\begin{minipage}{0.48\textwidth}
\resizebox{\hsize}{!}{\includegraphics{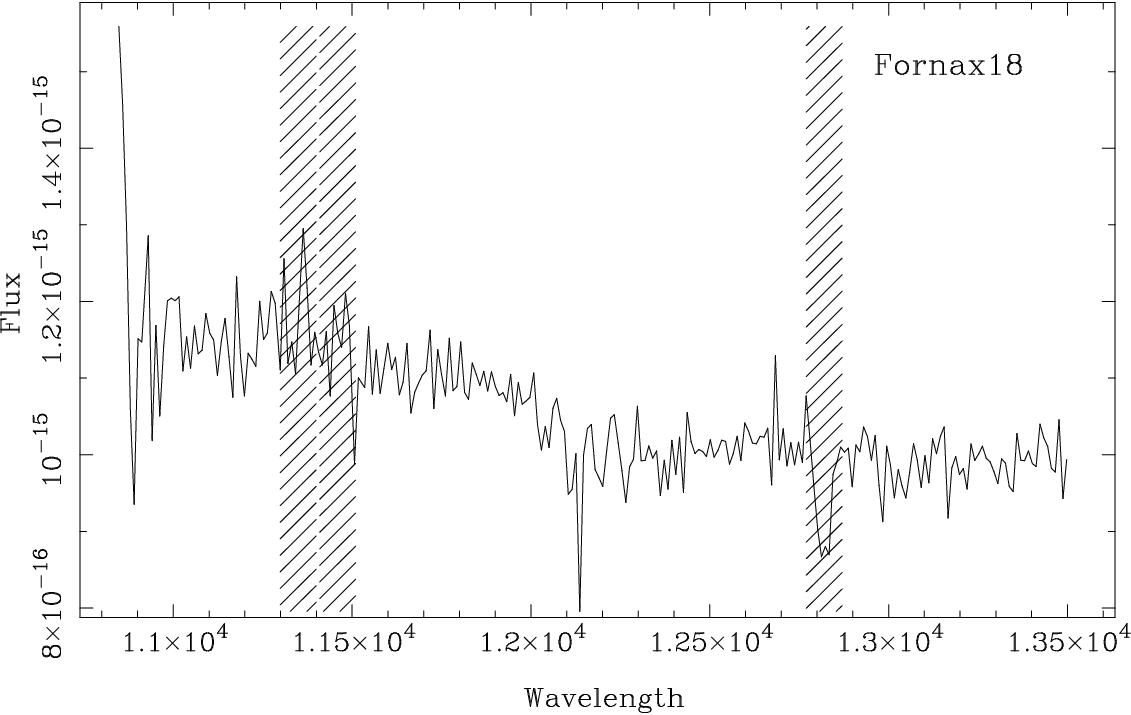}} 
\end{minipage}
\begin{minipage}{0.48\textwidth}
\resizebox{\hsize}{!}{\includegraphics{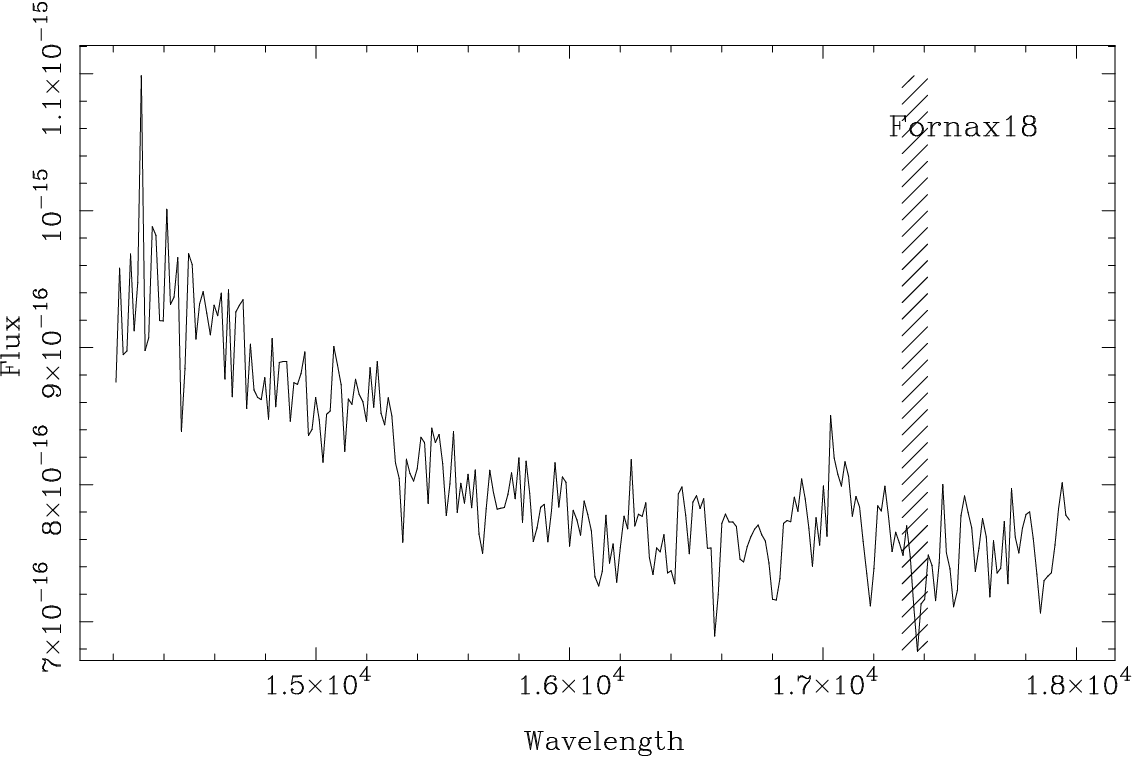}} 
\end{minipage}

\begin{minipage}{0.48\textwidth}
\resizebox{\hsize}{!}{\includegraphics{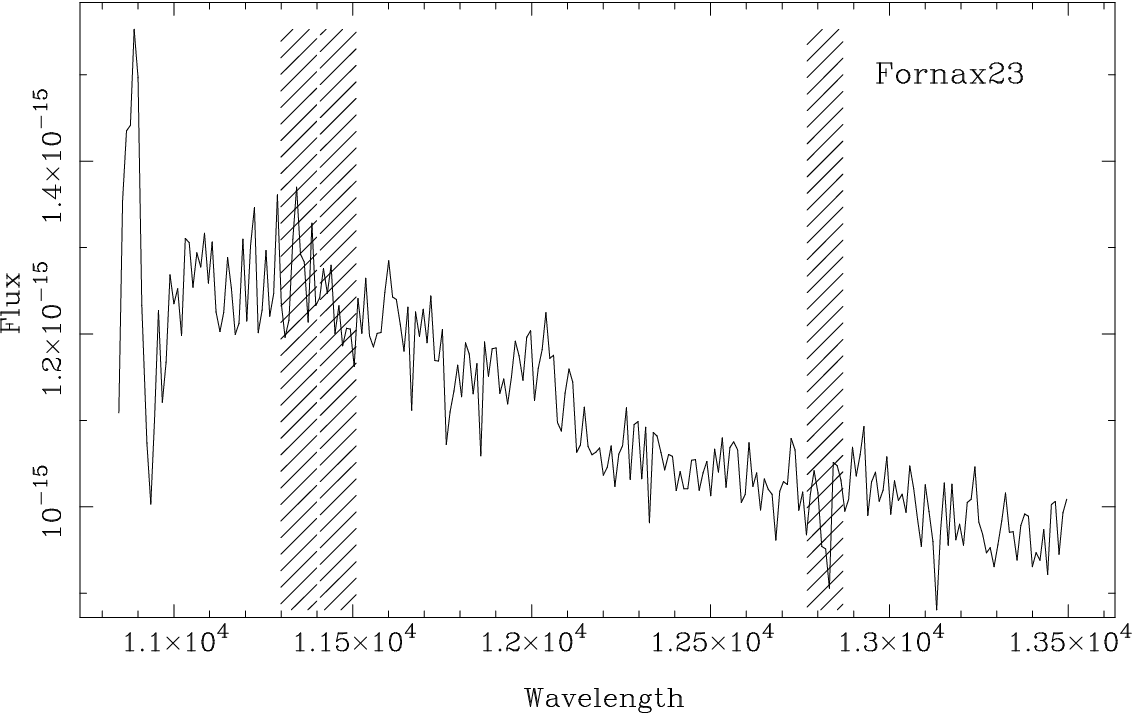}} 
\end{minipage}
\begin{minipage}{0.48\textwidth}
\resizebox{\hsize}{!}{\includegraphics{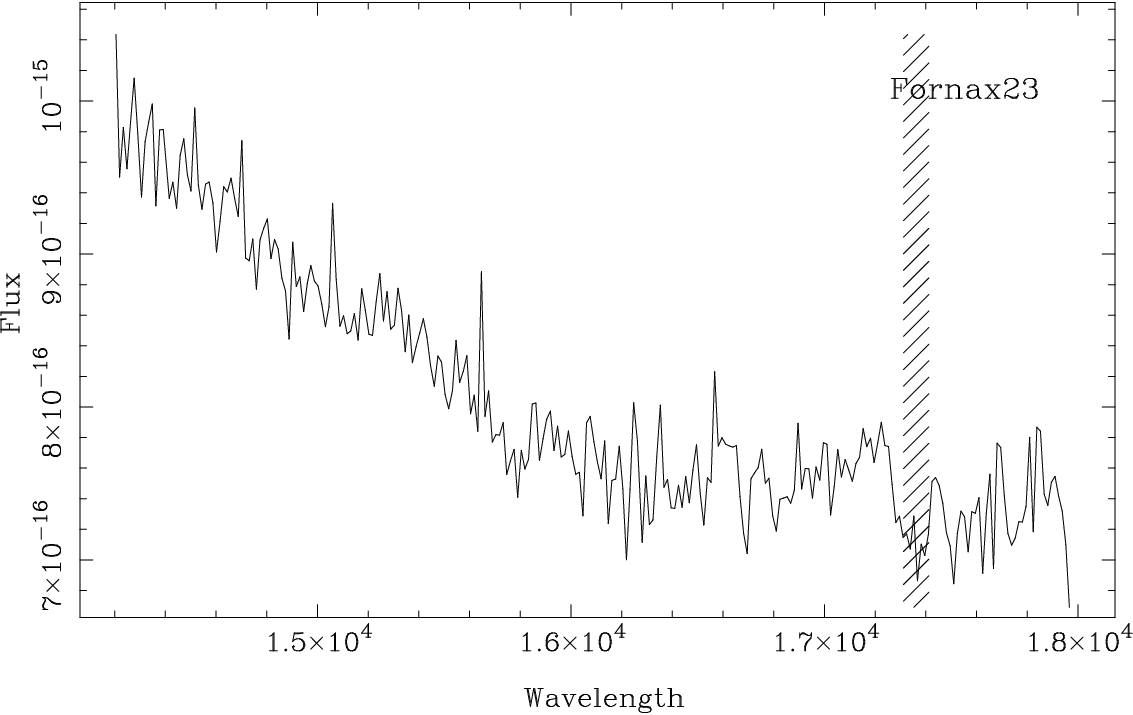}} 
\end{minipage}

\begin{minipage}{0.48\textwidth}
\resizebox{\hsize}{!}{\includegraphics{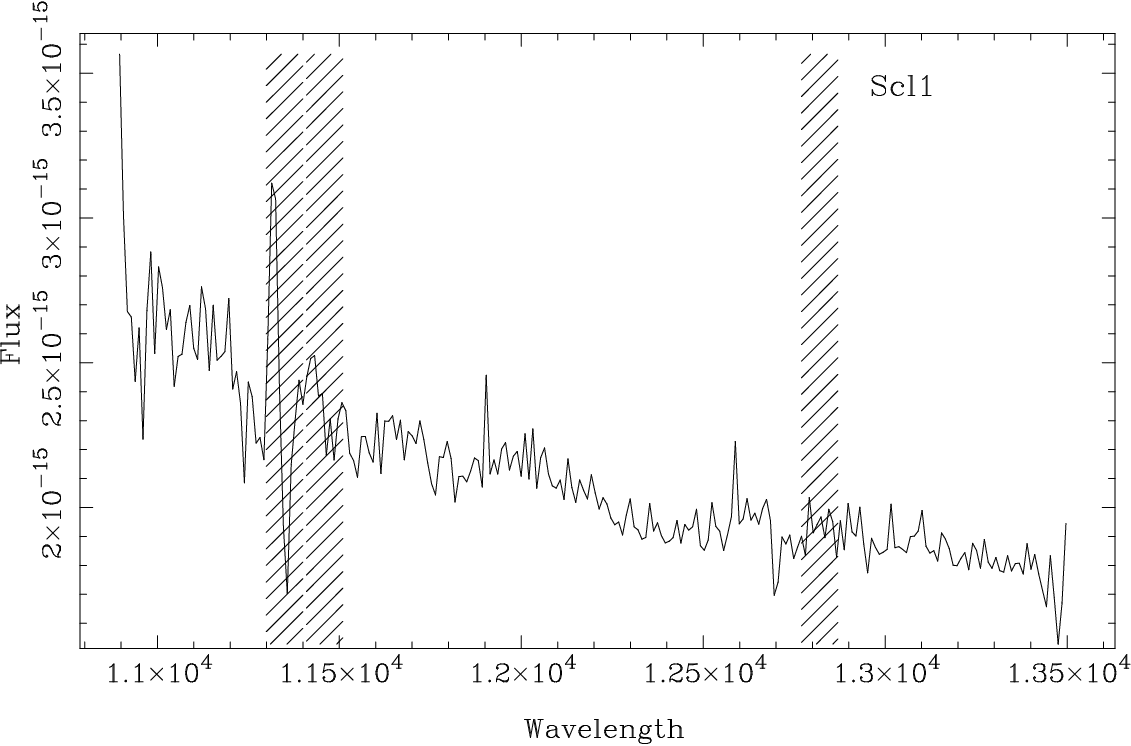}} 
\end{minipage}
\begin{minipage}{0.48\textwidth}
\resizebox{\hsize}{!}{\includegraphics{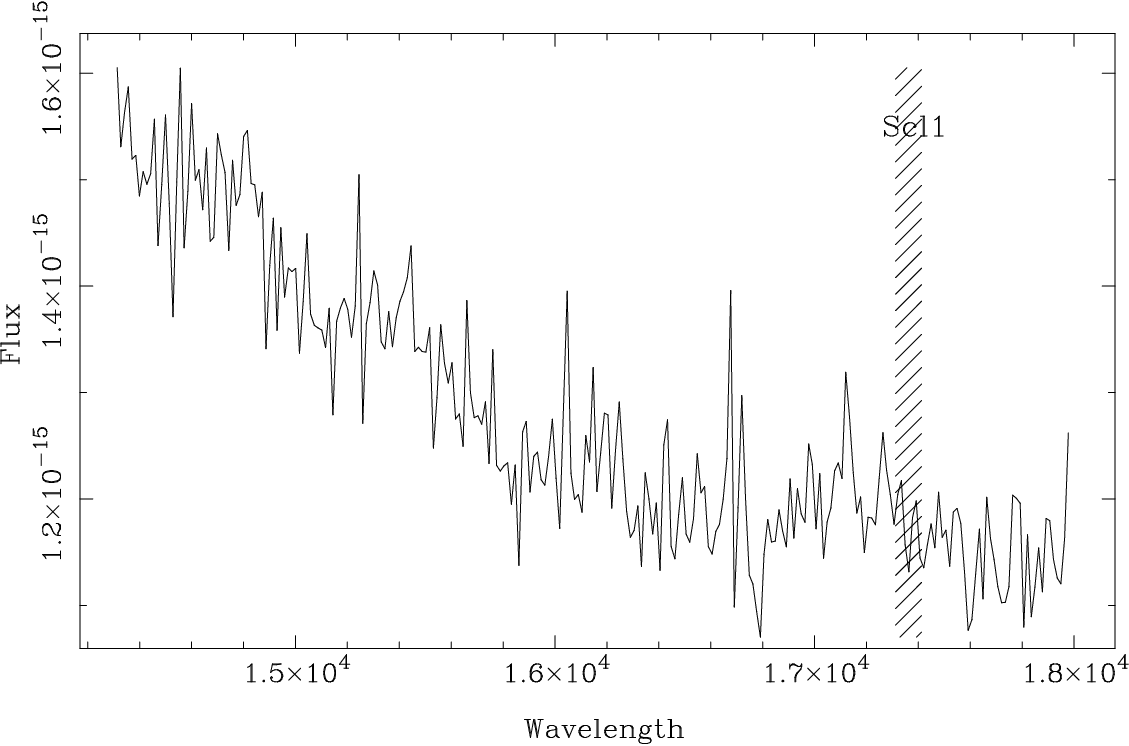}} 
\end{minipage}

\begin{minipage}{0.48\textwidth}
\resizebox{\hsize}{!}{\includegraphics{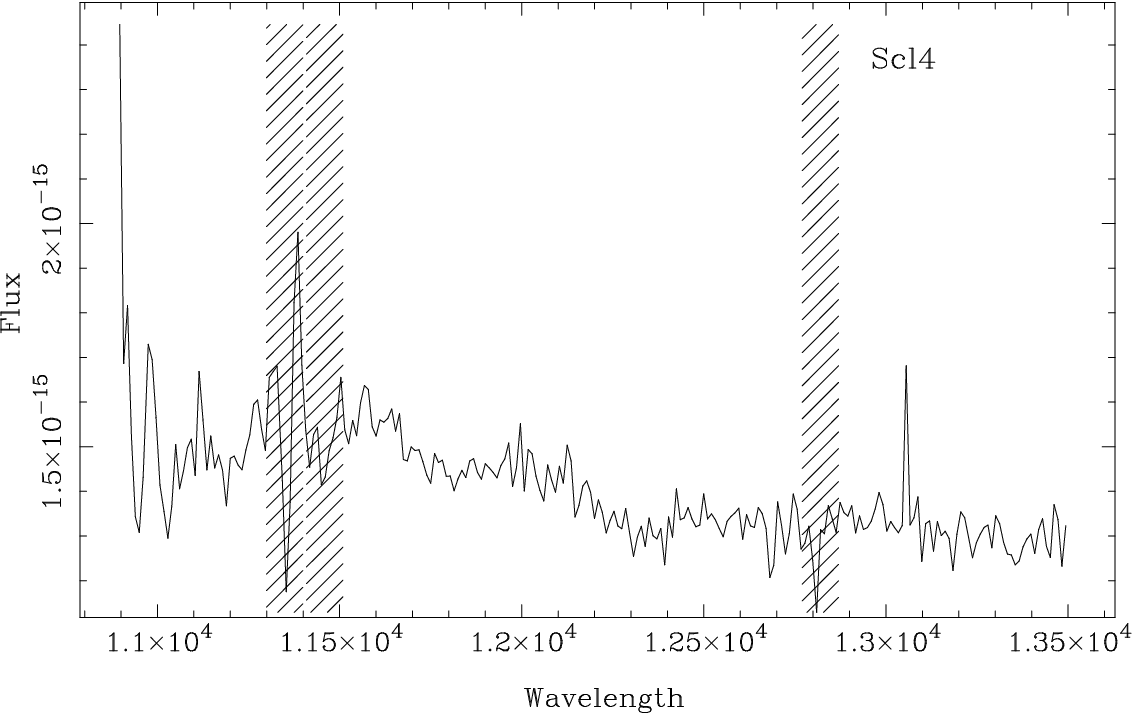}} 
\end{minipage}
\begin{minipage}{0.48\textwidth}
\resizebox{\hsize}{!}{\includegraphics{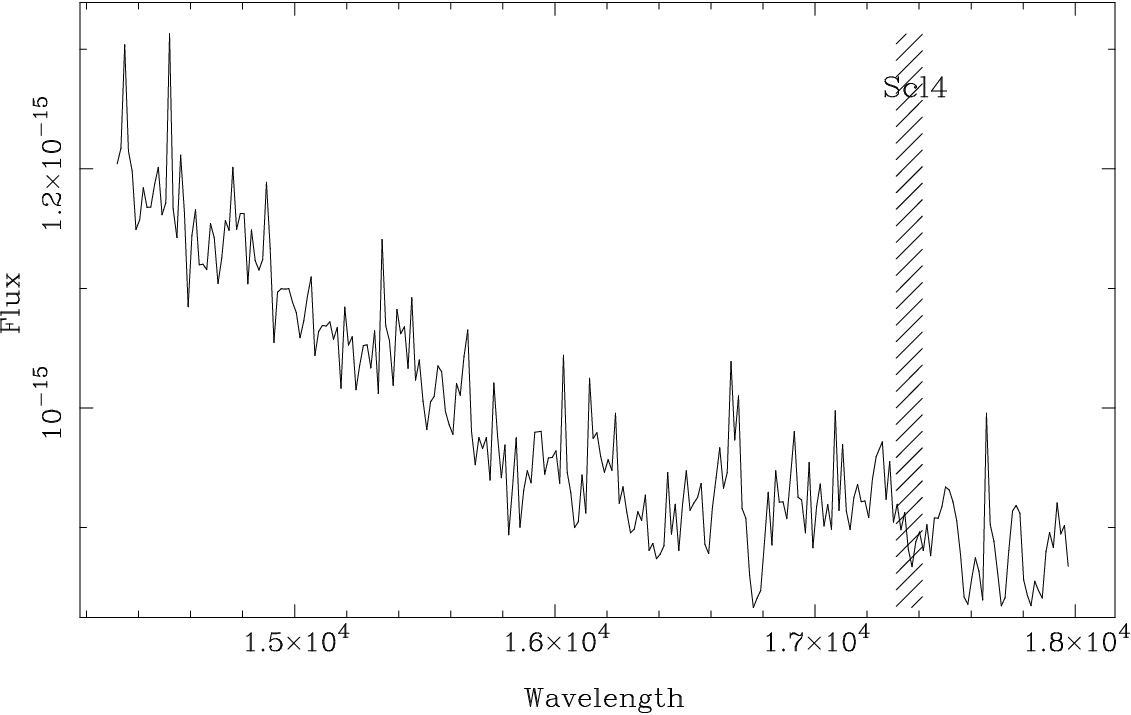}} 
\end{minipage}

\end{figure*}
}

\onlfig{6}{
\begin{figure*}[H]

\begin{minipage}{0.48\textwidth}
\resizebox{\hsize}{!}{\includegraphics{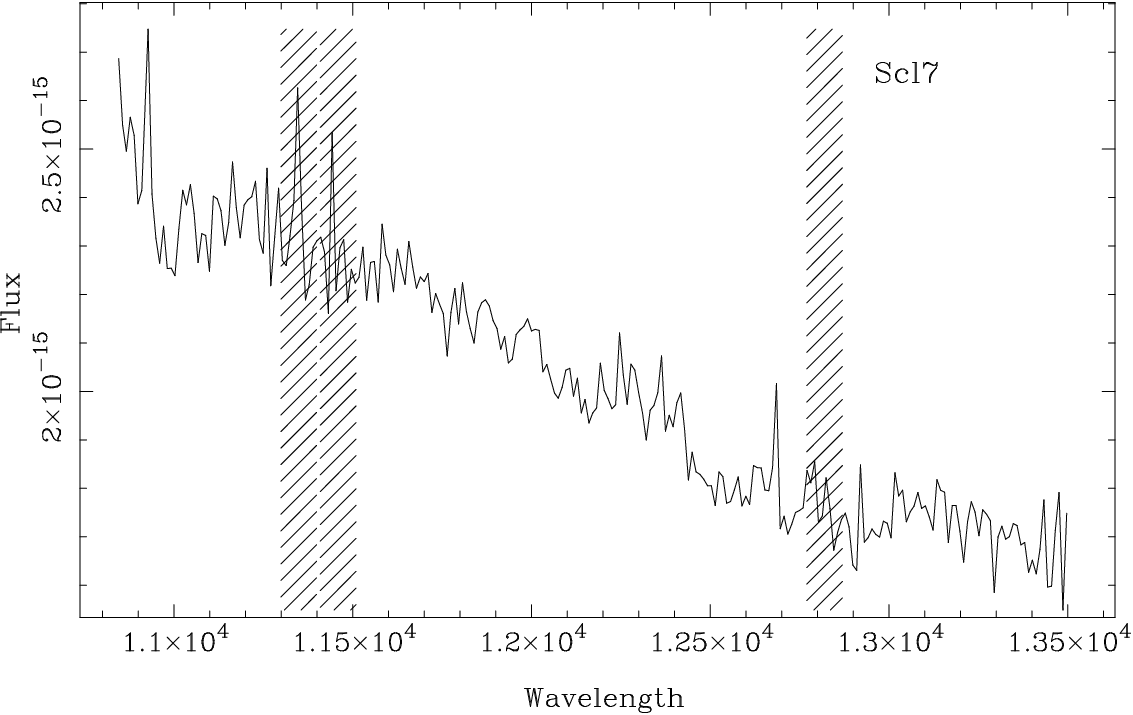}} 
\end{minipage}
\begin{minipage}{0.48\textwidth}
\resizebox{\hsize}{!}{\includegraphics{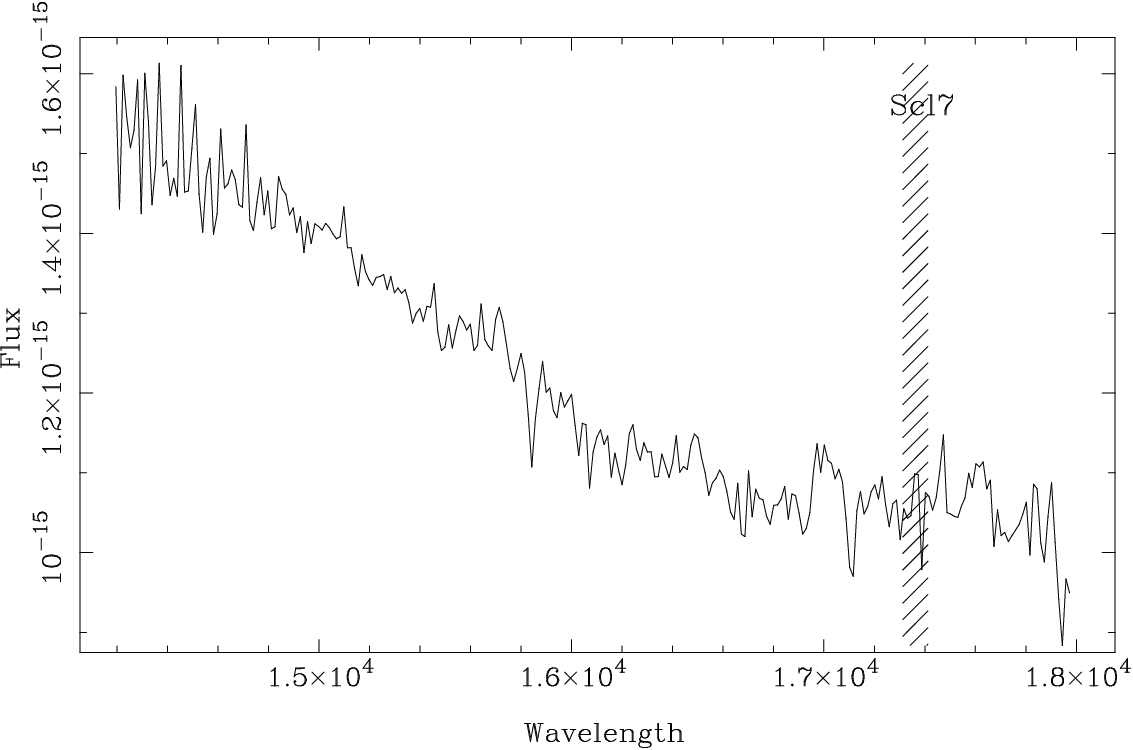}} 
\end{minipage}

\begin{minipage}{0.48\textwidth}
\resizebox{\hsize}{!}{\includegraphics{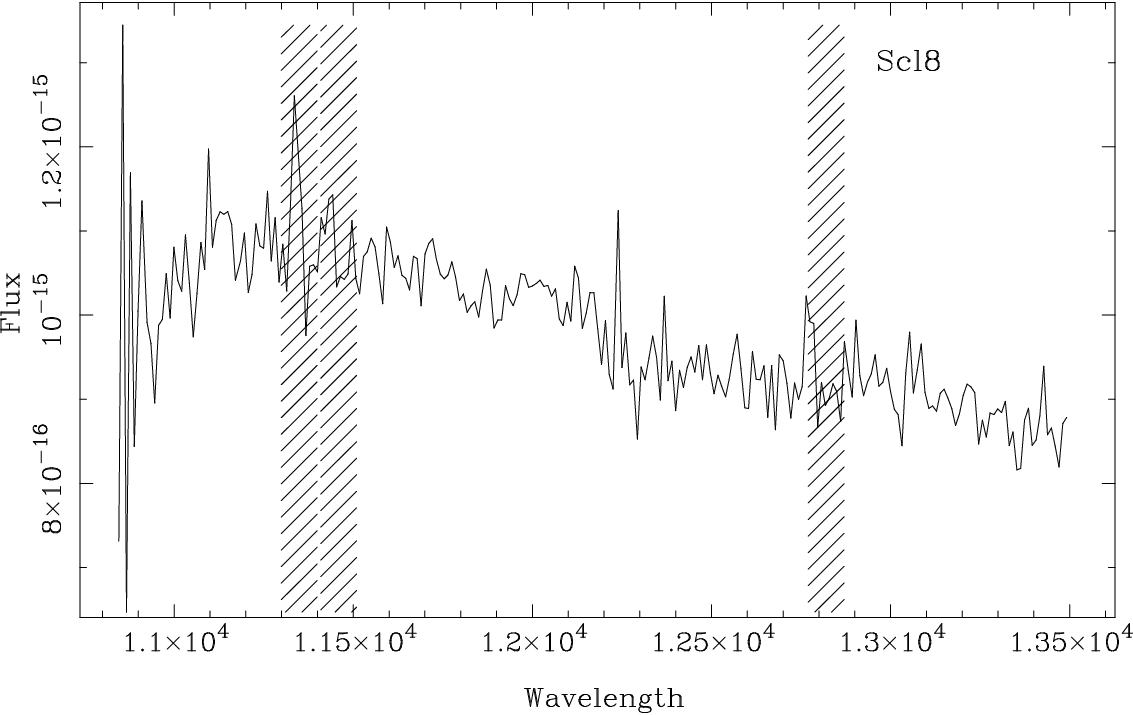}} 
\end{minipage}
\begin{minipage}{0.48\textwidth}
\resizebox{\hsize}{!}{\includegraphics{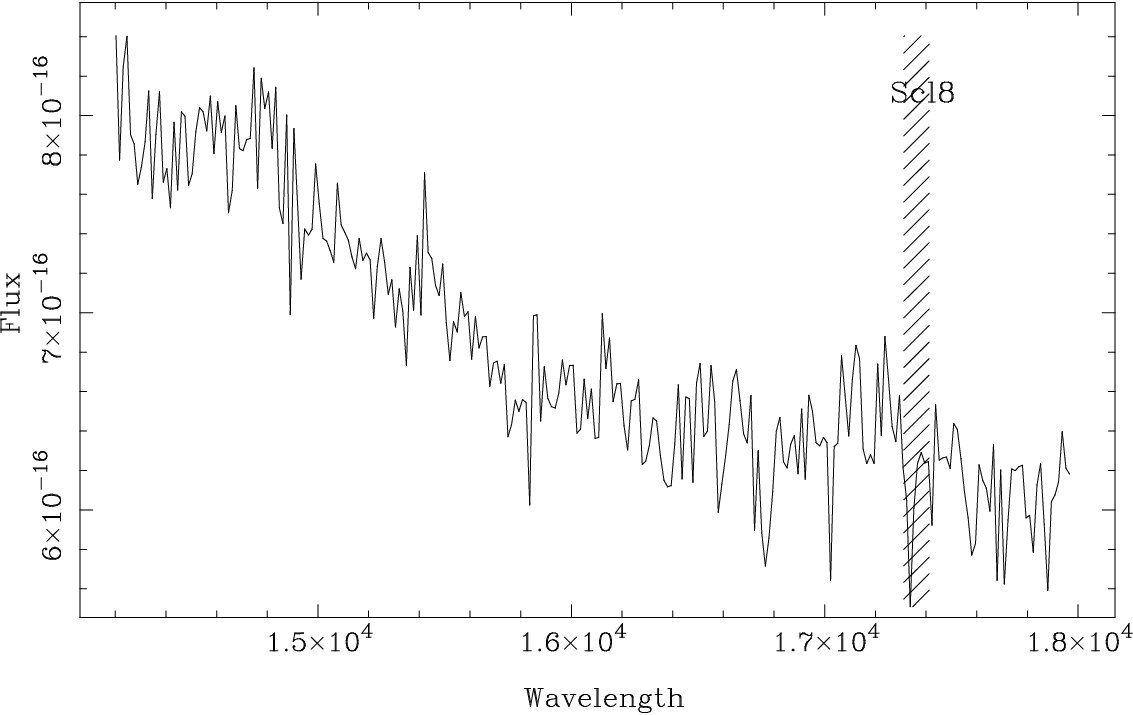}} 
\end{minipage}

\end{figure*}
}

\end{document}